\documentclass[11pt,aps,prd,preprintnumbers,superscriptaddress,nofootinbib]{revtex4}
\usepackage{bm,latexsym,amsmath,amssymb,amsfonts,mathrsfs}
\usepackage{xcolor}
\allowdisplaybreaks[1]
\usepackage{tikz}
\usetikzlibrary{decorations.pathmorphing}
\usepackage{array}[=2016-10-06]
\usepackage{hyperref}
\definecolor{toclink}{HTML}{1F4E79}
\hypersetup{
    colorlinks=true,
    citecolor=cyan,
}
\newcommand*{\D}{{\rm d}}
\newcommand*{\mpl}{M_{\rm Pl}}
\newenvironment{manuscripttableblock}[1]{
  \begin{minipage}{#1}
  \renewcommand{\baselinestretch}{1}\small
  \renewcommand{\arraystretch}{1.18}
  \setlength{\parindent}{0pt}
  \setlength{\abovecaptionskip}{0pt}
  \setlength{\belowcaptionskip}{6pt}
  \centering
}{\end{minipage}}
\newcommand{\manuscripttablenote}[1]{
  \par\vspace{5pt}{\footnotesize\raggedright #1\par}
}
\begin{document}

\title{Black Holes and Quasinormal Modes in Scalar-Tensor Effective Field Theory}

\author{Shin'ichi Hirano}
\email[Email: ]{shirano``at''hyogo-u.ac.jp}
\affiliation{Hyogo University of Teacher Education, 942-1 Shimokume, Kato, Hyogo 673-1494, Japan}
\affiliation{Department of Physics, Rikkyo University, Toshima, Tokyo 171-8501, Japan}

\author{Masashi Kimura}
\affiliation{Department of Information, Artificial Intelligence and Data Science,
Daiichi Institute of Technology, Tokyo 110-0005, Japan}
\affiliation{Department of Physics, Rikkyo University, Toshima, Tokyo 171-8501, Japan}

\author{Masahide Yamaguchi}
\affiliation{Cosmology, Gravity and Astroparticle Physics Group, Center for Theoretical Physics of the Universe, Institute for Basic Science, Daejeon 34126, Korea}
\affiliation{Department of Physics, Institute of Science Tokyo,
2-12-1 Ookayama, Meguro-ku, Tokyo 152-8551, Japan}
\affiliation{Department of Physics and IPAP, Yonsei University, 50 Yonsei-ro, Seodaemun-gu, Seoul 03722, Korea}

\preprint{RUP-26-35}

\begin{abstract}
We study black hole solutions and quasinormal modes in a covariant scalar-tensor effective field theory with scalar-curvature operators up to mass dimension eight, focusing on black hole perturbations and their implications for ringdown.
We construct static and slowly rotating black hole solutions and examine how these interactions affect the black hole geometry.
We then analyze the coupled scalar and gravitational perturbations of the static solutions, examining how different interactions modify the quasinormal mode spectrum.
For a light scalar, we give an example in which the background remains Schwarzschild but the coupled mode frequencies depend on the scalar mass.
In the heavy scalar limit, we show how integrating out the scalar generates higher curvature terms in an effective field theory of pure gravity.
We also investigate how mixing between scalar and gravitational modes affects the tensor polarization of black hole ringdown and how a linear conformal coupling to matter allows these modes to contribute to the breathing polarization.
\end{abstract}
\maketitle

\clearpage
\setcounter{tocdepth}{2}
\begingroup
\renewcommand{\baselinestretch}{1}\small
\hypersetup{linktoc=page,linkcolor=toclink}
\tableofcontents
\endgroup
\clearpage

\section{Introduction}\label{sec: introduction}

Observations of gravitational waves from binary black hole (BH) coalescences probe general relativity (GR) in strong-field, dynamical regimes. Current observations are consistent with GR and place increasingly stringent constraints on possible deviations~\cite{LVK:2026GWTC4Tests}. During ringdown, the frequencies and damping times of the remnant's quasinormal modes (QNMs) probe its geometry and the dynamics of its perturbations. Recent observations, including GW250114, have advanced black hole spectroscopy through measurements of multiple QNMs~\cite{LVK:2025GW250114Spectroscopy}.

In effective field theory (EFT), corrections to GR are described by additional terms in the action~\cite{Burgess:2007EFT}. Once the low-energy degrees of freedom and symmetries are specified, their interactions can be organized by mass dimension, with a finite number of independent operators at each order. The corrections to black hole solutions and their QNM frequencies can then be expanded in the operator coefficients within the regime of validity of the EFT.

When no additional light degrees of freedom are present, a pure gravity EFT, in which the metric is the only dynamical field, describes corrections to GR through higher curvature interactions~\cite{Endlich:2017tqa}. Representative studies have examined their effects on BH geometries~\cite{Cardoso:2018ptl,Cano:2022wwo} and gravitational QNMs~\cite{deRham:2020ejn,Cano:2021myl}, including those of rotating BHs~\cite{Cano:2024ezp,Boyce:2026rnn}. Applications to gravitational wave observations include constraints from inspiral and ringdown signals, as in Refs.~\cite{Sennett:2019bpc,Silva:2022srr,Maenaut:2024oci}.

If an additional scalar field is light on BH scales, it must be included as a dynamical degree of freedom. Such fields are motivated both by particle physics and by scalar-tensor extensions of GR~\cite{Berti:2015TestsGR}. Ultralight bosons, for example, can be probed through superradiance around rotating BHs, as reviewed in Refs.~\cite{Brito:2020Superradiance,Brito:2025Ultralight}. Nonminimal scalar-curvature couplings can generate scalar profiles and modify both BH geometries and their perturbations. Conversely, a scalar that is heavy compared with the relevant curvature and perturbation scales can be integrated out, yielding an effective description in terms of the metric alone.

BH solutions and perturbations with scalar-curvature interactions have been investigated in scalar Gauss--Bonnet (sGB) and dynamical Chern--Simons (dCS) gravity, for example in Refs.~\cite{Molina:2010fb,BlazquezSalcedo:2016EdGBQNM}. At weak coupling, both rotating black hole backgrounds and their QNM frequencies have been calculated~\cite{YunesPretorius:2009SlowRotation,Pani:2011SlowRotation,Wagle:2022dCSQNM,Chung:2024sGBMETRICS}. Covariant EFT analyses incorporating scalar-curvature couplings have also addressed rotating BH backgrounds~\cite{Cano:2019ore} and coupled perturbations with independent linear and quadratic sGB interactions~\cite{Nie:2024pby}. In our previous work, we studied a scalar coupled linearly to both the Gauss--Bonnet and Pontryagin densities and showed how these interactions induce gravitational mode mixing~\cite{Hirano:2024pmk}. The effects of a finite scalar mass on BH backgrounds and their perturbations have also been studied in Refs.~\cite{Molina:2010fb,Macedo:2018txb,Staykov:2022MassiveGBQNM,vanGemeren:2024MassiveGB}.

More general scalar-tensor interactions have been organized into covariant EFT operator bases, for example in Refs.~\cite{Weinberg:2008hq,Ruhdorfer:2020GravityEFT,Babichev:2025dpb}. A complementary approach constructs an EFT directly for perturbations around BH backgrounds with scalar hair, as in Ref.~\cite{Franciolini:2019BHEFT}. Here we ask which scalar-curvature couplings enter the BH metric and the QNM frequencies, and which curvature operators they induce when the scalar is heavy.

In this paper, we study BHs in a covariant scalar-tensor effective field theory (STEFT) containing scalar-curvature interactions of both parities through mass dimension eight. We restrict attention to the operators relevant to our calculation and set independent pure curvature couplings to zero. For a massless scalar, we construct static and slowly rotating backgrounds to second order in the couplings and first order in spin. To the same order in the couplings, we derive the coupled master equations on the static backgrounds and compute the fundamental quadrupolar QNM frequencies of the scalar-led and gravitational-led branches. We then expand in the mass of a light scalar in dCS gravity supplemented by a quadratic scalar-curvature coupling. In the limit of large scalar mass, we match the backgrounds and low-energy gravitational QNMs in the dCS and sGB examples to the induced pure gravity EFT and obtain the first subleading corrections in the inverse scalar mass. We also match the covariant actions at tree level through mass dimension sixteen, including higher powers of the couplings so that each scalar-curvature interaction contributes.

The paper is organized as follows. Section~\ref{sec: basic equations} constructs the scalar-tensor EFT and derives its equations of motion. Section~\ref{sec: BH} presents static and slowly rotating black hole solutions, and Sec.~\ref{sec: perturbation} studies perturbations and QNMs of the static solutions. Sections~\ref{sec: light mass limit} and~\ref{sec: integrating out} treat light and heavy scalars, respectively. Section~\ref{sec: discussion} examines the complementary information provided by these results and the excitation and detection of scalar-led modes, and Sec.~\ref{sec: summary} summarizes our findings. Technical details and the regime of validity are collected in the Appendices.

We use $c=\hbar=1$ and $\mpl^2=(8\pi G)^{-1}$, while black hole masses and angular momenta are expressed in geometrized units.

\section{Basic equations}\label{sec: basic equations}

\subsection{Construction of scalar-tensor effective field theory around black holes}\label{sec: theory}

Our four-dimensional STEFT contains a real scalar field $\phi$ and the metric $g_{\mu\nu}$. We construct the Lagrangian covariantly to relate black hole solutions and their perturbations to the underlying interactions. For representative covariant constructions of scalar-tensor EFTs, see Refs.~\cite{Weinberg:2008hq,Ruhdorfer:2020GravityEFT,Babichev:2025dpb}.

We study perturbative corrections to asymptotically flat, stationary black hole solutions of vacuum GR. At zeroth order, the action consists of the Einstein--Hilbert term and a free scalar of mass $m$. We take the zeroth-order scalar field to vanish, in which case the background metric is Ricci flat. In the static, spherically symmetric case, this choice is consistent with the no-hair results of Refs.~\cite{Heusler:1992ss,Bekenstein:1995un,Sudarsky:1995zg}.

We construct covariant interactions, including parity-violating terms, through mass dimension eight, with dimensions counted using the canonically normalized scalar. Field redefinitions, integration by parts, curvature identities, and commutators of covariant derivatives reduce the operators to an independent basis, as described in App.~\ref{sec: construction of Lagrangian}. The coefficients are defined in this reduced basis, including the higher-order terms induced by the field redefinitions at the working order~\cite{Criado:2018sdb}. For simplicity, we omit scalar self-interactions other than the mass term. We denote by $\Lambda$ the common scale suppressing the EFT interactions and attach a formal bookkeeping parameter $\epsilon$ to each interaction. The power of $\epsilon$ counts how many times the interactions enter a given term in the expansion. We hold the couplings, $\Lambda$, and $m$ fixed. Perturbativity is determined by the couplings together with their scale suppression. The couplings need not be numerically of order unity and can compensate for different powers of $\Lambda$, allowing these interactions to have comparable small effects.

We calculate the black hole solutions and their linear perturbations through ${\cal O}(\epsilon^2)$. Although each interaction carries one power of $\epsilon$, its effects on the fields can enter at higher orders. Interactions linear in $\phi$ source a background field $\bar\phi={\cal O}(\epsilon)$. This counting restricts the operators needed here. For scalar-curvature terms proportional to $\phi^n$ with $n\geq2$, the contribution quadratic in scalar perturbations scales as $\epsilon\bar\phi^{n-2}={\cal O}(\epsilon^{n-1})$. Thus, scalar-curvature interactions cubic in $\phi$ can affect linear perturbations at ${\cal O}(\epsilon^2)$, whereas terms with $n\geq4$ do not contribute at this order. Together with the truncation at mass dimension eight, these restrictions select a subset of the scalar-tensor operators classified through six derivatives in Ref.~\cite{Babichev:2025dpb}. The resulting Lagrangian is
\begin{align}
    \frac{{\cal L}}{\mpl^2}
    &= {\cal L}_{\rm free} +\epsilon\left({\cal L}^{\rm ST}_{\rm EFT} +{\cal L}^{\rm Curv}_{\rm EFT}\right).\label{eq: total Lagrangian}
\end{align}
Here $\epsilon$ is a bookkeeping parameter, not an additional physical coupling, and can be set to unity after the perturbative expansion. The terms in Eq.~(\ref{eq: total Lagrangian}) are
\begin{align}
    {\cal L}_{\rm free}
    &= \frac{1}{2}R -\frac{1}{2}(\nabla\phi)^2 -\frac{1}{2}m^2\phi^2,\label{eq: free Lagrangian}
\\
    {\cal L}^{\rm ST}_{\rm EFT}
    &= \frac{1}{\Lambda^2}\Big[(b_1 \phi +c_1\phi^2 +d_1\phi^3) {\cal C}_2 +(b_2 \phi +c_2\phi^2 +d_2\phi^3)\tilde{\cal C}_2\Big]
\notag\\
    &\quad 
     +\frac{1}{\Lambda^4}\Big[(e_1\phi+f_1\phi^2 ) {\cal C}_3 +(e_2\phi+f_2\phi^2 ) \tilde{\cal C}_3
    +f_3(\nabla\phi)^2{\cal C}_2 
    +f_4(\nabla\phi)^2\tilde{\cal C}_2 \Big],
    \label{eq: STEFT Lagrangian}
\\
    {\cal L}^{\rm Curv}_{\rm EFT} &= 
    \frac{1}{\Lambda^4}\Big(g_1{\cal C}_3
    +g_2\tilde{\cal C}_3\Big)
    +\frac{1}{\Lambda^6}\Big[h_1 ({\cal C}_2)^2
    +h_2{\cal C}_2 \tilde{\cal C}_2
    +h_3(\tilde{\cal C}_2)^2
    \Big],
    \label{eq: curvEFT Lagrangian}
\end{align}
with
\begin{align}
    {\cal C}_2 &=  R_{\mu\nu\rho\sigma}R^{\mu\nu\rho\sigma},~
    \tilde{\cal C}_2 = \tilde{R}_{\mu\nu\rho\sigma}R^{\mu\nu\rho\sigma},~ 
    {\cal C}_3 = R_{\mu\nu\rho\sigma}R^{\rho\sigma\alpha\beta}R_{\alpha\beta}{}^{\mu\nu},~
    \tilde{\cal C}_3 = \tilde{R}_{\mu\nu\alpha\beta}R^{\alpha\beta \rho\sigma}R_{\rho\sigma}{}^{\mu\nu}.
\end{align}
Here, $b_1,\ldots,h_3$ are coupling constants. We define ${\tilde R}^{\mu\nu}{}_{\rho\sigma}=\varepsilon^{\mu\nu\alpha\beta}R_{\alpha\beta\rho\sigma}$ and $\varepsilon^{\mu\nu\rho\sigma}=\epsilon^{\mu\nu\rho\sigma}/\sqrt{-g}$, where $\epsilon^{\mu\nu\rho\sigma}$ is the antisymmetric symbol with $\epsilon^{0123}=-1$. Our definition of the dual Riemann tensor does not include a factor of $1/2$. We factor out $\mpl^2$ from the action and divide the canonically normalized scalar by the Planck mass, so that $\phi$ is dimensionless.

The two lines of Eq.~(\ref{eq: STEFT Lagrangian}) give the scalar-tensor interactions with four and six derivatives in our basis, respectively. The two leading interactions linear in $\phi$ are the scalar Gauss--Bonnet (sGB) coupling~\cite{Kanti:1995vq}\footnote{The standard sGB coupling is $\phi(R_{\mu\nu\rho\sigma}^2 -4R_{\mu\nu}^2 +R^2)$. Here its Ricci-dependent terms are removed by field redefinitions, with induced contributions absorbed into the coefficients of the reduced basis at the working order (App.~\ref{sec: field redefinitions}).} and the dynamical Chern--Simons (dCS) coupling~\cite{Alexander:2009tp}. Such couplings are motivated, for example, by low-energy string effective actions~\cite{Metsaev:1987zx,Antoniadis:1992sa}. The $c_1,c_2,f_1$, and $f_2$ terms break the scalar shift symmetry and can generate curvature-dependent mass terms. The $f_3$ and $f_4$ interactions contain derivatives of the scalar and preserve the shift symmetry. The $d_1$ and $d_2$ terms contain one more scalar field than the $c_1$ and $c_2$ terms, while the $e_1$ and $e_2$ terms contain one more curvature tensor than the $b_1$ and $b_2$ terms.

${\cal L}_{\rm EFT}^{\rm Curv}$ contains only curvature tensors and no scalar field. Such interactions belong to a pure gravity EFT, whose only field is the metric. Their effects on black hole geometries and QNMs have been studied, for example, in Refs.~\cite{Cardoso:2018ptl,deRham:2020ejn,Cano:2021myl}. To focus on scalar-curvature interactions, we set the independent coefficients in ${\cal L}_{\rm EFT}^{\rm Curv}$ to zero in the following calculations.

For the classical black hole solutions considered here, the black hole radius provides the characteristic length scale. A nonzero scalar mass introduces the additional scale $m^{-1}$. We use a common interaction scale $\Lambda$. For an alternative normalization with separate interaction scales, see App.~\ref{sec: LEEFT}.

We restrict the calculation to the branch perturbatively connected to GR and do not address spontaneously scalarized black holes (for a review, see Ref.~\cite{Doneva:2022ewd}). Although the onset of scalarization can be identified as a linear instability, determining the resulting scalarized black hole requires a nonlinear treatment. The regime of validity of the EFT is discussed in App.~\ref{sec: validity and causality}.

\subsection{Equations of motion}

We continue to set ${\cal L}_{\rm EFT}^{\rm Curv}=0$. Varying the action with respect to $\phi$, we obtain the scalar field equation
\begin{align}
    \left(\Box - m^2\right)\phi      
    &= -\frac{\epsilon}{\Lambda^2}\Big[
    b_1 {\cal C}_2  
    +b_2\tilde{\cal C}_2 
    +2\phi\left(c_1 {\cal C}_2  
    +c_2\tilde{\cal C}_2 \right)
    +3\phi^2\left(d_1 {\cal C}_2  
    +d_2\tilde{\cal C}_2 \right)\Big]
\notag\\
    &\quad -\frac{\epsilon}{\Lambda^4}\Big[
    e_1 {\cal C}_3 
    +e_2 \tilde{\cal C}_3 
    +2\phi\left(
    f_1 {\cal C}_3 
    +f_2 \tilde{\cal C}_3 \right)
    -2f_3\nabla_\alpha(\nabla^\alpha\phi {\cal C}_2) -2f_4\nabla_\alpha(\nabla^\alpha\phi \tilde{\cal C}_2)\Big].
    \label{eq: scalarEoM}
\end{align}

Variation with respect to the metric gives
\begin{align}
    G_{\mu\nu}
    &= \nabla_{\mu}\phi\nabla_{\nu}\phi
    -\frac{1}{2}g_{\mu\nu}\left[(\nabla\phi)^2 +m^2\phi^2\right]
    -2\epsilon{\cal E}_{\mu\nu}.
    \label{eq: gravityEoM}
\end{align}
The first two terms on the right-hand side arise from the free scalar Lagrangian, while ${\cal E}_{\mu\nu}$ is defined from ${\cal L}_{\rm EFT}^{\rm ST}$ without its overall factor $\epsilon$. Using the formulas for the variation of the curvature terms of Refs.~\cite{Padmanabhan:2011ex,Bueno:2016ypa} and including the explicit dependence on scalar derivatives, we obtain
\begin{align}
    {\cal E}_{\mu\nu} &= 
    P_{\mu \rho\sigma\alpha}R_{\nu}{}^{\rho\sigma\alpha} 
    -\frac{1}{2}g_{\mu\nu}{\cal L}^{\rm ST}_{\rm EFT} 
    -2\nabla^{\rho}\nabla^{\sigma} P_{\mu\rho\sigma\nu} 
    +\frac{1}{2}\frac{\partial {\cal L}^{\rm ST}_{\rm EFT}}{\partial \nabla^{\mu}\phi}\nabla_{\nu}\phi,\label{eq: E-tensor}
\\
    P_{\mu\nu\rho\sigma} &:= 
    \frac{\partial {\cal L^{\rm ST}_{\rm EFT}}}{\partial R^{\mu\nu\rho\sigma}}
\notag\\
    &= \frac{1}{\Lambda^2}\Big[
    2b_1\phi R_{\mu\nu\rho\sigma} 
    +b_2\phi (\tilde{R}_{\mu\nu\rho\sigma}+\tilde{R}_{\rho\sigma\mu\nu})
    +2c_1\phi^2 R_{\mu\nu\rho\sigma} 
    +c_2\phi^2 (\tilde{R}_{\mu\nu\rho\sigma}+\tilde{R}_{\rho\sigma\mu\nu})
\notag\\
    &\quad\quad +2d_1\phi^3 R_{\mu\nu\rho\sigma} 
    +d_2\phi^3 (\tilde{R}_{\mu\nu\rho\sigma}+\tilde{R}_{\rho\sigma\mu\nu})\Big]
\notag\\
    &\quad +\frac{1}{\Lambda^4} \Big[
    3e_1\phi R_{\mu\nu\alpha\beta}R^{\alpha\beta}{}_{\rho\sigma} 
    +e_2\phi(R_{\mu\nu\alpha\beta}\tilde{R}^{\alpha\beta}{}_{\rho\sigma} +R_{\mu\nu\alpha\beta}\tilde{R}_{\rho\sigma}{}^{\alpha\beta} +R_{\rho\sigma\alpha\beta}\tilde{R}_{\mu\nu}{}^{\alpha\beta})
\notag\\    
    &\quad\quad +3f_1\phi^2 R_{\mu\nu\alpha\beta}R^{\alpha\beta}{}_{\rho\sigma} 
    +f_2\phi^2(R_{\mu\nu\alpha\beta}\tilde{R}^{\alpha\beta}{}_{\rho\sigma} +R_{\mu\nu\alpha\beta}\tilde{R}_{\rho\sigma}{}^{\alpha\beta} +R_{\rho\sigma\alpha\beta}\tilde{R}_{\mu\nu}{}^{\alpha\beta})
\notag\\    
    &\quad\quad +2f_3(\nabla\phi)^2R_{\mu\nu\rho\sigma} 
    +f_4(\nabla\phi)^2(\tilde{R}_{\mu\nu\rho\sigma}+\tilde{R}_{\rho\sigma\mu\nu})\Big].
\end{align}
The tensor $P_{\mu\nu\rho\sigma}$ collects the coefficients of the curvature variation in $\delta{\cal L}_{\rm EFT}^{\rm ST}$. Integrating this variation by parts gives the term involving derivatives of $P_{\mu\nu\rho\sigma}$ in Eq.~(\ref{eq: E-tensor}). The last term in Eq.~(\ref{eq: E-tensor}) accounts for the explicit dependence on scalar derivatives and arises here from the $f_3$ and $f_4$ interactions.

\section{Black hole solutions with a massless scalar field}\label{sec: BH}

In this section, we obtain black hole solutions with a massless scalar field, $m=0$, for which the perturbative background can be calculated analytically.

\subsection{Static spherically symmetric black hole}\label{sec: static BH}

First, we consider corrections to the static, spherically symmetric Schwarzschild solution. We use the Schwarzschild radius $r_{\rm g}=2M$, where $M$ is the black hole mass in geometrized units. The expansion concerns the coupling-weighted interaction strengths on this scale, rather than the powers of $\Lambda^{-1}$ alone. For example, the linear scalar-curvature interactions with four and six derivatives are controlled by $b_i/(\Lambda^2r_{\rm g}^2)$ and $e_i/(\Lambda^4r_{\rm g}^4)$, respectively. Their relative sizes depend on the couplings as well as on $\Lambda r_{\rm g}$. We therefore introduce the dimensionless parameters
\begin{align}
    \epsilon_{2} := \frac{1}{\Lambda^2 r_{\rm g}^2},~ \epsilon_{4} := \frac{1}{\Lambda^4 r_{\rm g}^4}
    .\label{eq: epsilon}
\end{align}
These parameters alone do not determine the perturbative order. We assume that the dimensionless combinations $b_i\epsilon_2$ and $e_i\epsilon_4$ are small. Both $\epsilon b_i\epsilon_2$ and $\epsilon e_i\epsilon_4$ are first order in $\epsilon$. The same convention applies to $c_i\epsilon_2$, $d_i\epsilon_2$, and $f_j\epsilon_4$.
For a static, spherically symmetric spacetime, we write
\begin{align}
     ds^2 &= \bar{g}_{\mu\nu}dx^\mu dx^\nu 
     = -A(r)dt^2 +\frac{1}{B(r)}dr^2 +r^2(d\theta^2 +\sin^2{\theta}d\varphi^2),
     \label{eq: BG metric}
\\
    \phi &= \bar{\phi}(r).\label{eq: BG scalar field}
\end{align}
We choose the zeroth-order scalar to vanish.\footnote{A nonzero zeroth-order constant $\phi_0$ would generate the cubic curvature terms
\begin{equation*}
    \frac{\epsilon\mpl^2}{\Lambda^4}\left[(e_1\phi_0+f_1\phi_0^2){\cal C}_3+(e_2\phi_0+f_2\phi_0^2)\tilde{\cal C}_3\right]
\end{equation*}
in the Lagrangian. These terms can be absorbed into the pure curvature coefficients when shifting the scalar origin. Here we choose $\phi_0=0$ and ${\cal L}_{\rm EFT}^{\rm Curv}=0$.}

The scalar profile starts at ${\cal O}(\epsilon)$ and modifies the metric at ${\cal O}(\epsilon^2)$~\cite{Hirano:2024pmk}. We therefore write
\begin{align}
    A(r) &= 1 -\frac{r_{\rm g}}{r} +\epsilon^2 a_{\rm 2nd}(r),\label{eq: static metric expansion}\\
    B(r) &= 1 -\frac{r_{\rm g}}{r} +\epsilon^2 b_{\rm 2nd}(r),\\
    \bar{\phi}(r) &= 0 +\epsilon\, \pi_{\rm 1st}(r) +\epsilon^2 \pi_{\rm 2nd}(r).\label{eq: BG ansatz}
\end{align}
We impose regularity at the horizon and $\bar\phi\to0$ at spatial infinity, normalize the time coordinate by $A(\infty)=1$, and keep the asymptotic mass $M$ fixed. Substituting the ansatz into Eqs.~(\ref{eq: scalarEoM}) and~(\ref{eq: gravityEoM}), we determine the corrections order by order in $\epsilon$. The parity-odd invariants $\tilde{\cal C}_2$ and $\tilde{\cal C}_3$ vanish on Schwarzschild, so the first-order scalar profile is sourced only by the parity-even interactions with coefficients $b_1$ and $e_1$. We obtain
\begin{align}
    \pi_{\rm 1st}(r) &= b_1\epsilon_{2}\pi^{(b_1)}_{\rm 1st} +e_1\epsilon_{4}\pi^{(e_1)}_{\rm 1st},\label{eq: 1st pi}
\\
    \pi^{(b_1)}_{\rm 1st} &= 
    \frac{2 r_{\rm g} (6 r^2+3 r r_{\rm g}+2 r_{\rm g}^2)}{3 r^3},
\\
    \pi^{(e_1)}_{\rm 1st} &= 
    \frac{r_{\rm g}(60 r^5+30 r^4 r_{\rm g}+20 r^3 r_{\rm g}^2+15 r^2 r_{\rm g}^3+12 r r_{\rm g}^4+10 r_{\rm g}^5)}{30 r^6}.
\end{align}
The first-order scalar profile and its nonminimal curvature couplings generate the metric corrections at ${\cal O}(\epsilon^2)$. These corrections depend on $b_1^2$, $b_1e_1$, and $e_1^2$, and are given by\footnote{In pure gravity EFTs~\cite{Cardoso:2018ptl}, metric corrections can arise at first order in $\epsilon$.}
\begin{align}   
    a_{\rm 2nd}(r) &= b_1^2\epsilon_{2}^2a^{(b_1b_1)}_{\rm 2nd} +b_1e_1\epsilon_2\epsilon_4 a^{(b_1e_1)}_{\rm 2nd} +e_1^2\epsilon_{4}^2a^{(e_1e_1)}_{\rm 2nd},\label{eq: 2nd a}
\\
     b_{\rm 2nd}(r) &= b_1^2\epsilon_{2}^2b^{(b_1b_1)}_{\rm 2nd} +b_1e_1\epsilon_2\epsilon_4 b^{(b_1e_1)}_{\rm 2nd} +e_1^2\epsilon_{4}^2b^{(e_1e_1)}_{\rm 2nd},\label{eq: 2nd b}
\end{align}
where
\begin{align}
    a^{(b_1b_1)}_{\rm 2nd} &= 
    \frac{2r_{\rm g}^3}{15 r^7}(10r^4+130 r^3 r_{\rm g}+33 r^2 r_{\rm g}^2+24 r r_{\rm g}^3-50 r_{\rm g}^4),
\\    
    a^{(b_1e_1)}_{\rm 2nd} &=
    \frac{r_{\rm g}^3}{315r^{10}}( 420r^7 +2940r^6r_{\rm g} +882r^5r_{\rm g}^2 +672r^4r_{\rm g}^3 +2700r^3r_{\rm g}^4 +13410r^2r_{\rm g}^5
\notag\\
    &\quad +6965rr_{\rm g}^6 +2240r_{\rm g}^7),
\\
    a^{(e_1e_1)}_{\rm 2nd} &=
    \frac{r_{\rm g}^3}{13860r^{13}}(4620r^{10} +4620r^9r_{\rm g}+ 4158r^8r_{\rm g}^2 +3696r^7r_{\rm g}^3 +50820r^6r_{\rm g}^4 +288090r^5r_{\rm g}^5 
\notag\\
    &\quad    
    +147455r^4r_{\rm g}^6 +99792r^3r_{\rm g}^7 +75600r^2r_{\rm g}^8 +60900rr_{\rm g}^9 +25410r_{\rm g}^{10}),
\\
    b^{(b_1b_1)}_{\rm 2nd} &= \frac{2r_{\rm g}^2}{15 r^7}(60r^5+30 r^4r_{\rm g} +260 r^3r_{\rm g}^2 +15r^2r_{\rm g}^3 +12 rr_{\rm g}^4 -230r_{\rm g}^5),
\\    
    b^{(b_1e_1)}_{\rm 2nd} &=
    \frac{r_{\rm g}^2}{315 r^{10}}(2520 r^8+1260 r^7 r_{\rm g}+5880 r^6 r_{\rm g}^2+630 r^5 r_{\rm g}^3+504 r^4 r_{\rm g}^4+166740 r^3 r_{\rm g}^5
\notag\\ 
    &\quad -55080 r^2 r_{\rm g}^6-16065 r r_{\rm g}^7-76160 r_{\rm g}^8),
\\
    b^{(e_1e_1)}_{\rm 2nd} &= 
    \frac{r_{\rm g}^2}{13860 r^{13}}(27720 r^{11}+13860 r^{10} r_{\rm g}+9240 r^9 r_{\rm g}^2+6930 r^8 r_{\rm g}^3 +5544 r^7 r_{\rm g}^4+3663660 r^6 r_{\rm g}^5
\notag\\
    &\quad
    -1215720 r^5 r_{\rm g}^6-356895 r^4 r_{\rm g}^7-154000 r^3 r_{\rm g}^8-77616 r^2 r_{\rm g}^9-41832 r r_{\rm g}^{10} -1115730 r_{\rm g}^{11}).
\end{align}
Solving $A(r_{\rm H})={\cal O}(\epsilon^3)$ for the horizon radius gives 
\begin{align}
    r_{\rm H} &= r_{\rm g}\left[ 1-\epsilon^2\left(b_{1}^2 \epsilon_{2}^2\frac{98}{5} +b_{1} e_{1}\epsilon_{2} \epsilon_{4}\frac{30229}{315} +e_{1}^2\epsilon_{4}^2\frac{765161}{13860}\right)\right].\label{eq: static horizon radius}
\end{align}
At second order in $\epsilon$, the $c_1,f_1$, and $f_3$ interactions act on the first-order scalar profile on the Schwarzschild background. They therefore enter the scalar solution in combination with $b_1$ or $e_1$, but do not affect the metric through ${\cal O}(\epsilon^2)$. We find
\begin{align}
    \pi_{\rm 2nd}(r) &= 
    \epsilon_2^2 b_1c_1 \pi^{(b_1c_1)}_{\rm 2nd} +\epsilon_2\epsilon_4\Big[b_1f_1 \pi^{(b_1f_1)}_{\rm 2nd} +b_1f_3 \pi^{(b_1f_3)}_{\rm 2nd} +c_1e_1 \pi^{(c_1e_1)}_{\rm 2nd}\Big] 
\notag\\    
    &\quad +\epsilon_4^2\Big[e_1f_1 \pi^{(e_1f_1)}_{\rm 2nd} +e_1f_3 \pi^{(e_1f_3)}_{\rm 2nd}\Big],
\end{align}
where
\begin{align}
    \pi^{(b_1c_1)}_{\rm 2nd} &= \frac{2r_{\rm g}}{225 r^6}(4380 r^5+2190 r^4 r_{\rm g}+1460 r^3 r_{\rm g}^2+1095 r^2 r_{\rm g}^3+336 r r_{\rm g}^4+100 r_{\rm g}^5),
\\
    \pi^{(b_1f_1)}_{\rm 2nd} &= \frac{r_{\rm g}}{79380 r^9}(1847160 r^8+923580 r^7 r_{\rm g}+615720 r^6 r_{\rm g}^2+461790 r^5 r_{\rm g}^3+369432 r^4 r_{\rm g}^4
\notag\\
    &\quad +307860 r^3 r_{\rm g}^5+263880 r^2 r_{\rm g}^6+94815 r r_{\rm g}^7+31360 r_{\rm g}^8),
\\
    \pi^{(b_1f_3)}_{\rm 2nd} &= \frac{4r_{\rm g}^7(72 r^2+63 r r_{\rm g}+56 r_{\rm g}^2)}{21 r^9},
\\
    \pi^{(c_1e_1)}_{\rm 2nd} &= \frac{r_{\rm g}}{79380 r^9}(1847160r^8+923580 r^7 r_{\rm g}+615720 r^6 r_{\rm g}^2+461790 r^5 r_{\rm g}^3+178920 r^4 r_{\rm g}^4
\notag\\
    &\quad +85596 r^3 r_{\rm g}^5+43128 r^2 r_{\rm g}^6+20727 r r_{\rm g}^7+7840 r_{\rm g}^8),
\\
    \pi^{(e_1f_1)}_{\rm 2nd} &= \frac{r_{\rm g}}{96049800 r^{12}}(1380650040r^{11}+690325020 r^{10} r_{\rm g}+460216680 r^9 r_{\rm g}^2+345162510r^8 r_{\rm g}^3
\notag\\
    &\quad +276130008 r^7 r_{\rm g}^4+230108340 r^6 r_{\rm g}^5+197235720 r^5 r_{\rm g}^6+90252855 r^4 r_{\rm g}^7+48208160 r^3r_{\rm g}^8
\notag\\
    &\quad +26311824 r^2 r_{\rm g}^9+13441680 r r_{\rm g}^{10}+5336100 r_{\rm g}^{11}),
\\
    \pi^{(e_1f_3)}_{\rm 2nd} &= \frac{2r_{\rm g}^7}{1155 r^{12}}(3960 r^5+3465 r^4 r_{\rm g}+3080 r^3 r_{\rm g}^2+2772 r^2 r_{\rm g}^3+2520 r r_{\rm g}^4+2310 r_{\rm g}^5).
\end{align}
The same procedure gives metric corrections at higher orders. Here we retain terms through ${\cal O}(\epsilon^2)$ in Eqs.~(\ref{eq: 2nd a}) and~(\ref{eq: 2nd b}).

\subsection{Slowly rotating black hole}\label{sec: rotating BH}

Following earlier slow-rotation constructions~\cite{PaniCardoso:2009SlowRotation,YunesPretorius:2009SlowRotation,Konno:2009SlowRotation,Pani:2011SlowRotation}, we consider stationary, axisymmetric $\ell=1$ perturbations of the static solutions, Eqs.~(\ref{eq: BG metric}) and~(\ref{eq: BG scalar field}), at first order in rotation. Parity-violating interactions couple the odd and even metric components and the scalar component within this sector~\cite{Hirano:2024pmk}. We impose horizon regularity and asymptotic flatness in coordinates nonrotating at infinity. We take the scalar ansatz
\begin{align}
    \phi = \bar{\phi} +\delta\phi(r)\cos{\theta},\label{eq: ansatz rot scalar}
\end{align}
where $\bar\phi$ is the static profile in Eq.~(\ref{eq: BG scalar field}). The metric ansatz is
\begin{align}
    g_{\mu\nu}dx^\mu dx^\nu &= -A(r)[1-H_0(r)\cos{\theta}]dt^2 +2\cos{\theta}H_1(r) dtdr 
    -2h_0(r)\sin^2{\theta}dtd\varphi 
\notag\\
    &\quad +\frac{1+H_2(r)\cos{\theta}}{B(r)}dr^2 +r^2(d\theta^2+\sin^2{\theta}d\varphi^2),\label{eq: ansatz rot metric}
\end{align}
where $A(r)$ and $B(r)$ are the metric functions of the static, spherically symmetric solution in Eq.~(\ref{eq: BG metric}).
Expanding the perturbations in $\epsilon$ as $G = G^{\rm 0th} +\epsilon G^{\rm 1st} +\epsilon^2G^{\rm 2nd}~(G = \delta\phi,~h_0,~H_0,~H_1,~H_2)$ and substituting the ansatz~(\ref{eq: ansatz rot scalar}) and~(\ref{eq: ansatz rot metric}) into the equations of motion~(\ref{eq: scalarEoM}) and~(\ref{eq: gravityEoM}), we obtain the slowly rotating solution. We define the dimensionful spin parameter by $\chi:=J/M$, where $J$ is the asymptotic angular momentum in geometrized units, and assume $|\chi|/M\ll1$. The mass and angular momentum are held fixed in the $\epsilon$ expansion, with $g_{t\varphi}=-2M\chi\sin^2\theta/r+{\cal O}(r^{-2})$ at infinity. At ${\cal O}(\epsilon^0\chi)$, we recover the Kerr solution to first order in spin, with
\begin{align}
    \delta\phi^{\rm 0th} = 0,~
    h_0^{\rm 0th} = \frac{r_{\rm g}}{r}\chi,~
    H_0^{\rm 0th} = H_1^{\rm 0th} = H_2^{\rm 0th} = 0.
\end{align}
Keeping only terms linear in $\chi$, we determine the corrections at successive orders in $\epsilon$ from the lower-order solutions. At ${\cal O}(\epsilon\chi)$, the rotational scalar dipole is sourced by $b_2$ and $e_2$, in contrast to the static scalar profile sourced by $b_1$ and $e_1$. The metric receives no correction at this order. Rotational EFT corrections to the metric first appear at ${\cal O}(\epsilon^2\chi)$.
We define $\delta g_{\mu\nu}=g_{\mu\nu}-\bar g_{\mu\nu}$ relative to the EFT-corrected static metric and $f(r)=1-r_{\rm H}/r$, using the static horizon radius in Eq.~(\ref{eq: static horizon radius}). Thus, $\delta g_{\mu\nu}$ includes the Kerr frame-dragging term. Through second order in $\epsilon$ and first order in $\chi$, its nonzero components are
\begin{align}
    \frac{1}{\chi f(r)\cos{\theta}}\delta g_{tt} 
    &= 
    \epsilon^2\Big\{b_1b_2\epsilon_2^2\delta g_{tt}^{(b_1b_2)} +\epsilon_2\epsilon_4 \Big[b_1e_2\delta g_{tt}^{(b_1e_2)} +b_2e_1\delta g_{tt}^{(b_2e_1)}\Big]
    +e_1e_2\epsilon_4^2\delta g_{tt}^{(e_1e_2)}\Big\},\label{eq: rotating tt correction}
\\
     \frac{f(r)}{\chi\cos{\theta}}\delta g_{rr} &= \epsilon^2\Big\{b_{1} b_{2}\epsilon_{2}^2\delta g_{rr}^{(b_1b_2)} +\epsilon_2\epsilon_4 \Big[b_1e_2\delta g_{rr}^{(b_1e_2)} +b_2e_1\delta g_{rr}^{(b_2e_1)}\Big]
    +e_1e_2\epsilon_4^2\delta g_{rr}^{(e_1e_2)}\Big\},\label{eq: rotating rr correction}
\\ 
    \frac{1}{\chi\sin^2{\theta}}\delta g_{t\varphi} &= -\frac{r_{\rm g}}{r}+\epsilon^2\Big\{\epsilon_{2}^2\Big[b_{1}^2 \delta g_{t\varphi}^{(b_1b_1)} +b_{2}^2\delta g_{t\varphi}^{(b_2b_2)}\Big] 
    +\epsilon_{2}\epsilon_{4}\Big[b_{1}e_1\delta g_{t\varphi}^{(b_1e_1)} +b_{2}e_2\delta g_{t\varphi}^{(b_2e_2)}\Big]
\notag\\
    &\quad +\epsilon_{4}^2\Big[e_{1}^2\delta g_{t\varphi}^{(e_1e_1)} +e_{2}^2\delta g_{t\varphi}^{(e_2e_2)}\Big]\Big\}.\label{eq: rotating frame dragging}
\end{align}
The explicit functions $\delta g_{tt}^{(ij)},\delta g_{rr}^{(ij)}$, and $\delta g_{t\varphi}^{(ij)}$, where $ij$ labels the coupling products, are provided in the supplementary Mathematica file~\cite{STEFT:Mathematica}. The same file gives the scalar dipole $\delta\phi(r)$ through ${\cal O}(\epsilon^2\chi)$. The correction to frame dragging involves products within the parity-even sector $(b_1,e_1)$ or within the parity-odd sector $(b_2,e_2)$. In contrast, the $tt$ and $rr$ corrections involve the mixed products $b_1b_2,b_1e_2,b_2e_1$, and $e_1e_2$, and are odd under the equatorial reflection $\theta\to\pi-\theta$. Thus, the reflection asymmetry at this order requires couplings of both parities, so neither $b_2$ nor $e_2$ alone produces the asymmetry. Equatorial reflection asymmetry in rotating black hole geometries was demonstrated in higher curvature EFT in Ref.~\cite{Cardoso:2018ptl}, and has also been studied in higher-derivative theories and their multipole moments~\cite{Cano:2019ore,Cano:2022wwo}. The theory with both sGB and dCS couplings provides a scalar-tensor example~\cite{Hirano:2024pmk}.

\section{Perturbations of static, spherically symmetric black holes with a massless scalar field}\label{sec: perturbation}

In this section, we consider perturbations with $\ell\geq2$ around the static, spherically symmetric black hole solutions of Sec.~\ref{sec: static BH}, Eqs.~(\ref{eq: BG metric}) and~(\ref{eq: BG scalar field}). We work to first order in the perturbation amplitudes and through ${\cal O}(\epsilon^2)$ in the EFT couplings.
We write the perturbed scalar field as
\begin{align}
    \phi(t,r,\theta) = \bar{\phi}(r) +e^{-i\omega t}\Pi(r)Y_{\ell0}(\theta).
    \label{eq: scalar field}
\end{align}
We decompose the metric perturbation into parts of odd and even parity in Regge--Wheeler gauge~\cite{Regge:1957td},
\begin{align}
    g_{\mu\nu} &= \bar{g}_{\mu\nu} +h_{\mu\nu}^{-} + h_{\mu\nu}^{+},
\notag\\
    h_{\mu\nu}^{-}dx^\mu dx^\nu &= 2e^{-i\omega t}\sin{\theta}\partial_{\theta}Y_{\ell0}d\varphi(h_0 dt+h_1dr),
\notag\\
    h_{\mu\nu}^{+}dx^\mu dx^\nu &=
    e^{-i\omega t}Y_{\ell0}\{AH_0dt^2+2H_1dtdr+B^{-1}H_2dr^2+r^2K(d\theta^2+\sin^2{\theta}d\varphi^2)\}.
    \label{eq: metric perturbations}
\end{align}
Here, $h_{\mu\nu}^{-}$ and $h_{\mu\nu}^{+}$ denote the metric perturbations of odd and even parity, respectively. Spherical symmetry allows us to set the azimuthal harmonic number to zero without changing the frequencies.

\subsection{Master equations}\label{sec: master eqs}

Following the procedure of Ref.~\cite{Hirano:2024pmk}, we eliminate higher derivatives iteratively using the lower-order equations, retaining all induced corrections through ${\cal O}(\epsilon^2)$. For the QNM calculation, we redefine the master variables perturbatively in $\epsilon$ while preserving the Schr\"odinger form of the coupled equations.\footnote{We choose the coefficients of the first radial derivatives in this redefinition so that $\omega^2$ multiplies the identity matrix in the usual tortoise coordinate defined by $dr/F$, where $F=\sqrt{AB}$.}
We then use the radial coordinate $r_*$ defined by $dr_*=dr/f$, where $f=1-r_{\rm H}/r$ (see, e.g., App.~B of Ref.~\cite{McManus:2019ulj}).
The resulting equations have a symmetric potential matrix and take the form
\begin{align} 
    &\frac{\D^2{\bf\Psi}}{\D r_{*}^2} 
    +\Big[\omega^2{\bf c}^2
    -f(r)({\bf V}_{\rm GR} +{\bf V}_{\rm EFT})\Big]{\bf\Psi} =0,\label{eq: master eq}
\end{align}
where 
\begin{align}
    {\bf \Psi} &= \begin{pmatrix}
    \Psi_{\rm s} \\
    \Psi_{-} \\
    \Psi_{+} \\
\end{pmatrix},~
    dr_{*} = \frac{dr}{f},
\\
    {\bf c}^2
    &= \frac{1}{Z(r)^2}{\bf I},~
    Z(r) = \frac{F}{f},~
   {\bf V}_{\rm GR} = 
   \begin{pmatrix}
    V_{\rm KG}
    & 0
    & 0 \\
    0 
    & V_{\rm RW}
    & 0\\
    0 
    & 0
    & V_{\rm Z}
    \end{pmatrix},\label{eq: speed and VGR}
\\
   {\bf V}_{\rm EFT} &= 
   \epsilon 
   \begin{pmatrix}
    V^{\rm s}_{\rm 1st} 
    & V^{\rm s-}_{\rm 1st} 
    & V^{\rm s+}_{\rm 1st} \\
     V^{\rm s-}_{\rm 1st} 
    & 0 
    & 0\\
    V^{\rm s+}_{\rm 1st} 
    & 0
    & 0
\end{pmatrix}
    +\epsilon^2 
   \begin{pmatrix}
    V^{\rm s}_{\rm 2nd} 
    & V^{\rm s-}_{\rm 2nd} 
    & V^{\rm s+}_{\rm 2nd} \\
    V^{\rm s-}_{\rm 2nd} 
    & V^{\rm -}_{\rm 2nd} 
    & V^{\rm -+}_{\rm 2nd}\\
    V^{\rm s+}_{\rm 2nd} 
    & V^{\rm -+}_{\rm 2nd} 
    & V^{\rm +}_{\rm 2nd}
\end{pmatrix}\label{eq: EFT potential}
.
\end{align}
Here, ${\bf I}$ is the $3\times 3$ identity matrix. The indices denote the scalar ($\mathrm{s}$), odd-parity gravitational ($-$), and even-parity gravitational ($+$) components. The factors $\epsilon_2$ and $\epsilon_4$ are included in the correction functions and potentials, while $\epsilon$ alone labels the interaction order. The master variables reduce to the scalar, Regge--Wheeler, and Zerilli variables in the GR limit~\cite{Regge:1957td,Zerilli:1970se}. In Eq.~(\ref{eq: master eq}), the common factor $f$ multiplies the potentials. We use the Schwarzschild forms of $V_{\rm KG},V_{\rm RW},V_{\rm Z}$ with horizon radius $r_{\rm H}$ and include the remaining potential corrections in $\mathbf V_{\rm EFT}$. The common coefficient ${\bf c}^2$ of $\omega^2$ depends on the choice of radial coordinate and master variables.

The scalar-curvature interactions in Eq.~(\ref{eq: STEFT Lagrangian}) produce couplings between the perturbation components, represented by the off-diagonal entries of $\mathbf V_{\rm EFT}$. At first order in $\epsilon$, the scalar potential is modified and the scalar perturbation couples to the gravitational perturbations. At second order, the diagonal gravitational potentials also receive corrections, and the odd and even gravitational perturbations couple directly. The interactions modify the perturbation equations both directly and through the corrected background. Mixing between scalar and gravitational perturbations can therefore be present even when the background scalar vanishes, as in the dCS example considered below.

Compared with the model with both sGB and dCS couplings of Ref.~\cite{Hirano:2024pmk}, the additional $c_1$ and $f_1$ interactions produce curvature-dependent scalar mass terms, while $f_3$ modifies the scalar kinetic sector. After the redefinition of the master variables, all three contribute to $V^{\rm s}_{\rm 1st}$. The $e_1$ and $e_2$ interactions mix the scalar perturbation with the even and odd gravitational perturbations, respectively, at first order in $\epsilon$, in the same way as $b_1$ and $b_2$. At second order, $d_1$ contributes to the diagonal potential in the scalar sector through the background scalar profile, while $c_2,f_2$, and $f_4$ can mix the scalar and odd gravitational perturbations.

\subsection{Quasinormal frequencies}

We compute the frequency shifts of the fundamental $\ell=2$ modes through ${\cal O}(\epsilon^2)$ at fixed BH mass $M$. We require the perturbations to be purely ingoing at the horizon and purely outgoing at spatial infinity. We truncate the series in Leaver's method~\cite{Leaver:1985ax}, following Sec.~IV of Ref.~\cite{Hirano:2024pmk} for the coupled system. Perturbative approaches to QNM frequency shifts from potential deformations are developed in Refs.~\cite{Cardoso:2019mqo,McManus:2019ulj,Hatsuda:2023geo}.

The scalar and two gravitational sectors are generally coupled. We follow the scalar-led and gravitational-led branches from their respective Schwarzschild frequencies at $\epsilon=0$. In the Schwarzschild limit, the scalar QNM frequency differs from the gravitational one, while the gravitational modes of odd and even parity are isospectral~\cite{Chandrasekhar:1975qn}. We therefore treat the two cases separately when solving for the frequency corrections.
We use the series ansatz
\begin{align}
    {\bf \Psi } &= e^{iq r_\ast}\left(\frac{r}{r_{\rm H}}\right)^pf^{n}\sum^{\infty}_{j=0}{\bf a}_{j}f^{j},\label{eq: QNM series ansatz}
\end{align}
where ${\bf a}_j$ are three-component coefficient vectors and $f=1-r_{\rm H}/r$. Matching the prefactor to the ingoing behavior at the horizon and the outgoing phase at infinity gives
\begin{align}
    q = \omega,~p = i\omega(r_{\rm g} -r_{\rm H}),~n = -i\omega r_{\rm H}\left[1+\frac{1}{Z(r_{\rm H})}\right].
\end{align}
Here $Z(r_{\rm H})$ denotes the limit of $F/f$ as $r\to r_{\rm H}^{+}$. Substituting the ansatz (\ref{eq: QNM series ansatz}) into Eq.~(\ref{eq: master eq}) gives recurrence relations for the coefficients ${\bf a}_j$. Iterating the recurrence gives ${\bf a}_j={\bf M}_j(\omega){\bf a}_0$. At a sufficiently large truncation order $j_{\max}$, we impose
\begin{align}
    \det{{\bf M}_{j_{\max}}}(\omega) = 0, \label{eq: QNM condition}
\end{align}
to determine the approximate QNM frequencies. The prefactor imposes the ingoing boundary condition at the horizon. At finite $j_{\max}$, Eq.~(\ref{eq: QNM condition}) approximates the outgoing condition at infinity. We follow the roots connected to the fundamental Schwarzschild modes as $j_{\max}$ increases. We expand the frequency about the fundamental Schwarzschild value as
\begin{align}
    \omega = \frac{1}{2M}\left[\Omega_{\rm Sch} +\epsilon\,\delta\Omega_{\rm 1st} +\epsilon^2\delta\Omega_{\rm 2nd}\right],
\end{align}
where $\Omega_{\rm Sch}$ and the corrections $\delta\Omega$ are dimensionless. We substitute this expansion into Eq.~(\ref{eq: QNM condition}) and expand the resulting frequency condition to the order needed to resolve each root.
We first consider the scalar Schwarzschild frequency and then the doubly degenerate gravitational frequency.

For the nondegenerate scalar-led mode, we solve successively for $\delta\Omega^{\rm scalar\text{-}led}_{\rm 1st}$ and $\delta\Omega^{\rm scalar\text{-}led}_{\rm 2nd}$, using a linear equation at each order. The new $c_1,f_1$, and $f_3$ interactions modify the diagonal potential in the scalar sector at first order and can therefore shift the scalar-led frequency at that order, which is absent in the model of Ref.~\cite{Hirano:2024pmk}. Using $\Omega^{\rm scalar}_{\rm Sch} = 0.967287744\cdots-i 0.193517552\cdots $, we obtain
\begin{align}
    \delta\Omega^{\rm scalar\text{-}led}_{\rm 1st} &=(-0.407263-0.0256371 i) c_{1}\epsilon_2
\notag\\ 
    &\quad +\epsilon_4\Big[ (-0.162492-0.0847550i) f_{1} -(0.351501 -0.415060i)f_3\Big],\label{eq: delomega^s_1st}
\\
    \delta\Omega^{\rm scalar\text{-}led}_{\rm 2nd} &=
    \epsilon_2^2\Big[(0.184154\, -0.206331 i)c_1^2
\notag\\
    &\quad\quad +(45.2251 +4.24848 i)b_1^2 +(148.386 +20.2922 i)b_2^2
    -(5.51157 +1.44477i)b_1d_1\Big]
\notag\\
    &\quad +\epsilon_2\epsilon_4\Big[(0.353167\, -0.267907 i)c_1f_1 -(2.42348 +2.15038i)c_1f_3
\notag\\
    &\quad\quad +(56.3581 +30.2886 i)b_1e_1 +(188.161 +108.693 i)b_2e_2
    -(3.10812 +1.05472i)d_1e_1\Big]
\notag\\
    &\quad +\epsilon_4^2\Big[(0.167531\, -0.0750214 i)f_1^2
    -(1.19824 +2.01133i)f_1f_3 -(5.23752 -3.31308i)f_3^2
\notag\\
    &\quad\quad +(19.3445 +19.8789 i)e_1^2 +(58.5546 +71.1231 i)e_2^2\Big].\label{eq: scalar QNM}
\end{align}
In the potential matrix of Eq.~(\ref{eq: EFT potential}), $V^{\rm s}_{\rm 2nd}$ contributes directly to the shift in the scalar-led frequency at second order, and $V^{\rm s}_{\rm 1st}$ contributes upon iteration. The off-diagonal potentials $V^{\rm s\pm}_{\rm 1st}$ couple the two sectors. The scalar perturbation sources a gravitational perturbation at first order, which in turn sources the scalar at second order. In Eq.~(\ref{eq: scalar QNM}), terms involving products of $c_1,f_1$, and $f_3$ are grouped first within each power of $\epsilon_2$ and $\epsilon_4$, followed by the $b_i,e_i$ terms and, where present, the $d_1$ terms.

For the gravitational-led modes, the GR degeneracy requires solving a quadratic equation for the corrections in the two-dimensional gravitational subspace, as illustrated by the matrix model in App.~B of Ref.~\cite{Hirano:2024pmk}. The determinant must be expanded far enough to resolve the double root, although the master equations are needed only through ${\cal O}(\epsilon^2)$. The first-order frequency shift vanishes, and the second-order corrections depend only on $b_i,e_i$ and generally lift the GR degeneracy. Using $\Omega^{\rm grav}_{\rm Sch} = 0.74734337\cdots -i0.17792463\cdots$, we write the result as
\begin{align}
    \delta\Omega^{\rm grav\text{-}led}_{\rm 1st}=0,~
    \delta\Omega^{\rm grav\text{-}led}_{\rm 2nd}={\cal A}\pm\sqrt{{\cal B}}.
    \label{eq: grav QNM}
\end{align}
The two signs give the frequency shifts of the two gravitational-led branches. When the odd and even sectors mix, the signs cannot in general be assigned to definite parities. The common correction ${\cal A}$ is the average of the two frequency shifts, while $2\sqrt{{\cal B}}$ gives their splitting. The explicit expressions are
\begin{align}
    {\cal A} &= \epsilon_2^2 \Big[(-5.30125-2.19115 i) b_{1}^2 -(31.5063\, +16.0767 i) b_{2}^2\Big]
\notag\\
    &\quad
    +\epsilon_2\epsilon_4 \Big[(12.5300\, +6.37929 i) b_{1} e_{1} -(20.4591\, +49.8444 i) b_{2}e_{2}\Big]
\notag\\
    &\quad 
    +\epsilon_4^2 \Big[(10.3554 +7.04967 i) e_{1}^2+(6.90408 -24.8157 i) e_{2}^2\Big],\label{eq: grav QNM mean}
\\
    {\cal B} &= \epsilon^4_2 \Big[(75.8256 +47.7592 i) b_{1}^4+(734.185 +1013.03 i) b_{2}^4 \notag\\
    &\quad\quad +(532.006\, +469.859 i) b_{1}^2 b_{2}^2\Big]
\notag\\
    &\quad    
    +\epsilon^3_2\epsilon_4 \Big[(2522.56\, -152.074 i) b_{1}^3 e_{1}+(19739.2\, -445.781 i) b_{1}^2 b_{2} e_{2} \notag\\
    &\quad\quad -(9008.96\, -660.753 i) b_{1} b_{2}^2 e_{1}
    -(313.482\, -3798.65 i) b_{2}^3 e_{2}\Big]
\notag\\
    &\quad    
    +\epsilon^2_2\epsilon_4^2 \Big[b_{1}^2 ((15493.0 -11171.6 i) e_{1}^2+(55808.8\, -45383.1 i) e_{2}^2) \notag\\
    &\quad\quad +(13824.8\, +349.810 i) b_{1} b_{2} e_{1} e_{2}
\notag\\
    &\quad\quad         
    +b_{2}^2 (-(6369.96+824.604 i) e_{1}^2-(3298.84\, -3381.26 i) e_{2}^2)\Big]
\notag\\
    &\quad    
    +\epsilon_2\epsilon_4^3 \Big[(19647.1 -12946.8 i) b_{1} e_{1}^3+(79750.4 -53026.4 i) b_{1} e_{1} e_{2}^2 \notag\\
    &\quad\quad+(411.957 -1023.09 i) b_{2} e_{1}^2 e_{2}
    -(2756.35-327.157 i) b_{2} e_{2}^3\Big]
\notag\\
    &\quad
    +\epsilon^4_4 \Big[(6880.45\, -3606.54 i) e_{1}^4+(28145.1\, -15203.8 i) e_{1}^2 e_{2}^2 \notag\\
    &\quad\quad -(568.153\, +342.659 i) e_{2}^4\Big].\label{eq: grav QNM discriminant}
\end{align}
Setting $e_1=e_2=0$ reproduces the result for the theory with both sGB and dCS couplings of Ref.~\cite{Hirano:2024pmk}.\footnote{In Eq.~(41) of Ref.~\cite{Hirano:2024pmk}, the imaginary part of the $b_1^2$ coefficient outside the square root should be negative, as follows from the pure sGB frequencies in Eqs.~(45) and (46) there.}

The splitting of the degenerate gravitational frequency depends on the difference between the diagonal gravitational corrections, on direct mixing between the odd and even sectors, and on mixing mediated by the scalar. The quantity ${\cal B}$ in Eq.~(\ref{eq: grav QNM discriminant}) depends on all three effects, so the direct coupling between the two gravitational sectors alone does not determine the splitting. The potentials $V^{\rm s\pm}_{\rm 2nd}$ mix scalar and gravitational perturbations at second order but do not shift the frequencies at this order. Away from a resonance between scalar and gravitational modes, the leading contribution of these terms to a frequency shift also involves a first-order mixing term. The static metric and the gravitational-led frequencies both first change at second order in $\epsilon$, through different combinations of $b_i$ and $e_i$. The $c_1,f_1$, and $f_3$ interactions can shift the scalar-led frequency at first order.

\section{Expansion of quasinormal frequencies for a light scalar field}\label{sec: light mass limit}

We now extend the preceding massless analysis to a massive scalar field. A nonzero scalar mass introduces an additional frequency scale and allows for quasibound states in addition to QNMs, as studied in massive dCS gravity in Ref.~\cite{Macedo:2018txb}. The mass dependence of scalar-led and gravitational-led QNMs in this theory was studied numerically in Ref.~\cite{Molina:2010fb}.

Here we focus on the QNM branches continuously connected to the massless modes, with $\omega r_{\rm g}={\cal O}(1)$, for small scalar mass, $\alpha:=m^2r_{\rm g}^2\ll1$. For the fundamental mode with $\ell=2$, we expand the frequencies through second order in $\epsilon$ and each coefficient through second order in $\alpha$, keeping the factors of $\epsilon_2$ in the coefficients. As an example, we add the $c_1$ interaction to massive dCS gravity and use the Lagrangian
\begin{align}
    \frac{{\cal L}}{\mpl^2}
    &= \frac{1}{2}R -\frac{1}{2}(\nabla\phi)^2 -\frac{1}{2}m^2\phi^2
    +\frac{\epsilon}{\Lambda^2}\Big(
    b_2\phi \tilde{R}_{\mu\nu\rho\sigma}R^{\mu\nu\rho\sigma}
    +c_1\phi^2R_{\mu\nu\rho\sigma}^2\Big).
\end{align}
The $b_2$ interaction couples scalar and odd-parity gravitational perturbations, while $c_1$ produces a curvature-dependent scalar mass term. The Schwarzschild metric with $\bar\phi=0$ is an exact solution of this model, so the scalar mass and couplings enter only through the perturbation equations on this background. The even-parity mode decouples and obeys the unmodified Zerilli equation, so only the coupled scalar and odd gravitational equations need to be solved.

The scalar and odd-parity master equations extend the massive dCS system~\cite{Molina:2010fb,Kimura:2018nxk,Macedo:2018txb} by the $c_1$ term. With $\epsilon_2=(\Lambda r_{\rm g})^{-2}$ and the coupling constants included in the potentials as in Sec.~\ref{sec: perturbation}, the equations read
\begin{align}    
    \frac{\D^2 \Psi_{\rm s}}{\D r_{\ast}^2} 
    +\omega^2\Psi_{\rm s}
    -f(m^2 +V_{\rm KG} +\epsilon V^{\rm s}_{\rm 1st} +\epsilon^2 V^{\rm s}_{\rm 2nd})\Psi_{\rm s} -\epsilon f V^{\rm s-}_{\rm 1st}\Psi_{\rm -} &= 0,\label{eq: scalar dCS master eq}
\\
    \frac{\D^2\Psi_{\rm -}}{\D r_{\ast}^2} 
    +\omega^2\Psi_{\rm -}
    -fV_{\rm RW} \Psi_{\rm -} -\epsilon f V^{\rm s-}_{\rm 1st}\Psi_{\rm s} &= 0,
    \label{eq: odd dCS master eq}
\end{align}
where
\begin{align}
    dr_{\ast} &= \frac{dr}{f(r)},~f(r) = 1 -\frac{r_{\rm g}}{r},~r_{\rm H}=r_{\rm g},
\\
    V_{\rm KG} &= \frac{(\lambda+2)r +r_{\rm g}}{r^3},
    ~V_{\rm RW} = \frac{(\lambda+2)r-3r_{\rm g}}{r^3},
\\
    V^{\rm s}_{\rm 1st} &=
    -\frac{24 r_{\rm g}^4}{r^6}c_{1}\epsilon_2,
~
    V^{\rm s-}_{\rm 1st} = -\frac{24\sqrt{
    2\lambda(\lambda+2)} r_{\rm g}^3}{r^5}b_{2}\epsilon_2,
\\
    V^{\rm s}_{\rm 2nd} &= \frac{1152(\lambda+2)r_{\rm g}^6}{r^8}b_2^2\epsilon_2^2.
\end{align}
Here, $\lambda:=(\ell-1)(\ell+2)$.
At zeroth order in the couplings, the scalar-led frequency is that of a free massive scalar on Schwarzschild and already depends on $\alpha$, whereas the gravitational-led frequency remains the GR value. At infinity, the scalar wave number is $k=\sqrt{\omega^2-m^2}$, while the gravitational wave number is $\omega$. Using a different prefactor for each field would introduce the ratio of the two prefactors into the coupling terms. We therefore use a common prefactor, which gives recurrence relations with finitely many terms.

For the scalar-led solution, the scalar component starts at zeroth order and the gravitational component at first order in the coupling. The truncation condition ${\bf a}_{j_{\max}}=0$ selects the solution that is purely outgoing in both components as $j_{\max}$ increases (see App.~\ref{sec: prefactor and purely outgoing condition}). With the common massive scalar prefactor, the series are
\begin{align}
     \Psi_{\rm s}^{\rm scalar\text{-}led} &= e^{ik r_\ast}\left(\frac{r}{r_{\rm g}}\right)^{i\frac{m^2}{2k}r_{\rm g}}f^{-i(k+\omega)r_{\rm g}}\sum^{\infty}_{j=0}{a}^{\rm s}_{j}f^{j}, \label{eq: massive ansatz1}
\\
    \Psi_{\rm -}^{\rm scalar\text{-}led} &= \epsilon\, e^{ik r_\ast}\left(\frac{r}{r_{\rm g}}\right)^{i\frac{m^2}{2k}r_{\rm g}}f^{-i(k+\omega)r_{\rm g}}\sum^{\infty}_{j=0}{a}^{\rm -}_{j}f^{j},  \label{eq: massive ansatz2}
\\
    k &= \sqrt{\omega^2 -m^2}.
\end{align}
We choose the branch of $k$ by analytic continuation from $k=\omega$ at $m=0$. After substituting the ansatz with its mass-dependent prefactor, we expand the radial equations through order $\alpha^2$ and use the resulting recurrence relations to determine the QNM frequencies perturbatively. The scalar-led frequency receives corrections proportional to $c_1$ at first order and to $c_1^2$ and $b_2^2$ at second order. For the fundamental mode with $\ell=2$, we obtain
\begin{align}
    \omega &= \frac{1}{2M}\left[\Omega^{\rm s}_{\rm Sch}(\alpha) +\epsilon\,\delta\Omega^{\rm s}_{\rm 1st} +\epsilon^2\delta\Omega^{\rm s}_{\rm 2nd}\right],
\\
    \delta\Omega^{\rm s}_{\rm 1st} &= \epsilon_2 c_1 \big[(-0.407263 -0.0256371i) +\alpha(0.220179 +0.0578730i)
\notag\\
    &\quad\quad\quad -\alpha^2(0.0230111 +0.00941562i)\big],\label{eq: light scalar QNM first}
\\
    \delta\Omega^{\rm s}_{\rm 2nd}  &=
    \epsilon_2^2\Big\{b_2^2\big[(148.386+20.2922i) -\alpha(137.299+39.4323i) +\alpha^2(52.8071 +48.8277i)\big]
\notag\\*
    &\quad +c_1^2 \big[(0.184154 -0.206331i) +\alpha(0.112678 +0.153725i) 
\notag\\*
    &\quad\quad\quad\quad -\alpha^2(0.116819 +0.0474727i)\big]\Big\}.\label{eq: light scalar QNM second}
\end{align}
Here, $\Omega^{\rm s}_{\rm Sch}(\alpha)$ is the dimensionless QNM frequency of a free massive scalar on Schwarzschild, obtained by setting $b_2=c_1=0$. Its expansion for small scalar mass is~\cite{Hatsuda:2023geo}
\begin{align}
    \Omega^{\rm s}_{\rm Sch}(\alpha)
    &=\Omega^{\rm s}_{\rm Sch}(0)
      +\alpha(0.157816+0.0540776i)
\notag\\
    &\quad+\alpha^2(0.00442646+0.00327611i)
      +{\cal O}(\alpha^3).
    \label{eq: free massive scalar QNM}
\end{align}
Here $\Omega^{\rm s}_{\rm Sch}(0)=\Omega^{\rm scalar}_{\rm Sch}$ denotes the massless reference eigenfrequency introduced in Sec.~\ref{sec: perturbation}. This mass dependence is present even at $\epsilon=0$ and is separate from the coupling-dependent corrections above.

We next consider the branch that reduces to the odd-parity gravitational mode as $\epsilon\to0$. The even-parity frequency equals its Schwarzschild value. For the gravitational-led branch we use the massless gravitational prefactor for both components, with the scalar component starting at first order in the coupling. The series ansatz is
\begin{align}
    \Psi_{\rm s}^{\rm grav\text{-}led} 
    &= \epsilon\, e^{i\omega r_{\ast}}f^{-2i\omega r_{\rm g}}\sum^{\infty}_{j=0}{a}^{\rm s}_{j}f^{j}, \label{eq: massless ansatz1}
\\
    \Psi_{\rm -}^{\rm grav\text{-}led} 
    &= e^{i\omega r_{\ast}}f^{-2i\omega r_{\rm g}}\sum^{\infty}_{j=0}{a}^{\rm -}_{j}f^{j}. \label{eq: massless ansatz2}
\end{align}
As for the scalar-led mode, we substitute the ansatz into Eqs.~(\ref{eq: scalar dCS master eq}) and~(\ref{eq: odd dCS master eq}) to obtain the recurrence relations, then solve the resulting frequency condition perturbatively as in Sec.~\ref{sec: perturbation}. The gravitational frequency shift vanishes at first order. At second order, the correction is proportional to $b_2^2$. The $c_1$ correction to the induced scalar affects the gravitational frequency only at higher order in the couplings. For the fundamental $\ell=2$ mode, we obtain
\begin{align}
    \omega 
    &= \frac{1}{2M}\left[\Omega^{\rm grav}_{\rm Sch} +\epsilon^2\delta\Omega^{-}_{\rm 2nd}\right],
\\
    \delta\Omega^{-}_{\rm 2nd} 
    &= \epsilon_2^2 b_2^2\big[(-63.0125 -32.1533i) +\alpha(86.6187 +26.5273i) -\alpha^2(55.3579 +71.0783i)\big].\label{eq: light gravitational QNM}
\end{align}
At $\alpha=0$, the frequency corrections reproduce the scalar and odd-parity results of Eqs.~(\ref{eq: delomega^s_1st})--(\ref{eq: grav QNM discriminant}) when all couplings except $b_2$ and $c_1$ are set to zero. For $c_1=0$, these massless limits reduce to the dCS results of Ref.~\cite{Hirano:2024pmk}. For nonzero mass and $c_1=0$, the equations reduce to the massive dCS system of Ref.~\cite{Molina:2010fb}, for which Ref.~\cite{Lui:2026ultralight} computed the gravitational-led $\ell=2$ frequency numerically.

\section{Integrating out the heavy scalar and matching to the pure gravity EFT}\label{sec: integrating out}

We next study the limit of large scalar mass, $\beta:=(m^2r_{\rm g}^2)^{-1}\ll1$, and its relation to a pure gravity EFT, containing only the metric with interactions constructed from curvature tensors and their covariant derivatives. The intermediate regime $mr_{\rm g}\sim1$ lies outside the range of both the small-mass expansion of Sec.~\ref{sec: light mass limit} and the large-mass expansion used here. We first establish this correspondence through the background and perturbation equations in the dCS and sGB examples. We then derive the effective gravitational action induced by the scalar-curvature interactions in our Lagrangian and consider a Horndeski example.

We keep the multipole number $\ell$ and overtone number $n$ fixed at low values and consider frequencies with $\omega r_{\rm g}={\cal O}(1)$ in the limit $m r_{\rm g}\gg1$, so that $|\omega|/m={\cal O}(\sqrt{\beta})\ll1$. We assume a regular expansion in nonnegative integer powers of $\beta$. This low-frequency expansion includes the scalar perturbation sourced by the gravitational perturbations. Scalar modes with $|\omega|\sim m$ lie outside its range of validity. As in the preceding section, we expand in $\epsilon$ with the factors of $\epsilon_2$ in the coefficients.

\subsection{The limit of large scalar mass in dCS gravity}\label{sec: heavy dCS}
The dCS example is particularly simple because its background remains Schwarzschild. In contrast to the case of a light scalar, $mr_{\rm g}\ll1$, the scalar can be eliminated perturbatively from the low-frequency gravitational branch when $mr_{\rm g}\gg1$. Setting $c_1=0$, we rewrite Eqs.~(\ref{eq: scalar dCS master eq}) and~(\ref{eq: odd dCS master eq}) as
\begin{align}    
    \left(\frac{\D^2 }{\D r_{\ast}^2} 
    +\omega^2 -fV_{\rm KG}\right)\Psi_{\rm s} -fm^2\Psi_{\rm s}
    &= \epsilon f V^{\rm s-}_{\rm 1st}\Psi_{\rm -} +\epsilon^2 fV^{\rm s}_{\rm 2nd}\Psi_{\rm s},\label{eq: scalar dCS master eq2}
\\
    \left(\frac{\D^2}{\D r_{\ast}^2} 
    +\omega^2  -f V_{\rm RW}\right)\Psi_{\rm -}
    &= \epsilon f V^{\rm s-}_{\rm 1st}\Psi_{\rm s}.
    \label{eq: odd dCS master eq2}
\end{align}
The potentials and the perturbative field coefficients include their factors of $\epsilon_2$, as in the preceding sections. The expansions in $\epsilon$ are
\begin{align}
    \Psi_{\rm s} &= \Psi_{\rm s}^{(0)} +\epsilon \Psi_{\rm s}^{(1)} +\epsilon^2\Psi_{\rm s}^{(2)} +\cdots,
\\
    \Psi_{\rm -} &= \Psi_{\rm -}^{(0)} +\epsilon \Psi_{\rm -}^{(1)} +\epsilon^2\Psi_{\rm -}^{(2)} +\cdots.
\end{align}
For a putative scalar-led branch, $\Psi_{\rm s}^{(0)}\neq0$ and $\Psi_-^{(0)}=0$.
At ${\cal O}(\epsilon^0)$, the system reduces to
\begin{align}    
    \left[ \beta r_{\rm g}^2\left(\frac{\D^2}{\D r_{\ast}^2} +\omega^2 -f  V_{\rm KG}\right) -f \right]\Psi_{\rm s}^{(0)}
    &= 0.\label{eq: scalar field-led branch in heavy limit}
\end{align}
We expand both the scalar perturbation and the frequency in nonnegative integer powers of $\beta$ as
\begin{align}
    \Psi^{(0)}_{\rm s} &= \Psi^{(0,0)}_{\rm s} +\beta\Psi^{(0,1)}_{\rm s} +\beta^2\Psi^{(0,2)}_{\rm s} +\cdots,
\\
    \omega &= \omega^{(0,0)}_{\rm s} +\beta \omega^{(0,1)}_{\rm s} +\beta^2\omega^{(0,2)}_{\rm s} +\cdots,
\\
    \omega^{(0,0)}_{\rm s} &= {\cal O}(r_{\rm g}^{-1}).
\end{align}
At ${\cal O}(\beta^0)$, the scalar master equation gives, in the exterior region $r>r_{\rm g}$,
\begin{align}
   -f\Psi^{(0,0)}_{\rm s} =0.
\end{align}
Since $f\neq0$ in the exterior region, we obtain $\Psi_{\rm s}^{(0,0)}=0$. At ${\cal O}(\beta^1)$, the equation becomes
\begin{align}
   r_{\rm g}^2\left(\frac{\D^2}{\D r_{\ast}^2} +\omega^2 -f  V_{\rm KG}\right)\Psi_{\rm s}^{(0,0)}-f\Psi^{(0,1)}_{\rm s} =0.
\end{align}
Substituting $\Psi_{\rm s}^{(0,0)}=0$ then gives $\Psi_{\rm s}^{(0,1)}=0$. Continuing recursively gives $\Psi_{\rm s}^{(0,n)}=0$ for every nonnegative integer $n$. Thus, there is no independent scalar-led solution within this regular low-frequency expansion. Modes outside this expansion, including those with $\omega\sim m$ or nonanalytic dependence on $\beta$, are not addressed here.

On the gravitational-led branch, $\Psi_{\rm s}^{(0)}=0$ and $\Psi_-^{(0)}\neq0$. The leading odd-parity perturbation induces a scalar response at ${\cal O}(\epsilon\beta)$, which in turn acts as a source for the odd-parity gravitational perturbation at ${\cal O}(\epsilon^2\beta)$. We write $\omega=\omega_0+\epsilon^2\omega_2+\cdots$, where $\omega_0=\Omega^{\rm grav}_{\rm Sch}/(2M)$ is the Schwarzschild frequency.
At first order in $\epsilon$, the system reduces to
\begin{align}
    \left[ \beta r_{\rm g}^2\left(\frac{\D^2}{\D r_{\ast}^2} +\omega_0^2 -f  V_{\rm KG}\right) -f \right]\Psi_{\rm s}^{(1)} 
    &= \beta f r_{\rm g}^2V^{\rm s-}_{\rm 1st}\Psi_{\rm -}^{(0)},\label{eq: g-led 1st scalar EoM}
\\
    \left(\frac{\D^2}{\D r_{\ast}^2} 
    +\omega_0^2  -f V_{\rm RW}\right)\Psi_{\rm -}^{(1)} &= 0.
\end{align}
On the regular gravitational-led branch, the solutions at first order take the form
\begin{align}
    \Psi_{\rm s}^{(1)} 
    &= \beta s_{11}+\beta^2s_{12} +\cdots,\label{eq: beta expansion in 1st scalar soln}
\\
    \Psi_{\rm -}^{(1)} &= 0.
\end{align}
The homogeneous first-order gravitational correction can be absorbed into the normalization of $\Psi_-^{(0)}$. The scalar coefficients $s_{11},s_{12},\ldots$ are determined successively from Eq.~(\ref{eq: g-led 1st scalar EoM})
\begin{align}
    s_{11}
    = -r_{\rm g}^2V^{\rm s-}_{\rm 1st}\Psi_{\rm -}^{(0)},~
    s_{12} = -\frac{1}{f}{\cal D}_{\rm s}\!\left[r_{\rm g}^2V^{\rm s-}_{\rm 1st}\Psi_{\rm -}^{(0)}\right],~
    \cdots,
\end{align}
where
\begin{align}
    {\cal D}_{\rm s} = r_{\rm g}^2\left(\frac{\D^2}{\D r_{\ast}^2} +\omega^2 -fV_{\rm KG}\right).\label{eq: Ds}
\end{align}
Evaluating ${\cal D}_{\rm s}$ at $\omega=\omega_0$, we can express the first-order scalar response by a formal inverse operator. In $f^{-1}{\cal D}_{\rm s}$, the differential operator acts on the full product to its right before multiplication by $f^{-1}$
\begin{align}
    \Psi_{\rm s}^{(1)} 
    &= -\beta\left(1 -\beta \frac{1}{f}{\cal D}_{\rm s}\right)^{-1}r_{\rm g}^2V^{\rm s-}_{\rm 1st}\Psi_{\rm -}^{(0)} = -\beta\left(1 +\beta \frac{1}{f}{\cal D}_{\rm s} +\cdots\right)r_{\rm g}^2V^{\rm s-}_{\rm 1st}\Psi_{\rm -}^{(0)} .\label{eq: heavy scalar 1st soln}
\end{align}
At each order in $\beta$, the induced scalar perturbation is a local function of the gravitational perturbation and its radial derivatives. We impose QNM boundary conditions order by order, expanding the frequency and the boundary behavior consistently as in Ref.~\cite{Hatsuda:2023geo}. The induced scalar perturbation inherits the exponential factor of the ingoing gravitational wave at the horizon and that of the outgoing gravitational wave at infinity. The coupling potential and its derivatives enter the radial prefactors.
Since $\Psi_{\rm s}^{(0)}=\Psi_{\rm -}^{(1)}=0$, the second-order equations reduce to
\begin{align}    
    \left[\beta r_{\rm g}^2\left(\frac{\D^2}{\D r_{\ast}^2} +\omega_0^2 -f  V_{\rm KG}\right) -f \right]\Psi_{\rm s}^{(2)} 
    &= 0,
\\
    \left(\frac{\D^2}{\D r_{\ast}^2} 
    +\omega_0^2  -f V_{\rm RW}\right)\Psi_{\rm -}^{(2)}
    &= f V^{\rm s-}_{\rm 1st}\Psi_{\rm s}^{(1)}-2\omega_0\omega_2\Psi_{\rm -}^{(0)}.\label{eq: odd dCS master eq2 2nd}
\end{align}
The solutions at second order have the schematic form
\begin{align}
    \Psi_{\rm s}^{(2)} 
    &= 0,
\\
    \Psi_{\rm -}^{(2)} &= \beta\Psi_{\rm -}^{(2,1)}+\beta^2\Psi_{\rm -}^{(2,2)} +\cdots.
\end{align}
Collecting the odd powers of $\epsilon$ in the scalar field and the even powers in the gravitational field gives
\begin{align}
    \Psi_{\rm s} &= \epsilon\Psi_{\rm s}^{(1)}(\beta) +\epsilon^3\Psi_{\rm s}^{(3)}(\beta) +\cdots,
\\
    \Psi_{\rm -} &= \Psi_{\rm -}^{(0)} +\epsilon^2\Psi_{\rm -}^{(2)}(\beta) +\epsilon^4\Psi_{\rm -}^{(4)}(\beta) +\cdots.
\end{align}
Combining the $\epsilon$ and $\beta$ expansions and restoring the full $\omega$ in ${\cal D}_{\rm s}$ gives the scalar response
\begin{align}
    \Psi_{\rm s} = 
    -\epsilon\beta\left(1 +\beta \frac{1}{f}{\cal D}_{\rm s}\right)r_{\rm g}^2V^{\rm s-}_{\rm 1st}\Psi_{\rm -}+\cdots.\label{eq: heavy scalar leading soln}
\end{align}
Substituting this response into Eq.~(\ref{eq: odd dCS master eq2}) gives the effective odd-parity equation
\begin{align}
    \left(\frac{\D^2}{\D r_{\ast}^2} 
    +\omega^2  -f V_{\rm RW}\right)\Psi_{\rm -}
    -\epsilon^2\beta f \left[-r_{\rm g}^2(V^{\rm s-}_{\rm 1st})^2 -\beta r_{\rm g}^2V^{\rm s-}_{\rm 1st}\frac{1}{f}{\cal D}_{\rm s}V^{\rm s-}_{\rm 1st} +\cdots\right]\Psi_{\rm -} = 0.
\end{align}
Here ${\cal D}_{\rm s}$ acts on the product $V^{\rm s-}_{\rm 1st}\Psi_{\rm -}$.
For comparison with the quartic curvature EFT, we set $\epsilon=1$ only when matching the couplings. The leading potential correction in $\beta$ then agrees with that of quartic curvature gravity~\cite{Cardoso:2018ptl}\footnote{The $b_2$ coupling induces $(\tilde{\cal C}_2)^2$. Our $\epsilon_{\rm q 2}$ corresponds to $\epsilon_2$ of Ref.~\cite{Cardoso:2018ptl}, with $\epsilon_1=\epsilon_3=0$ there.}
\begin{align}
   \beta [-r_{\rm g}^2(V^{\rm s-}_{\rm 1st})^2] &=
   \epsilon_{\rm q 2}\frac{4608 M^8(\ell-1)\ell(\ell+1)(\ell+2)}{r^{10}},
\\
    \epsilon_{\rm q 2} &= \frac{-b_2^2}{m^2\Lambda^4M^6}.
\end{align}
With the couplings matched in this way, we compute the frequency of the fundamental $\ell=2$ mode as in Sec.~\ref{sec: perturbation}. Writing the second-order frequency correction in Eq.~(\ref{eq: odd dCS master eq2 2nd}) as $\omega_2=\beta\,\delta\Omega^{(2)}_{-}(\beta)/(2M)$ and restoring $\epsilon$, we obtain
\begin{align}
    \omega
    &= \frac{1}{2M}\left[\Omega^{\rm grav}_{\rm Sch} +\epsilon^2\beta\delta\Omega^{(2)}_{-}(\beta)+\cdots\right],
\\
   \delta\Omega^{(2)}_{-}(\beta)
    &= \epsilon_2^2 b_2^2 [-(101.892+180.388i)+\beta(624.794 +876.432i) +\cdots].\label{eq: heavy QNM correctionin dCS}
\end{align}
With this identification, the leading frequency shift in Eq.~(\ref{eq: heavy QNM correctionin dCS}) is consistent with the results for the fundamental odd-parity $\ell=2$ mode of quartic curvature gravity in Refs.~\cite{Cardoso:2018ptl,Cardoso:2019mqo}. The next term is the first correction in the inverse scalar mass predicted by this dCS model beyond the leading quartic curvature approximation.

\subsection{The limit of large scalar mass in sGB gravity}\label{sec: heavy sGB}

In the sGB example, we compare both the corrected black hole background and its gravitational QNMs with those in the induced quartic curvature theory. The Lagrangian is given by
\begin{align}
    \frac{{\cal L}}{\mpl^2}
    &= \frac{1}{2}R -\frac{1}{2}(\nabla\phi)^2 -\frac{1}{2}m^2\phi^2
    +\frac{\epsilon}{\Lambda^2}b_1 \phi R_{\mu\nu\rho\sigma}^2.\label{eq: sGB-dCS}
\end{align}
Unlike the dCS example, the sGB interaction modifies the Schwarzschild background. Scalar and even-parity perturbations are directly coupled. The odd sector remains decoupled from them, although both gravitational sectors receive corrections from the modified background.

To obtain the background solution for the perturbation analysis, we use the static ansatz of Eqs.~(\ref{eq: static metric expansion})--(\ref{eq: BG ansatz}), with asymptotically normalized time and fixed mass $M$, so that $r_{\rm g}=2M$. We select the scalar solution that is regular at the horizon and tends to zero at infinity. Including the factors of $\epsilon_2$ in the scalar and metric correction functions, we write
\begin{align}
    A(r) &= 1 -\frac{r_{\rm g}}{r} +\epsilon^2 a_{\rm 2nd}(r),\label{eq: BG ansatz2-1}\\
    B(r) &= 1 -\frac{r_{\rm g}}{r} +\epsilon^2 b_{\rm 2nd}(r),\\
    \bar{\phi}(r) &= \epsilon\, \pi_{\rm 1st}(r)+{\cal O}(\epsilon^3).\label{eq: BG ansatz2-3}
\end{align}
The scalar profile is odd in the interaction strength, whereas the metric corrections are even. The first-order scalar equation on Schwarzschild is
\begin{align}
    \frac{r-r_{\rm g}}{r}\pi_{\rm 1st}'' +\frac{2r -r_{\rm g}}{r^2}\pi_{\rm 1st}' -m^2\pi_{\rm 1st} +\frac{12b_1\epsilon_2 r_{\rm g}^4}{r^6} =0.\label{eq: heavy scalar EoM}
\end{align}
The prime denotes a derivative with respect to $r$. Expanding the scalar profile in $\beta$ and solving successively, as in the dCS example, gives
\begin{align}
    \pi_{\rm 1st} &= 
    b_1\epsilon_2\beta\left[\frac{12 r_{\rm g}^6}{r^6}
    +\beta\frac{72(5r -6r_{\rm g})r_{\rm g}^8}{r^9}\right]+{\cal O}(\beta^3).\label{eq: 1st heavy scalar}
\end{align}
The first-order scalar profile, Eq.~(\ref{eq: 1st heavy scalar}), sources the second-order metric corrections. Expanding them in $\beta$ gives
\begin{align}
    a_{\rm 2nd} &= b_1^2\epsilon_2^2\beta \left[\frac{8(16r -11r_{\rm g})r_{\rm g}^9}{r^{10}} +\beta\frac{72(640r^2 -1242rr_{\rm g} +561r_{\rm g}^2)r_{\rm g}^{11}}{11r^{13}}\right],
\\
    b_{\rm 2nd} &= b_1^2\epsilon_2^2\beta \left[\frac{8(72r -67r_{\rm g})r_{\rm g}^9}{r^{10}} +\beta\frac{72(3520r^2 -8016rr_{\rm g} +4455r_{\rm g}^2)r_{\rm g}^{11}}{11r^{13}}\right].
\end{align}
The leading metric corrections in $\beta$ match those in quartic curvature gravity~\cite{Cardoso:2018ptl}, with the couplings matched as in Eq.~(\ref{eq: sGB quartic coupling}) below. The event horizon is determined by $A(r_{\rm H})={\cal O}(\epsilon^3)$, with the coefficients expanded through $\beta^2$. This gives $r_{\rm H}-r_{\rm g}=-\epsilon^2r_{\rm g}a_{\rm 2nd}(r_{\rm g})$ and hence
\begin{align}
    r_{\rm H} &= r_{\rm g}\left[1 -\epsilon^2\beta b_{1}^2\epsilon_2^2\left(40 -\frac{2952}{11}\beta\right)\right].
\end{align}

As in the background calculation, we first solve for the scalar perturbation at first order in $\epsilon$ and use this solution to derive the gravitational equations at second order. On the gravitational-led branch, $\Psi_{\rm s}^{(0)}=0$. At first order in $\epsilon$, the scalar perturbation is sourced by the even-parity perturbation $\Psi_+^{(0)}$ at zeroth order.
\begin{align}
    \left[r_{\rm g}^2\left(\frac{\D^2}{\D r_*^2}+\omega_0^2-fV_{\rm KG}\right)
    -\frac{f}{\beta}\right]\Psi_{\rm s}^{(1)}
    &= f r_{\rm g}^2 V_{\rm s+}\Psi_+^{(0)}.
    \label{eq: scalar sGB 1st master eq}
\end{align}
At this order we use the Schwarzschild background, with $f=1-r_{\rm g}/r$ and $\D r_*=\D r/f$. As in Eq.~(\ref{eq: g-led 1st scalar EoM}), $\omega_0$ is the Schwarzschild gravitational QNM frequency. The potential $V_{\rm s+}$ denotes the coupling between the scalar and even-parity perturbations in the expansion in the inverse scalar mass. Expanding in $\beta$ as in Eq.~(\ref{eq: heavy scalar 1st soln}) gives
\begin{align}
    \Psi_{\rm s}^{(1)}
    &= -\beta\left(1+\beta\frac{1}{f}{\cal D}_{\rm s}\right)
    \left[r_{\rm g}^2 V_{\rm s+}\Psi_+^{(0)}\right]+{\cal O}(\beta^3),
    \label{eq: heavy scalar leading soln sGB}
\end{align}
where ${\cal D}_{\rm s}$ is the operator in Eq.~(\ref{eq: Ds}) evaluated at $\omega=\omega_0$.

At second order in $\epsilon$, we substitute Eq.~(\ref{eq: heavy scalar leading soln sGB}) into the even-parity gravitational equation. The odd-parity equation decouples from the scalar perturbation. We write both equations in terms of $F=\sqrt{AB}$. As in Sec.~\ref{sec: master eqs}, we use the Schwarzschild potentials $V_{\rm RW}$ and $V_{\rm Z}$ with the corrected horizon radius $r_{\rm H}$ (App.~\ref{sec: appendix frequency and GR potentials}). Since $(r_{\rm H}-r_{\rm g})/r_{\rm g}={\cal O}(\epsilon^2\beta)$, we may replace $r_{\rm g}$ by $r_{\rm H}$ in terms already proportional to $\epsilon^2\beta$. Keeping the leading corrections in $\beta$, we obtain
\begin{align}
    \left\{F\frac{\D}{\D r}\left(F\frac{\D}{\D r}\right)+\omega^2 c^2
    -F\left[V_{\rm RW}+\epsilon^2\beta V_-+{\cal O}(\epsilon^2\beta^2)\right]\right\}\Psi_- &= 0,
    \label{eq: odd sGB master eq2}
\\
    \left\{F\frac{\D}{\D r}\left(F\frac{\D}{\D r}\right)+\omega^2 c^2
    -F\left[V_{\rm Z}+\epsilon^2\beta\left(V_+-r_{\rm H}^2V_{\rm s+}^2\right)+{\cal O}(\epsilon^2\beta^2)\right]\right\}\Psi_+ &= 0.
    \label{eq: even sGB master eq2}
\end{align}
To the same order, $F$ and $c^2$ are given by
\begin{align}
    F(r) &= 1 -\frac{r_{\rm H}}{r} -\epsilon^2\beta b_1^2\epsilon_2^2\frac{8r_{\rm H}(5r^9 -44rr_{\rm H}^8 +39r_{\rm H}^9)}{r^{10}},
\\
    c^2 &= 1- \epsilon^2\beta b_1^2\epsilon_2^2 F\frac{4032r_{\rm H}^8}{r^{8}}.
\end{align}
The functions $V_-$ and $V_+$ are the corrections to the gravitational potentials before substitution of the scalar solution. This substitution gives the additional term $-r_{\rm H}^2V_{\rm s+}^2$ in the even-parity potential at leading order in $\beta$. Setting $\epsilon=1$ to match the couplings, we recover the leading potential corrections of quartic curvature gravity in Ref.~\cite{Cardoso:2018ptl}.\footnote{The $b_1$ coupling induces $(\mathcal C_2)^2$. Our $\epsilon_{\rm q 1}$ corresponds to $\epsilon_1$ of Ref.~\cite{Cardoso:2018ptl}, with $\epsilon_2=\epsilon_3=0$ there.}
\begin{align}
    \epsilon_{\rm q 1} &= \frac{-b_1^2}{m^2\Lambda^4M^6},\label{eq: sGB quartic coupling}
\\
    \beta V_{-} &= -\epsilon_{\rm q 1}
    \frac{r_{\rm H} \left[-15 r^9+288 (77-2 \lambda ) r^2 r_{\rm H}^7+4 (146 \lambda -13217) r r_{\rm H}^8+31122 r_{\rm H}^9\right]}{8r^{12}}
    ,
\\
    \beta\{V_{+} -r_{\rm H}^2V_{\rm s+}^2\} 
    &= \epsilon_{\rm q 1}\frac{r_{\rm H}}{8r^{12} (\lambda  r+3 r_{\rm H})^3}\Big[15 \lambda ^2 (\lambda +4) r^{12}-45 \lambda ^2 r^{11} r_{\rm H}-135 \lambda  r^{10} r_{\rm H}^2-135 r^9 r_{\rm H}^3
\notag\\    
    &\quad +36 \lambda ^3 (\lambda ^2 +18 \lambda +176) r^5 r_{\rm H}^7 -20 \lambda ^2 (13 \lambda ^2+521 \lambda -1512) r^4 r_{\rm H}^8 
\notag\\    
    &\quad +18 \lambda  (321 \lambda ^2 -3698 \lambda +2016) r^3 r_{\rm H}^9 +18 \lambda  (2249 \lambda -5634) r^2 r_{\rm H}^{10}
\notag\\
    &\quad +36 (1984 \lambda -675) r r_{\rm H}^{11} +28836 r_{\rm H}^{12}\Big].
\end{align}
In computing the QNM frequencies, we include corrections through ${\cal O}(\epsilon^2\beta^2)$ in $F$, $c^2$, and the gravitational potentials. We use the redefinitions of the master variables introduced in Sec.~\ref{sec: master eqs} to write the radial operator in terms of $f=1-r_{\rm H}/r$. We impose purely ingoing boundary conditions at the corrected horizon and purely outgoing boundary conditions at infinity. We keep $M$ fixed and define $\beta$ and $\epsilon_2$ using $r_{\rm g}=2M$. Restoring $\epsilon$, we obtain the fundamental ($\ell=2,n=0$) frequencies
\begin{align}
    \omega
    &= \frac{1}{2M}\left[\Omega^{\rm grav}_{\rm Sch} +\epsilon^2\beta\delta\Omega^{(2)}_{\mp}(\beta)+\cdots\right],
\\
   \delta\Omega^{(2)}_{\rm -}(\beta)
    &= \epsilon_2^2 b_1^2 [(-10.1608 -7.37379i)  +\beta(140.199 +316.156i)  +\cdots],\label{eq: odd heavy QNM correctionin sGB}
\\
    \delta\Omega^{(2)}_{\rm +}(\beta)
    &= \epsilon_2^2 b_1^2 [(-21.2658 -31.2792i) -\beta(27.8717 +73.5804i)+\cdots].\label{eq: even heavy QNM correctionin sGB}
\end{align}
The leading frequency shifts in the odd and even sectors are consistent with the numerical results for the fundamental $\ell=2$ modes in Ref.~\cite{Cardoso:2018ptl}. The next terms in Eqs.~(\ref{eq: odd heavy QNM correctionin sGB}) and~(\ref{eq: even heavy QNM correctionin sGB}) give the first corrections in the inverse scalar mass beyond the quartic curvature approximation for this sGB model. The frequencies of the odd and even modes shift by different amounts, breaking the isospectrality of the Schwarzschild gravitational perturbations.

\subsection{Integrating out the heavy scalar at the action level}\label{sec: action matching}

The heavy scalar can be integrated out at tree level by solving its equation of motion in an expansion in the inverse scalar mass and substituting the solution into the original action.

For the interactions linear in $\phi$, write the scalar-curvature coupling as $\epsilon J\phi$, with
\begin{align}
    J &= \frac{1}{\Lambda^2}(b_1\mathcal C_2+b_2\tilde{\mathcal C}_2)
    +\frac{1}{\Lambda^4}(e_1\mathcal C_3+e_2\tilde{\mathcal C}_3).\label{eq: linear scalar matching source}
\end{align}
The scalar equation $(m^2-\Box)\phi=\epsilon J$ gives, for $|\Box|/m^2\ll1$,
\begin{align}
    \phi = \epsilon(m^2-\Box)^{-1}J
    = \epsilon\left(\frac{J}{m^2}+\frac{\Box J}{m^4}+\cdots\right).
    \label{eq: linear scalar matching solution}
\end{align}
Substituting this solution into the scalar kinetic, mass, and interaction terms and integrating by parts yields
\begin{align}
    \frac{\mathcal L_{\rm eff}}{\mpl^2}
    &= \frac12R+\epsilon^2\left(\frac{J^2}{2m^2}+\frac{J\Box J}{2m^4}+\cdots\right).
    \label{eq: linear scalar induced action}
\end{align}
Substituting the explicit source $J$ in Eq.~(\ref{eq: linear scalar matching source}) into Eq.~(\ref{eq: linear scalar induced action}) gives, through operator mass dimension ten and up to total derivatives,
\begin{align}
    \frac{{\cal L}_{\rm eff}}{\mpl^2}
    &= \frac12R
    +\frac{\epsilon^2 b_1^2}{2m^2\Lambda^4}({\cal C}_2)^2
    +\frac{\epsilon^2 b_1b_2}{m^2\Lambda^4}{\cal C}_2\tilde{\cal C}_2
    +\frac{\epsilon^2 b_2^2}{2m^2\Lambda^4}(\tilde{\cal C}_2)^2
\notag\\
    &\quad +\frac{\epsilon^2(b_1{\cal C}_2+b_2\tilde{\cal C}_2)
    (e_1{\cal C}_3+e_2\tilde{\cal C}_3)}{m^2\Lambda^6}
\notag\\
    &\quad +\frac{\epsilon^2(b_1{\cal C}_2+b_2\tilde{\cal C}_2)
    \Box(b_1{\cal C}_2+b_2\tilde{\cal C}_2)}{2m^4\Lambda^4}+\cdots.
    \label{eq: LEEFT2}
\end{align}
Each correction in Eq.~(\ref{eq: LEEFT2}) involves two interactions linear in $\phi$ and is therefore of order $\epsilon^2$. The $J^2/(2m^2)$ term in Eq.~(\ref{eq: linear scalar induced action}) gives both the $b_i b_j$ terms, which reproduce the quartic curvature action underlying the leading matching in Secs.~\ref{sec: heavy dCS} and~\ref{sec: heavy sGB}~\cite{Endlich:2017tqa,Cardoso:2018ptl}, and the $b_i e_j$ cross terms. The quartic curvature part is proportional to $(b_1{\cal C}_2+b_2\tilde{\cal C}_2)^2$, so its three coefficients are correlated. The two terms shown at mass dimension ten have different dependences on the scalar mass. The $b_i e_j$ term is proportional to $m^{-2}$, while the term containing $\Box$ comes from $J\Box J/(2m^4)$ and gives the first subleading correction in the inverse scalar mass, corresponding to the subleading terms in Eqs.~(\ref{eq: heavy QNM correctionin dCS}), (\ref{eq: odd heavy QNM correctionin sGB}), and~(\ref{eq: even heavy QNM correctionin sGB}).

Through mass dimension ten, only the interactions linear in $\phi$ contribute to the induced action. To find where the remaining scalar-curvature interactions in Eq.~(\ref{eq: STEFT Lagrangian}) first enter, we extend the expansion to higher operator dimensions. Table~\ref{tab: STEFT matching orders} gives the lowest operator dimension at which each coupling contributes, assuming a generic nonzero curvature source from $b_1$ and $b_2$. We count each curvature tensor as dimension two and each covariant derivative as dimension one, excluding the coefficients multiplying the operators.

\begin{table}[htbp]
\centering
\begin{manuscripttableblock}{0.74\textwidth}
\caption{Lowest dimensions of induced pure gravity operators.}
\label{tab: STEFT matching orders}
\setlength{\tabcolsep}{11pt}
\begin{tabular}{cccc}
\hline\hline
Couplings & Example products & Mass factors & \shortstack{\strut Lowest operator\\\strut dimension}\\
\hline
$b_1,b_2$ & $b_1^2,\ b_1b_2$ & $\Lambda^{-4}m^{-2}$ & $8$\\
$e_1,e_2$ & $b_1e_1,\ b_1e_2$ & $\Lambda^{-6}m^{-2}$ & $10$\\
$c_1,c_2$ & $b_1^2c_1,\ b_1^2c_2$ & $\Lambda^{-6}m^{-4}$ & $12$\\
$f_1,f_2$ & $b_1^2f_1,\ b_1^2f_2$ & $\Lambda^{-8}m^{-4}$ & $14$\\
$f_3,f_4$ & $b_1^2f_3,\ b_1^2f_4$ & $\Lambda^{-8}m^{-4}$ & $14$\\
$d_1,d_2$ & $b_1^3d_1,\ b_1^3d_2$ & $\Lambda^{-8}m^{-6}$ & $16$\\
\hline\hline
\end{tabular}
\end{manuscripttableblock}
\end{table}

The interactions with coefficients $c_1,c_2,f_1$, and $f_2$ modify the quadratic scalar term, while those with $f_3,f_4$ modify the scalar kinetic term. The $d_1$ and $d_2$ interactions contain three powers of $\phi$ and two curvature tensors. Inserting the leading solution $\phi=\epsilon J/m^2$ from Eq.~(\ref{eq: linear scalar matching solution}), with $J$ starting at quadratic order in curvature, gives eight curvature tensors at the lowest order. The resulting operators have mass dimension sixteen. We therefore extend the expansion in App.~\ref{sec: complete action matching} to this dimension to include the first contribution from every coupling in Table~\ref{tab: STEFT matching orders}.

This result is the action induced by the specified STEFT interactions, rather than a complete basis of pure gravity operators through mass dimension sixteen. The STEFT describes massive scalar exchange explicitly and fixes the subleading corrections. In this limited sense, the STEFT provides a partial ultraviolet completion of the induced pure gravity EFT.

\subsection{An example from Horndeski theory}\label{sec: Horndeski matching}

As an example with second-order field equations, we consider a restricted set of Horndeski interactions~\cite{Horndeski:1974wa,Deffayet:2011gz,Kobayashi:2011nu}, including the scalar Gauss--Bonnet coupling. In this example we also allow scalar self-interactions omitted in Sec.~\ref{sec: theory}. Expanding about $\phi=X=0$ and taking a canonical scalar kinetic term and $m^2>0$, we write the Lagrangian as
\begin{align}
    \frac{\cal L}{\mpl^2} &= \frac{1}{2}R +X -\frac{1}{2}m^2\phi^2 
    +\frac{\epsilon}{\Lambda^2} (b_1\phi +c_1\phi^2 +d_1\phi^3){\cal R}_{\rm GB}
\notag\\
    &\quad +\epsilon\left[-m^2g_{21}\phi^3 +g_{22}\phi X +\frac{g_{31}}{\Lambda^2}X\Box\phi\right]+\cdots,\label{eq: Horndeski}
\end{align}
where $g_{21}$, $g_{22}$, and $g_{31}$ are dimensionless couplings for the scalar self-interactions and
\begin{align}
    X = -\frac{1}{2}(\nabla_{\mu}\phi)^2,~
    {\cal R}_{\rm GB} = R_{\mu\nu\rho\sigma}^2 -4R_{\mu\nu}^2 +R^2.
\end{align}
The ellipsis in Eq.~(\ref{eq: Horndeski}) denotes interactions omitted from the matching. The sourced scalar solution contains both a linear response and nonlinear corrections. The leading terms that exhibit these two contributions are
\begin{align}
    \phi = \frac{\epsilon b_1}{\Lambda^2m^2}\left({\cal R}_{\rm GB}+\frac{\Box{\cal R}_{\rm GB}}{m^2}\right)+\frac{\epsilon^2 b_1(2c_1-3\epsilon b_1g_{21})}{\Lambda^4m^4}{\cal R}_{\rm GB}^2+\cdots.\label{eq: Horndeski scalar profile}
\end{align}
The combination $2c_1-3\epsilon b_1g_{21}$ follows from the curvature-dependent quadratic coupling and the cubic scalar potential, respectively. In deriving the effective action, we include the higher derivative terms in the ellipsis of Eq.~(\ref{eq: Horndeski scalar profile}) that contribute through mass dimension sixteen. Substituting into Eq.~(\ref{eq: Horndeski}) and integrating by parts gives
\begin{align}
    \frac{{\cal L}_{\rm eff}}{\mpl^2}
    &= \frac12 R+\frac{\epsilon^2 b_1^2}{2\Lambda^4m^2}{\cal R}_{\rm GB}
      \sum_{n=0}^{4}\left(\frac{\Box}{m^2}\right)^n{\cal R}_{\rm GB}
      +\frac{\epsilon^3 b_1^2(c_1-\epsilon b_1g_{21})}{\Lambda^6m^4}{\cal R}_{\rm GB}^3
\notag\\
    &\quad +\frac{\epsilon^3 b_1^2}{\Lambda^6m^6}
      \left(2c_1-3\epsilon b_1g_{21}+\frac14\epsilon b_1g_{22}\right)
      {\cal R}_{\rm GB}^2\Box{\cal R}_{\rm GB}
\notag\\
    &\quad +\frac{\epsilon^3 b_1^2}{\Lambda^6m^8}
      \left(5c_1-9\epsilon b_1g_{21}+\epsilon b_1g_{22}\right)
      {\cal R}_{\rm GB}(\Box{\cal R}_{\rm GB})^2
\notag\\
    &\quad +\frac{\epsilon^3 b_1^2}{\Lambda^6m^8}
      \left(4c_1-6\epsilon b_1g_{21}+\frac12\epsilon b_1g_{22}\right)
      (\nabla_\mu{\cal R}_{\rm GB})^2\Box{\cal R}_{\rm GB}
\notag\\
    &\quad +\frac{\epsilon^4 b_1^2}{\Lambda^8m^6}
    \left[b_1d_1+\frac12(2c_1-3\epsilon b_1g_{21})^2\right]{\cal R}_{\rm GB}^4
    -\frac{\epsilon^4 b_1^3g_{31}}{2\Lambda^8m^6}(\nabla_\mu{\cal R}_{\rm GB})^2\Box{\cal R}_{\rm GB}+\cdots.
    \label{eq: reduced Horndeski effective action}
\end{align}
In Eq.~(\ref{eq: reduced Horndeski effective action}), the $c_1$ and $g_{21}$ interactions first contribute at $\Lambda^{-6}m^{-4}$, while $g_{22}$ first enters the derivative term at $\Lambda^{-6}m^{-6}$. At $\Lambda^{-8}m^{-6}$, the $\mathcal R_{\rm GB}^4$ coefficient combines a direct $d_1$ contribution with the iterated lower-order scalar nonlinearity. The cubic Galileon coupling $g_{31}$ first appears in $(\nabla\mathcal R_{\rm GB})^2\Box\mathcal R_{\rm GB}$, which also has mass dimension sixteen. These are the first contributions of all the interactions specified in Eq.~(\ref{eq: Horndeski}). Keeping operators through mass dimension sixteen gives terms through $\epsilon^6$ in this example, as distinct from the second-order QNM calculation.

This example shows how different scalar interactions contribute to the same pure gravity operator. Since $\mathcal R_{\rm GB}$ is quadratic in curvature, its powers $\mathcal R_{\rm GB}^3$ and $\mathcal R_{\rm GB}^4$ contain six and eight curvature tensors, respectively. They are distinct from the cubic Riemann interaction considered next.

\subsection{A remark on cubic curvature interactions}\label{sec: cubic remark}

Black holes with cubic curvature corrections have been studied, for example, in Refs.~\cite{deRham:2020ejn,Cano:2021myl}, while Ref.~\cite{Cardoso:2018ptl} provides the quartic curvature results used in our matching above.
The scalar sources in Secs.~\ref{sec: action matching} and~\ref{sec: Horndeski matching} start at quadratic order in curvature, so their tree-level exchange starts at quartic order. It therefore does not generate an independent cubic Riemann interaction within the matching considered here. This conclusion concerns the specified scalar interactions. Terms cubic in curvature can arise in other settings, including quantum corrections~\cite{Goroff:1985th,Bern:2021ppb}.

To illustrate how a cubic Riemann interaction can arise, consider a massive rank-four tensor field,
\begin{align}
    {\cal L} = \frac{\mpl^2}{2}R -\frac{1}{2}(\nabla_{\alpha}X_{\mu\nu\rho\sigma})^2 -\frac{1}{2}m_X^2X_{\mu\nu\rho\sigma}^2 +\epsilon g_X m_X X_{\mu\nu\rho\sigma}R^{\mu\nu\rho\sigma},
\end{align}
where $X_{\mu\nu\rho\sigma}$ has the algebraic symmetries of the Riemann tensor, $m_X$ is its mass, and $g_X$ is a dimensionless coupling constant. The kinetic term gives rise to ghost degrees of freedom, so we use this theory only as an algebraic toy model.
The curvature source is now linear in the Riemann tensor, giving $X_{\mu\nu\rho\sigma}=\epsilon g_XR_{\mu\nu\rho\sigma}/m_X+\cdots$. Expanding the propagator for exchange of $X$ between two such sources gives $\epsilon^2 g_X^2 R^{\mu\nu\rho\sigma}\Box R_{\mu\nu\rho\sigma}/(2m_X^2)$ at order $m_X^{-2}$. By Eq.~(\ref{eq: derivative curvature reduction}), the curvature contraction reduces to $-3{\cal C}_3$ up to Ricci-dependent terms, giving
\begin{align}
    \Delta\mathcal L_{\rm cubic} = -\frac{3\epsilon^2g_X^2}{2m_X^2}{\cal C}_3.
\end{align}

\section{Physical imprints of scalar-curvature interactions}\label{sec: discussion}

\subsection{Complementary information from black hole backgrounds and QNMs}\label{sec: observational implications}

Black hole backgrounds and QNM frequencies probe different combinations of the scalar-curvature couplings. Table~\ref{tab: coupling sensitivity} lists the combinations entering the leading corrections in the massless calculation.

\begin{table}[!htbp]
\centering
\begin{manuscripttableblock}{\textwidth}
\caption{Leading coupling dependence of black hole metrics and QNM frequencies for a massless scalar.}
\label{tab: coupling sensitivity}
\renewcommand{\arraystretch}{1.18}
\setlength{\tabcolsep}{4pt}
\begin{tabular*}{\textwidth}{@{\extracolsep{\fill}}>{\raggedright\arraybackslash}p{0.27\textwidth} >{\centering\arraybackslash}p{0.13\textwidth} >{\raggedright\arraybackslash}p{0.38\textwidth} >{\centering\arraybackslash}p{0.16\textwidth}@{}}
\hline\hline
Quantity & Order & Coupling dependence & Equations\\
\hline
Static metric & $\epsilon^2$ & $b_1^2,\ b_1e_1,\ e_1^2$ & (\ref{eq: 2nd a}), (\ref{eq: 2nd b})\\[3pt]
Frame dragging & $\epsilon^2$ & $b_1^2,\ b_2^2,\ b_1e_1,\ b_2e_2,$\newline $e_1^2,\ e_2^2$ & (\ref{eq: rotating frame dragging})\\[3pt]
Equatorial reflection asymmetry & $\epsilon^2$ & $b_1b_2,\ b_1e_2,\ b_2e_1,\ e_1e_2$ & (\ref{eq: rotating tt correction}), (\ref{eq: rotating rr correction})\\[3pt]
Gravitational-led frequencies & $\epsilon^2$ & $b_1,b_2,e_1,e_2$,\newline through the eigenvalues & (\ref{eq: grav QNM})--(\ref{eq: grav QNM discriminant})\\[3pt]
Scalar-led frequencies & $\epsilon$ & $c_1,\ f_1,\ f_3$ & (\ref{eq: delomega^s_1st})\\
\hline\hline
\end{tabular*}
\manuscripttablenote{Factors of $\epsilon_2$ and $\epsilon_4$ are suppressed in the coupling column. Rotational entries are linear in $\chi$. The QNM entries concern the fundamental $\ell=2$ modes on the static background. The scalar-led row lists the first-order shift. The second-order result is given in Eq.~(\ref{eq: scalar QNM}).}
\end{manuscripttableblock}
\end{table}

For a static black hole, the metric corrections through ${\cal O}(\epsilon^2)$ depend only on $b_1$ and $e_1$. For a slowly rotating black hole, $b_2$ and $e_2$ also enter the correction to frame dragging. The terms in the metric that break equatorial reflection symmetry involve products of one parity-even coupling and one parity-odd coupling. An equatorial reflection asymmetry therefore indicates that couplings of both parities are present.

The scalar-led QNM frequencies on a static background probe $c_1,f_1$, and $f_3$ already at ${\cal O}(\epsilon)$. These couplings can shift the scalar-led frequencies even when both the static metric and the gravitational-led frequencies agree with GR through ${\cal O}(\epsilon^2)$. The $d_1$ coupling first enters the scalar-led frequency at second order, through the products $b_1d_1$ and $e_1d_1$ in Eq.~(\ref{eq: scalar QNM}).

The gravitational-led frequencies first shift at ${\cal O}(\epsilon^2)$ and depend on $b_1,b_2,e_1$, and $e_2$. In particular, $b_2$ and $e_2$ can shift these frequencies even though they do not affect the static metric at this order. A Schwarzschild background therefore need not have the GR gravitational spectrum. The couplings $c_2,f_2,f_4$, and $d_2$ do not enter either the metric or the static QNM frequencies through this order. As discussed in Sec.~\ref{sec: perturbation}, $c_2,f_2$, and $f_4$ do mix the scalar and odd gravitational perturbations at second order, but their contribution to the frequencies enters at higher order. For rotating black holes, changes in the parity selection rules may allow scalar-led QNMs to probe additional couplings.

At fixed $M$ and $\Lambda$, a single complex scalar-led frequency gives at most two real constraints on the three coefficients $c_1,f_1$, and $f_3$ in Eq.~(\ref{eq: delomega^s_1st}). Black holes of different masses can provide further information because the terms in the frequency shift scale differently with $r_{\rm g}$. For fixed couplings, $\Lambda$, and scalar mass $m$, the relevant dimensionless parameters are
\begin{align}
    \epsilon_2 = (\Lambda r_{\rm g})^{-2},\quad
    \epsilon_4 = (\Lambda r_{\rm g})^{-4},\quad
    \alpha = m^2r_{\rm g}^2.
    \label{eq: mass and interaction scalings}
\end{align}
In Eq.~(\ref{eq: delomega^s_1st}), the $c_1$ contribution to the dimensionless frequency shift scales as $r_{\rm g}^{-2}$, while the $f_1$ and $f_3$ contributions scale as $r_{\rm g}^{-4}$. Comparing black holes of different masses can therefore help distinguish the $c_1$ contribution from the other two. Separating $f_1$ and $f_3$, which have the same mass scaling, requires their different contributions to the real and imaginary parts of the frequency or information from other modes.

For a light scalar of fixed mass, Eq.~(\ref{eq: mass and interaction scalings}) shows that $\alpha$ increases with the black hole mass. In the example of Sec.~\ref{sec: light mass limit}, the scalar-led frequency depends on $\alpha$ even for a free scalar on Schwarzschild. The scalar mass also changes the frequency corrections proportional to $c_1$, $c_1^2$, and $b_2^2$ in Eqs.~(\ref{eq: light scalar QNM first}) and~(\ref{eq: light scalar QNM second}). The odd-parity gravitational frequency depends on $\alpha$ through the correction proportional to $b_2^2$ in Eq.~(\ref{eq: light gravitational QNM}). The two branches thus provide different information about the scalar mass and its couplings on the same Schwarzschild background.

For a heavy scalar, the low-frequency gravitational QNMs probe the coefficients of the induced curvature operators. The quartic curvature coefficients in Eq.~(\ref{eq: LEEFT2}) are proportional to $b_i b_j/(m^2\Lambda^4)$, so the leading frequency shifts alone cannot determine $m$ separately from the scalar couplings. The first subleading corrections have a different dependence on $m$, as shown in the dCS and sGB examples of Sec.~\ref{sec: integrating out}. In the general theory, additional scalar-curvature interactions also contribute at higher operator dimensions. The quartic curvature terms in Eq.~(\ref{eq: LEEFT2}) are proportional to $(b_1{\cal C}_2+b_2\tilde{\cal C}_2)^2$. The relation among their three coefficients therefore provides a consistency check on this scalar exchange interpretation.

\subsection{Mode mixing and matter coupling in black hole ringdown}\label{sec: scalar-led observability}

We consider the fundamental $\ell=2$ modes of the static black holes with a massless scalar studied in Sec.~\ref{sec: perturbation}. To determine how these modes enter the measured strain, we specify the coupling to matter and use the coupled master equations to estimate each contribution to leading order in $\epsilon$.

A detector responds to the metric to which matter couples. As a representative choice, we supplement the vacuum action with
\begin{align}
    S_{\rm m} &= \int \D^4x\sqrt{-\tilde g}\,
    {\cal L}_{\rm m}(\psi_{\rm m},\tilde g_{\mu\nu}),
    \quad \tilde g_{\mu\nu}=A_{\rm m}^2(\phi)g_{\mu\nu}.
    \label{eq: conformal matter coupling}
\end{align}
Here $\psi_{\rm m}$ denotes the matter fields. The physical metric $\tilde g_{\mu\nu}$ governs the motion of the detector's freely falling test masses. The function $A_{\rm m}^2$ rescales all metric components by the same scalar-dependent factor, defining a conformal coupling. Setting $A_{\rm m}=1$ gives minimal coupling to $g_{\mu\nu}$. The matter sector specifies this function independently of the vacuum STEFT coefficients. When matter is treated as a test field, this coupling does not change the vacuum QNM frequencies.

The radiation contains tensor polarizations $h_+,h_\times$ and can also contain a scalar breathing polarization through the conformal matter coupling. We denote its strain by $h_{\rm S}$. For a given detector, the observed strain is
\begin{align}
    h_{\rm obs} = h_{\rm T}+F_{\rm S}h_{\rm S},\quad
    h_{\rm T} := F_{+}h_{+}+F_{\times} h_{\times}.
    \label{eq: ringdown detector strain}
\end{align}
The strains $h_+$ and $h_\times$ describe the two tensor polarizations of the physical metric, distinct from the odd and even master variables $\Psi_-,\Psi_+$. The antenna pattern functions $F_+, F_\times, F_{\rm S}$ project each polarization onto the detector and are treated as constant over the ringdown. The matter coupling enters through the polarization waveforms. In Eq.~(\ref{eq: ringdown detector strain}), $h_{\rm T}$ already includes the tensor antenna response, while $F_{\rm S}$ multiplies $h_{\rm S}$ separately.

Within linear perturbation theory, the ringdown signal is a superposition of eigenmodes of the coupled system, while QNMs dominate. The labels gravitational-led and scalar-led refer to the gravitational and Klein--Gordon modes recovered in the GR limit, respectively. Each coupled eigenmode generally contains both gravitational and scalar components sharing one complex frequency. Thus $h_{\rm S}$, like $h_{\rm T}$, can contain both gravitational-led and scalar-led contributions. Writing $\tau=t-t_0$, with $t_0$ a reference time for the ringdown model, the contributions of the fundamental quadrupolar modes are
\begin{align}
    h_{\rm T}(t) &\simeq \operatorname{Re}\left\{
    \sum_{\sigma=\pm}{\cal A}_{\rm T}^{({\rm g},\sigma)}e^{-i\omega^{{\rm grav},\sigma}\tau}
    +{\cal A}_{\rm T}^{({\rm s})}e^{-i\omega^{\rm s}\tau}\right\},
    \label{eq: ringdown tensor signal}\\
    h_{\rm S}(t) &\simeq \operatorname{Re}\left\{
    \sum_{\sigma=\pm}{\cal A}_{\rm S}^{({\rm g},\sigma)}e^{-i\omega^{{\rm grav},\sigma}\tau}
    +{\cal A}_{\rm S}^{({\rm s})}e^{-i\omega^{\rm s}\tau}\right\}.
    \label{eq: ringdown scalar signal}
\end{align}
Here $\sigma=\pm$ distinguishes the two gravitational-led branches of Sec.~\ref{sec: perturbation}, not tensor polarizations. The subscript on each complex amplitude ${\cal A}$ indicates whether the mode contributes to the tensor strain $h_{\rm T}$ or the breathing strain $h_{\rm S}$. For example, ${\cal A}_{\rm T}^{({\rm s})}$ is the amplitude of the scalar-led mode in the tensor strain.
$\omega^{\rm grav}$ and $\omega^{\rm s}$ denote the gravitational-led and scalar-led QNM frequencies, respectively.
Higher multipoles and overtones are omitted from these schematic signals.

The scalar-led term in $h_{\rm T}$ follows from the off-diagonal potentials in Eq.~(\ref{eq: EFT potential}). For example, the odd-parity equation in Eq.~(\ref{eq: master eq}), to first order, reads
\begin{align}
    \left[\frac{\D^2}{\D r_*^2}+\omega^2-fV_{\rm RW}\right]\Psi_-
    = \epsilon fV_{\rm 1st}^{\rm s-}\Psi_{\rm s}+{\cal O}(\epsilon^2).
    \label{eq: ringdown mixing source}
\end{align}
The even component is similarly sourced through $V_{\rm 1st}^{\rm s+}$. In a scalar-led eigenmode, $\omega=\omega^{\rm s}$, and the scalar perturbation sources a metric perturbation at the same frequency, producing the second term of Eq.~(\ref{eq: ringdown tensor signal}) even for minimal matter coupling. Conversely, these mixing potentials in the scalar equation induce a scalar component at $\omega^{{\rm grav},\sigma}$ within each gravitational-led eigenmode. Through matter coupling, this scalar component produces the first terms of Eq.~(\ref{eq: ringdown scalar signal}). The mixing potentials depend on $b_i,e_i$ and are given in App.~\ref{sec: appendix EFT potentials}. The contribution of scalar-led modes to the gravitational waveform has been studied in dCS~\cite{Molina:2010fb} and more general coupled systems~\cite{Lestingi:2025contamination}.

To relate the radiative scalar perturbation $\delta\phi$ to the breathing strain $h_{\rm S}$, we expand the conformal coupling in the wave zone, where the background metric approaches the Minkowski metric and the background scalar tends to zero. Choosing $A_{\rm m}(0)=1$, we write
\begin{align}
    A_{\rm m}(\phi) = 1+\alpha_{\rm m}\phi+{\cal O}(\phi^2),\quad
    \alpha_{\rm m} := \left.\frac{\D\ln A_{\rm m}}{\D\phi}\right|_{\phi=0}.
    \label{eq: conformal matter strength}
\end{align}
Thus $\alpha_{\rm m}$ measures the linear response of the physical metric to the scalar. To linear order in wave amplitudes, Eq.~(\ref{eq: conformal matter coupling}) gives
\begin{align}
    \tilde h_{\mu\nu} = h_{\mu\nu}+2\alpha_{\rm m}\delta\phi\,\eta_{\mu\nu},
    \label{eq: conformal scalar response}
\end{align}
where $h_{\mu\nu}$ and $\tilde h_{\mu\nu}$ perturb $g_{\mu\nu}$ and $\tilde g_{\mu\nu}$, respectively. For a massless scalar wave, the conformal term produces a transverse breathing polarization, an isotropic change in the separation of test masses in the transverse plane. With the usual normalization of the breathing strain, the $h_{\rm S}$ introduced in Eq.~(\ref{eq: ringdown detector strain}) is
\begin{align}
    h_{\rm S} = 2\alpha_{\rm m}\delta\phi,
    \label{eq: scalar breathing strain}
\end{align}
with the radiative field evaluated at the observer~\cite{Evstafyeva:2023scalar}. Brans--Dicke theory is a familiar example with $\alpha_{\rm m}\neq0$~\cite{Brans:1961sx}. Relating $\alpha_{\rm m}$ to the vacuum couplings requires an underlying theory. Field redefinitions removing Ricci-dependent operators can also induce nonconformal matter couplings, so both sectors must be transformed consistently~\cite{Flanagan:2004bz}.

We next estimate the amplitudes in Eqs.~(\ref{eq: ringdown tensor signal}) and~(\ref{eq: ringdown scalar signal}). The nondegenerate scalar and gravitational Schwarzschild frequencies allow a perturbative expansion about the GR modes for weak mixing~\cite{McManus:2019ulj}.\footnote{If complex QNM frequencies approach each other, mixing and excitation can be enhanced. Resonances associated with avoided crossings in coupled systems are studied in Ref.~\cite{Takahashi:2025coupledresonance}. The estimates for the scalar and gravitational components assume nondegenerate frequencies. The GR odd/even degeneracy is treated separately in Sec.~\ref{sec: perturbation}.} We normalize the dominant component of each eigenmode to be of order unity. To first order, the mixing then gives
\begin{align}
    \text{gravitational-led}\quad &\Psi_\pm = {\cal O}(1),\quad 
    \Psi_{\rm s} = {\cal O}(\epsilon),
    \notag\\
    \text{scalar-led}\quad &\Psi_{\rm s} = {\cal O}(1),\quad
    \Psi_\pm = {\cal O}(\epsilon).
    \label{eq: ringdown mode components}
\end{align}
Here $\Psi_\pm$ denotes the gravitational components collectively. Equation~(\ref{eq: ringdown mode components}) gives the relative sizes of the components of each normalized linear mode. How strongly each mode is excited depends on the initial data.

We assume that the merger initially excites only the gravitational perturbations of even and odd parity and that their initial data agree with those in GR through first order in $\epsilon$. The first-order mixing terms in the scalar equation then give
\begin{align}
    \delta\phi = {\cal O}(\epsilon).
    \label{eq: ringdown excitation assumption}
\end{align}
A time-dependent gravitational source can produce both a driven response at gravitational-led frequencies and transient ringing at scalar-led frequencies~\cite{Evstafyeva:2023scalar,Lestingi:2025contamination}. Thus the excitation coefficients multiplying the normalized gravitational-led and scalar-led eigenmodes are generically ${\cal O}(1)$ and ${\cal O}(\epsilon)$, respectively.

Combining excitation with Eq.~(\ref{eq: ringdown mode components}), and holding $\alpha_{\rm m}$ fixed in the $\epsilon$ expansion, gives
\begin{align}
    {\cal A}_{\rm T}^{({\rm g},\sigma)} &= {\cal O}(1),\quad
    {\cal A}_{\rm T}^{({\rm s})} = {\cal O}(\epsilon^2),
\notag\\
    {\cal A}_{\rm S}^{({\rm g},\sigma)} &= {\cal O}(\alpha_{\rm m}\epsilon),\quad
    {\cal A}_{\rm S}^{({\rm s})} = {\cal O}(\alpha_{\rm m}\epsilon).
    \label{eq: scalar signal counting}
\end{align}
The scalar-led contribution to $h_{\rm T}$ combines an ${\cal O}(\epsilon)$ excitation with an ${\cal O}(\epsilon)$ metric component. For such initial data, the scalar response is linear in the coupling and the scalar-led amplitude in the gravitational waveform is quadratic in the coupling, as shown in Eqs.~(8b), (9a), and~(9b) of Ref.~\cite{Macedo:2026coupled} and confirmed by the time-domain evolutions in that work. In the scalar polarization, the gravitational-led term combines an excitation of order unity with an ${\cal O}(\epsilon)$ scalar component, while the scalar-led term combines an ${\cal O}(\epsilon)$ excitation with a scalar component of order unity. Both branches therefore contribute to the scalar polarization at the same leading order. Setting $\alpha_{\rm m}=0$ removes the direct breathing response but leaves scalar-led tensor radiation from vacuum mixing possible.

Using the results of Sec.~\ref{sec: perturbation}, we write the QNM frequencies in the ringdown waveforms of Eqs.~(\ref{eq: ringdown tensor signal}) and~(\ref{eq: ringdown scalar signal}) as
\begin{align}
    r_{\rm g}\omega^{{\rm grav},\sigma}
    &= \Omega_{\rm Sch}^{\rm grav}+\epsilon^2\delta\Omega_{\rm 2nd}^{{\rm grav},\sigma}
    +{\cal O}(\epsilon^3),
    \notag\\
    r_{\rm g}\omega^{\rm s}
    &= \Omega_{\rm Sch}^{\rm scalar}+\epsilon\delta\Omega_{\rm 1st}^{\rm s}
    +\epsilon^2\delta\Omega_{\rm 2nd}^{\rm s}+{\cal O}(\epsilon^3).
    \label{eq: ringdown frequency connection}
\end{align}
Here $\delta\Omega_{\rm 1st}^{\rm s}$ and $\delta\Omega_{\rm 2nd}^{\rm s}$ are the scalar-led shifts in Eqs.~(\ref{eq: delomega^s_1st}) and~(\ref{eq: scalar QNM}), and $\delta\Omega_{\rm 2nd}^{{\rm grav},\sigma}={\cal A}+\sigma\sqrt{{\cal B}}$ gives the shift for branch $\sigma$ in Eq.~(\ref{eq: grav QNM}).
For these initial data, the scalar-led mode contributes to the tensor ringdown at ${\cal O}(\epsilon^2)$, the same order as the gravitational-led frequency shifts~\cite{Lestingi:2025contamination,Asuncion:2026contamination}. At leading order, the complex frequency of this scalar-led contribution to the tensor strain is $\Omega_{\rm Sch}^{\rm scalar}/r_{\rm g}$. Because its tensor amplitude is already ${\cal O}(\epsilon^2)$, the shift in the scalar-led frequency at first order enters $h_{\rm T}$ at ${\cal O}(\epsilon^3)$ for $\tau/r_{\rm g}={\cal O}(1)$. Analyses of simulated signals in dCS and sGB illustrate the effect of including this additional tensor component in ringdown parameter estimation~\cite{Crescimbeni:2025kxi}.

Conformal coupling makes both branches accessible through an additional polarization with leading amplitudes ${\cal O}(\alpha_{\rm m}\epsilon)$. The first scalar-led frequency shift affects this strain at ${\cal O}(\alpha_{\rm m}\epsilon^2)$, whereas gravitational-led frequency shifts enter it at ${\cal O}(\alpha_{\rm m}\epsilon^3)$ for the same range of $\tau/r_{\rm g}$. Frequency shifts change phase evolution and damping, while complex amplitudes set magnitude and phase at $t_0$. As in Sec.~\ref{sec: observational implications}, the gravitational-led frequency shifts depend on $b_i,e_i$, while the shift in the scalar-led frequency at first order depends on $c_1,f_1,f_3$.

The leading amplitudes in the breathing polarization depend on $\alpha_{\rm m}$, the $b_i,e_i$ sources, and excitation. With only $b_1$ nonzero, for example, the nonzero amplitudes in $h_{\rm S}$ are proportional to $\alpha_{\rm m}\epsilon b_1\epsilon_2$, while $\delta\Omega_{\rm 1st}^{\rm s}=0$. These leading amplitudes alone do not separate the vacuum coupling from matter coupling and excitation. We have computed the QNM frequencies and estimated how the amplitudes scale with $\epsilon$ for the initial data leading to Eq.~(\ref{eq: ringdown excitation assumption}). Determining the excitation coefficients, observable amplitude ratios, and detectability requires an initial-value or merger calculation and a signal analysis. Independent scalar excitation of order unity would change the amplitude hierarchy.

\section{Summary}\label{sec: summary}

We studied black hole solutions and their quasinormal modes in a covariant scalar-tensor EFT. In Sec.~\ref{sec: theory}, we identified an independent basis of scalar-curvature operators of both parities through mass dimension eight relevant to this calculation. We set the independent pure curvature couplings to zero to isolate the effects of the scalar-curvature interactions.

In Sec.~\ref{sec: BH}, we constructed static and slowly rotating black hole solutions with a massless scalar through second order in the couplings and first order in spin. Only parity-even scalar sources correct the static metric at this order. The correction to frame dragging also depends on parity-odd couplings, and equatorial reflection asymmetry requires products of one parity-even and one parity-odd coupling.

In Sec.~\ref{sec: perturbation}, we derived the coupled master equations on the static backgrounds and calculated the fundamental $\ell=2$ QNM frequencies. The $c_1,f_1$, and $f_3$ interactions shift the scalar-led frequencies at first order. The gravitational-led shifts begin at second order, depend on $b_i,e_i$, and generally lift the degeneracy of the two Schwarzschild gravitational modes. The $b_2$ and $e_2$ couplings, for example, enter the gravitational-led shifts even though they do not affect the static metric through this order.

In Sec.~\ref{sec: light mass limit}, we studied the limit of small scalar mass in dCS gravity with the additional $c_1$ interaction. The background remains Schwarzschild and the even-parity mode retains its GR frequency, while the scalar-led frequency and the gravitational-led frequency in the odd sector both shift with the scalar mass.

In Sec.~\ref{sec: integrating out}, we integrated out the heavy scalar in the dCS and sGB examples. The leading low-frequency gravitational QNM shifts agree with those of quartic curvature gravity, including the effect of the corrected background in sGB. We also obtained the first subleading corrections in the inverse scalar mass. At the action level, the scalar sources start at quadratic order in curvature, so scalar exchange at tree level first generates quartic curvature operators with correlated coefficients. For generic couplings, matching through mass dimension sixteen captures the first contribution from each scalar-curvature interaction in our Lagrangian.

In Sec.~\ref{sec: discussion}, we examined how the coupled modes enter the ringdown of a static black hole with a massless scalar. For purely gravitational initial data that agree with GR through first order in $\epsilon$, mixing produces a scalar-led contribution to the tensor ringdown at ${\cal O}(\epsilon^2)$, the same order as the gravitational-led frequency shifts. A linear conformal coupling to matter also gives a breathing polarization to which both branches contribute at ${\cal O}(\alpha_{\rm m}\epsilon)$. The QNM spectra of rotating black holes and the regime of intermediate scalar masses remain to be explored.

\begin{acknowledgments}

We thank T. Kobayashi and C.~M.~Yoo for fruitful discussions.
S.H. acknowledges support from internal research incentive funds at Hyogo University of Teacher Education. M.K. is supported by JSPS KAKENHI Grant Number JP22K03626. M.Y. is supported by the Institute for Basic Science (project code IBS-R018-D3) and by JSPS KAKENHI Grant Number JP23K20843.
The authors used ChatGPT (OpenAI, GPT-6.2 Astra) and Claude (Anthropic, Fable 5.2) to assist in checking derivations, identifying possible inconsistencies, and improving the presentation of the manuscript. All AI-assisted outputs were independently checked by the authors, who take full responsibility for the content.

\end{acknowledgments}

\appendix

\section{Construction of the Lagrangian}\label{sec: construction of Lagrangian}

This Appendix explains the reduction of the covariant scalar-tensor operators to the basis used in Eq.~(\ref{eq: total Lagrangian}). We first describe perturbative field redefinitions, then collect the curvature identities needed for the reduction, and finally impose the restrictions on mass dimension and perturbative order of Sec.~\ref{sec: theory}. General scalar-tensor EFT constructions are discussed in Refs.~\cite{Weinberg:2008hq,Babichev:2025dpb}.

In the notation of Sec.~\ref{sec: theory}, the curvature contractions are
\begin{align*}
    {\cal C}_2 &= R_{\mu\nu\rho\sigma}R^{\mu\nu\rho\sigma},\quad
    \tilde{\cal C}_2 =\tilde R_{\mu\nu\rho\sigma}R^{\mu\nu\rho\sigma},
\\
    {\cal C}_3 &= R_{\mu\nu\rho\sigma}R^{\rho\sigma\alpha\beta}R_{\alpha\beta}{}^{\mu\nu},\quad
    \tilde{\cal C}_3 = \tilde R_{\mu\nu\alpha\beta}R^{\alpha\beta\rho\sigma}R_{\rho\sigma}{}^{\mu\nu}.
\end{align*}
We use a canonically normalized scalar $\varphi$, with mass dimension $[\varphi]=1$, for the construction below. Its relation to the dimensionless field used in the main text is given at the end of this Appendix.

\subsection{Field redefinitions}\label{sec: field redefinitions}

Local perturbative field redefinitions remove operators proportional to the leading equations of motion, up to integration by parts. The coefficients in the main text are defined in the resulting reduced basis. When matching a specified action to this basis through ${\cal O}(\epsilon^2)$, one must also include the terms quadratic in the first-order field shifts and the changes they induce in the interaction terms. These higher-order effects are discussed systematically in Ref.~\cite{Criado:2018sdb}. The first-order formulas below illustrate the reduction but do not by themselves give the complete second-order matching.

Under the conformal transformation
\begin{align}
    g_{\mu\nu} \rightarrow e^{\epsilon Q(x)}g_{\mu\nu},
    \label{eq: conformal trans}
\end{align}
where $Q$ is a local dimensionless scalar, the volume element and Ricci scalar transform in four dimensions as
\begin{align}
    \sqrt{-g} &\rightarrow e^{2\epsilon Q}\sqrt{-g},\\
    R&\rightarrow R-\epsilon(QR+3\Box Q)+{\cal O}(\epsilon^2).
\end{align}
Consequently,
\begin{align}
    \sqrt{-g}(R-\epsilon RQ)
    \rightarrow \sqrt{-g}\left[R+{\cal O}(\epsilon^2)
    +\text{total derivatives}\right].
\end{align}
A suitable choice of $Q$ therefore removes a Ricci scalar operator at first order.

Similarly, consider the metric shift
\begin{align}
    g_{\mu\nu} \rightarrow g_{\mu\nu}-\epsilon Y_{\mu\nu},
    \label{eq: metric shift}
\end{align}
where $Y_{\mu\nu}$ is symmetric and $Y=g^{\mu\nu}Y_{\mu\nu}$. To first order,
\begin{align}
    \sqrt{-g} &\rightarrow \sqrt{-g}\left(1-\frac{\epsilon}{2}Y\right)
       +{\cal O}(\epsilon^2),\\
    R &\rightarrow R+\epsilon R_{\mu\nu}Y^{\mu\nu}
       -\epsilon\nabla_\mu\nabla_\nu Y^{\mu\nu}
       +\epsilon\Box Y+{\cal O}(\epsilon^2).
\end{align}
Combining this shift with the conformal transformation gives, up to total derivatives,
\begin{align}
    \sqrt{-g}(R-\epsilon R_{\mu\nu}Y^{\mu\nu})
    \rightarrow \sqrt{-g}\left[R+\epsilon R\left(Q-\frac{Y}{2}\right)
       +{\cal O}(\epsilon^2)\right].
\end{align}
The remaining Ricci scalar term is removed by choosing $Q=Y/2$.

These formulas give only the Einstein--Hilbert contribution. The scalar kinetic and potential terms also transform. For example, take $Q=q{\cal C}_2/\Lambda^4$, where $q$ is a dimensionless constant. For the free scalar Lagrangian, ${\cal L}_{\rm scalar}=-(\nabla\varphi)^2/2-m^2\varphi^2/2$, the conformal transformation gives
\begin{align}
    \frac{\Delta(\sqrt{-g}{\cal L}_{\rm scalar})}{\sqrt{-g}}
    &= -\frac{\epsilon q}{2\Lambda^4}(\nabla\varphi)^2{\cal C}_2
    -\frac{\epsilon q m^2}{\Lambda^4}\varphi^2{\cal C}_2
    +{\cal O}(\epsilon^2).
    \label{eq: conformal scalar example}
\end{align}
These terms have the structures multiplying $f_3$ and $c_1$, respectively. Thus, eliminating a Ricci operator can shift other coefficients already present in the basis. All induced terms within the truncation must be retained even when the background is Ricci flat at zeroth order.

For the scalar, the redefinition
\begin{align}
    \varphi\rightarrow\varphi-\epsilon Z(x)
    \label{eq: scalar field redefinition}
\end{align}
changes the free scalar Lagrangian, up to a total derivative, by
\begin{align}
    \Delta{\cal L}_{\rm scalar}
    = -\epsilon Z(\Box-m^2)\varphi+{\cal O}(\epsilon^2).
\end{align}
Here $Z$ is a local scalar of mass dimension one. This transformation removes operators proportional to the leading scalar equation. If one instead removes a term proportional to $\Box\varphi$, the accompanying $m^2\varphi Z$ contribution must be retained. These reductions specify a basis. Omitting independent scalar self-interactions and pure curvature couplings is a separate restriction on the model, as described in App.~\ref{sec: EFT operators}.

Metric transformations also provide a method for generating black hole solutions in the correspondingly transformed theories. Invertible conformal transformations with parity-violating factors yield exact rotating black hole solutions from Kerr~\cite{Tahara:2023pyg}. Disformal transformations have also been used to construct spherically symmetric hairy black holes~\cite{BenAchour:2019Disformal} and rotating solutions from stealth Kerr~\cite{Anson:2020Disforming,BenAchour:2020Rotating}. Such transformations can likewise generate solutions in EFT, provided that the action and the solution are transformed consistently through the same perturbative order.

\subsection{Useful identities for curvature tensors}\label{sec: prop of Riemann}

This subsection collects the identities needed to reduce the curvature structures in the operators considered here. Integration by parts, Bianchi identities, and commutators of covariant derivatives convert derivatives of curvature into algebraic curvature contractions, derivatives acting on the scalar, and Ricci-dependent terms. Four-dimensional identities then reduce the remaining algebraic contractions to the independent combinations listed in Table~\ref{tab: independent curvature basis}~\cite{Fulling:1992vm,Hervey:1995aa,Chung:2022ees,Babichev:2025dpb}.

Here, Ricci-dependent terms contain at least one Ricci tensor, Ricci scalar, or covariant derivative of either. Relations valid up to Ricci-dependent terms are written below with a remainder ${\cal O}(R_{\mu\nu})$. The remainder labels the tensor structures that have been omitted, independently of their order in $\epsilon$. Removing these terms from the action requires the field redefinitions of App.~\ref{sec: field redefinitions}, including all induced scalar interactions within this truncation.

The algebraic and differential Bianchi identities are
\begin{align}
    R_{\mu\nu\rho\sigma}+R_{\mu\rho\sigma\nu}+R_{\mu\sigma\nu\rho} &= 0,
    \label{eq: 1st Bianchi}\\
    \nabla_\alpha R_{\mu\nu\rho\sigma}
    +\nabla_\rho R_{\mu\nu\sigma\alpha}
    +\nabla_\sigma R_{\mu\nu\alpha\rho} &= 0.
    \label{eq: 2nd Bianchi}
\end{align}
Contracting the differential Bianchi identity, Eq.~(\ref{eq: 2nd Bianchi}), gives
\begin{align}
    \nabla^\alpha R_{\alpha\mu\nu\rho}
    &= \nabla_\nu R_{\mu\rho}-\nabla_\rho R_{\mu\nu}.
    \label{eq: contracted 2nd Bianchi 1}
\end{align}
Applying one more derivative to Eq.~(\ref{eq: contracted 2nd Bianchi 1}) and combining Eq.~(\ref{eq: 2nd Bianchi}) with commutators of covariant derivatives yields the exact relation
\begin{align}
    \Box R_{\mu\nu\rho\sigma}
    &= \nabla_\rho\nabla_\mu R_{\sigma\nu}
    -\nabla_\rho\nabla_\nu R_{\sigma\mu}
    -\nabla_\sigma\nabla_\mu R_{\rho\nu}
    +\nabla_\sigma\nabla_\nu R_{\rho\mu}
    \notag\\
    &\quad +R_{\alpha\rho}R^\alpha{}_{\sigma\mu\nu}
    -R_{\alpha\sigma}R^\alpha{}_{\rho\mu\nu}
    -2R^\beta{}_{\sigma\alpha\rho}R^\alpha{}_{\beta\mu\nu}
    \notag\\
    &\quad -2R^\beta{}_{\mu\alpha\rho}R^\alpha{}_{\sigma\beta\nu}
    -2R^\beta{}_{\nu\alpha\rho}R^\alpha{}_{\sigma\mu\beta}.
    \label{eq: contracted 2nd Bianchi 2}
\end{align}
The Ricci terms are explicit in these formulas. The remaining terms in Eq.~(\ref{eq: contracted 2nd Bianchi 2}) are quadratic in the Riemann tensor.

We use the convention for the dual tensor adopted in the main text, without a factor of $1/2$,
\begin{align}
    \tilde R_{\mu\nu\rho\sigma}
    :=\varepsilon_{\mu\nu}{}^{\alpha\beta}R_{\alpha\beta\rho\sigma}.
    \label{eq: appendix dual convention}
\end{align}
The dual tensor is not generally symmetric under exchange of the two index pairs. More precisely,
\begin{align}
    \tilde R_{\mu\nu\rho\sigma}-\tilde R_{\rho\sigma\mu\nu}
    &= 2\varepsilon_{\mu\nu\rho}{}^\alpha R_{\sigma\alpha}
    -2\varepsilon_{\mu\nu\sigma}{}^\alpha R_{\rho\alpha}
    -R\varepsilon_{\mu\nu\rho\sigma}.
    \label{eq: dual pair exchange}
\end{align}
Equation~(\ref{eq: dual pair exchange}) shows that symmetry under exchange of the index pairs holds on a Ricci-flat background. Independently of this property, the algebraic Bianchi identity, Eq.~(\ref{eq: 1st Bianchi}), implies $g^{\mu\rho}\tilde R_{\mu\nu\rho\sigma}=0$. Products containing an even number of duals can be reduced by contracting the Levi-Civita tensors. With the normalization in Eq.~(\ref{eq: appendix dual convention}), a useful example is
\begin{align}
    \tilde R_{\mu\nu}{}^{\alpha\beta}\tilde R_{\alpha\beta\rho\sigma}
    &= -4R_{\mu\nu}{}^{\alpha\beta}R_{\alpha\beta\rho\sigma} +{\cal O}(R_{\mu\nu}).
    \label{eq: tildeRiemann-tildeRiemann1}
\end{align}
The remainder in Eq.~(\ref{eq: tildeRiemann-tildeRiemann1}) is needed for a general metric and vanishes on a Ricci-flat metric.

Since $\nabla_\lambda\varepsilon_{\mu\nu\rho\sigma}=0$, dualizing the first index pair in Eq.~(\ref{eq: 2nd Bianchi}) gives the corresponding differential identity for $\tilde R_{\mu\nu\rho\sigma}$. Its divergences on the two index pairs satisfy
\begin{align}
    \nabla^\alpha\tilde R_{\alpha\mu\nu\rho} &= 0,\label{eq: dual first divergence}
\\
    \nabla^\alpha\tilde R_{\nu\rho\alpha\mu}
    &= 2\varepsilon_{\nu\rho}{}^{\alpha\beta}\nabla_\alpha R_{\mu\beta}.\label{eq: dual second divergence}
\end{align}
Equation~(\ref{eq: dual first divergence}) is exact without imposing Ricci flatness. For the second derivative, it is convenient to use the exact identity
\begin{align}
    \Box\tilde R_{\mu\nu\rho\sigma}
    &= \varepsilon_{\mu\nu}{}^{\alpha\beta}\Box R_{\alpha\beta\rho\sigma},\label{eq: contracted tR 2nd Bianchi}
\end{align}
with $\Box R_{\alpha\beta\rho\sigma}$ given by Eq.~(\ref{eq: contracted 2nd Bianchi 2}). This form keeps all Ricci terms and avoids exchanging the index pairs of a general dual Riemann tensor.

We next simplify the algebraic curvature contractions. In four dimensions, the Weyl tensor $W_{\mu\nu\rho\sigma}$ obeys the dimension-dependent identity~\cite{Babichev:2025dpb}
\begin{align}
    W^{[\rho\sigma}{}_{[\mu\nu}\delta^{\alpha]}{}_{\beta]} = 0.
    \label{eq: Weyl dimension identity}
\end{align}
Because the Weyl and Riemann tensors differ only by Ricci-dependent terms, identities for Weyl contractions also reduce Riemann contractions up to such terms.

First consider quadratic curvature products with two free indices, which can contract with derivatives of the scalar. Contracting Eq.~(\ref{eq: Weyl dimension identity}) with one Weyl tensor and rewriting the result in terms of the Riemann tensor gives the parity-even relation below. The parity-odd relation in Eq.~(\ref{eq: nontrivial tRR}) is exact.\footnote{The parity-odd relation follows from the four-dimensional two-form identity $\tilde F_{\mu\alpha}F_\nu{}^\alpha=\tfrac14g_{\mu\nu}\tilde F_{\alpha\beta}F^{\alpha\beta}$. Apply it to the first index pair of $R_{\mu\alpha\rho\sigma}$ and contract the remaining pair $\rho,\sigma$.}
\begin{align}
    R_{\mu\alpha\beta\gamma}R_\nu{}^{\alpha\beta\gamma}
    &= \frac14g_{\mu\nu}{\cal C}_2+{\cal O}(R_{\mu\nu}),\label{eq: nontrivial RR}
\\
    \tilde R_{\mu\alpha\beta\gamma}R_\nu{}^{\alpha\beta\gamma}
    &= \frac14g_{\mu\nu}\tilde{\cal C}_2.\label{eq: nontrivial tRR}
\end{align}
Contracting Eqs.~(\ref{eq: nontrivial RR}) and~(\ref{eq: nontrivial tRR}) with $\nabla^\mu\varphi\nabla^\nu\varphi$ gives $\tfrac14(\nabla\varphi)^2{\cal C}_2$ and $\tfrac14(\nabla\varphi)^2\tilde{\cal C}_2$, respectively, up to Ricci-dependent terms in the first contraction. These are the operator structures multiplying $f_3$ and $f_4$ in Eq.~(\ref{eq: STEFT Lagrangian}).

We now turn to fully contracted curvature products. At quadratic order, ${\cal C}_2$ and $\tilde{\cal C}_2$ span the contractions of both parities up to Ricci-dependent terms. At cubic order, contractions of Eq.~(\ref{eq: Weyl dimension identity}) with two Weyl tensors, together with the algebraic Bianchi identity~(\ref{eq: 1st Bianchi}), relate the different index contractions. A representative reduction is~\cite{Fulling:1992vm,Hervey:1995aa,Chung:2022ees,Babichev:2025dpb}
\begin{align}
    R_{\mu\nu\rho\sigma}R^\mu{}_{\alpha}{}^\rho{}_{\beta}R^{\nu\alpha\sigma\beta}
    &= \frac12{\cal C}_3+{\cal O}(R_{\mu\nu}).\label{eq: nontri cubic1}
\end{align}
Only one independent cubic contraction remains in each parity sector. We choose ${\cal C}_3$ and $\tilde{\cal C}_3$.

At quartic order, four-dimensional identities similarly leave two parity-even contractions and one parity-odd contraction up to Ricci-dependent terms~\cite{Fulling:1992vm,Hervey:1995aa,Chung:2022ees}. For example, in terms of the quadratic invariants, the chain contraction is
\begin{align}
    R_{\mu\nu}{}^{\rho\sigma}R_{\rho\sigma}{}^{\alpha\beta}
    R_{\alpha\beta}{}^{\gamma\delta}R_{\gamma\delta}{}^{\mu\nu}
    &= \frac14({\cal C}_2)^2-\frac1{16}(\tilde{\cal C}_2)^2 +{\cal O}(R_{\mu\nu}).\label{eq: quartic chain reduction}
\end{align}
The factor $1/16$ follows from the dual normalization in Eq.~(\ref{eq: appendix dual convention}). A convenient quartic basis is therefore $({\cal C}_2)^2$, $(\tilde{\cal C}_2)^2$, and ${\cal C}_2\tilde{\cal C}_2$.

The differential and algebraic identities can also be used together. Contracting Eq.~(\ref{eq: contracted 2nd Bianchi 2}) with $R^{\mu\nu\rho\sigma}$ and applying the cubic reduction~(\ref{eq: nontri cubic1}) gives
\begin{align}
    R^{\mu\nu\rho\sigma}\Box R_{\mu\nu\rho\sigma}
    &= -3{\cal C}_3+{\cal O}(R_{\mu\nu}).\label{eq: derivative curvature reduction}
\end{align}
Up to Ricci-dependent terms, the operator on the left therefore reduces to a cubic Riemann invariant.

The algebraic curvature basis in Table~\ref{tab: independent curvature basis} contains one invariant in each parity sector at quadratic and cubic orders, and two parity-even invariants and one parity-odd invariant at quartic order. These curvature structures are used to construct the scalar-curvature and pure curvature operators in App.~\ref{sec: EFT operators}.
\begin{table}[!htbp]
\centering
\begin{manuscripttableblock}{0.70\textwidth}
\centering
    \caption{Independent algebraic curvature invariants in four dimensions.}
    \label{tab: independent curvature basis}
    \setlength{\tabcolsep}{12pt}
    \renewcommand{\arraystretch}{1.15}
    \begin{tabular}{c c c}
        \hline\hline
        Curvature order&Parity even&Parity odd\\
        \hline
        $2$&${\cal C}_2$&$\tilde{\cal C}_2$\\
        $3$&${\cal C}_3$&$\tilde{\cal C}_3$\\
        $4$&$({\cal C}_2)^2,\ (\tilde{\cal C}_2)^2$&${\cal C}_2\tilde{\cal C}_2$\\
        \hline\hline
    \end{tabular}
\manuscripttablenote{Invariants are listed up to Ricci-dependent terms. Curvature order counts Riemann tensors. The labels give the parity of each contraction. We allow the theory to violate parity.}
\end{manuscripttableblock}
\end{table}

In four dimensions, ${\cal C}_2={\cal R}_{\rm GB}+4R_{\mu\nu}R^{\mu\nu}-R^2$, where ${\cal R}_{\rm GB}$ is the Gauss--Bonnet density, while $\tilde{\cal C}_2$ is proportional to the Pontryagin density. With constant coefficients, the Gauss--Bonnet and Pontryagin terms do not contribute to the local equations of motion. After the field redefinitions described in App.~\ref{sec: field redefinitions}, no independent pure curvature operator remains at quadratic order. All induced scalar interactions within this truncation must still be included. Coefficients depending on the scalar or its derivatives instead give nontrivial scalar-curvature interactions, as listed in Table~\ref{tab: appendix operator basis}.

\subsection{Independent EFT operators}\label{sec: EFT operators}

The free Lagrangian is $\mpl^2R/2-(\nabla\varphi)^2/2-m^2\varphi^2/2$. We work on the asymptotically flat branch perturbatively connected to GR with $\bar\varphi=\epsilon\varphi_{(1)}+{\cal O}(\epsilon^2)$, where $\varphi_{(1)}$ is the first-order background profile. As in Eq.~(\ref{eq: total Lagrangian}), each interaction carries one explicit factor of $\epsilon$, while its dimensionful coefficient and coupling are held fixed. The coefficients can have different numerical sizes. Scalar self-interactions other than the mass term are omitted in the reduced basis as a model assumption.

After eliminating redundant operators, the perturbative counting provides a separate restriction. For a scalar-curvature interaction proportional to $\epsilon\varphi^n$ with $n\geq2$ and fixed dimensionful coefficient, the term quadratic in the scalar perturbation carries the factor
\begin{align}
    \epsilon\bar\varphi^{\,n-2}
    = \epsilon^{n-1}\varphi_{(1)}^{\,n-2}+{\cal O}(\epsilon^n).\label{eq: appendix scalar power counting}
\end{align}
Thus scalar-curvature interactions cubic in $\varphi$ can enter linear perturbations at ${\cal O}(\epsilon^2)$, whereas terms with four or more scalar fields first enter beyond this order. The same counting applies to derivative interactions with four or more scalar factors, since derivatives of the background scalar are also ${\cal O}(\epsilon)$. For example, $(\nabla\varphi)^4$ is independent but contributes to the quadratic perturbation action only from ${\cal O}(\epsilon^3)$ when its coefficient carries one power of $\epsilon$.

Combining the curvature structures in Table~\ref{tab: independent curvature basis} with scalar fields and their derivatives, and applying the field redefinitions of App.~\ref{sec: field redefinitions} and the counting in Eq.~(\ref{eq: appendix scalar power counting}), gives the operator basis in Table~\ref{tab: appendix operator basis}. The table organizes the interactions in Eqs.~(\ref{eq: STEFT Lagrangian}) and~(\ref{eq: curvEFT Lagrangian}) by canonical mass dimension and identifies the corresponding couplings in the main text. The scalar-curvature terms are a subset of the general classification of interactions with four and six derivatives of Ref.~\cite{Babichev:2025dpb}. For example, terms with six derivatives and three differentiated canonical scalars have mass dimension nine and lie outside our truncation.

\begin{table}[!htbp]
\centering
\begin{manuscripttableblock}{0.84\textwidth}
\centering
    \setlength{\tabcolsep}{10pt}
    \renewcommand{\arraystretch}{1.15}
    \caption{Scalar-curvature and pure curvature operators through mass dimension eight.}
    \label{tab: appendix operator basis}
    \normalsize
    \begin{tabular}{c l l}
        \hline\hline
        Mass dimension&Independent operators&Couplings in the main text\\
        \hline
        $5$&$\varphi{\cal C}_2,\ \varphi\tilde{\cal C}_2$&$b_1,b_2$\\
        $6$&$\varphi^2{\cal C}_2,\ \varphi^2\tilde{\cal C}_2$&$c_1,c_2$\\
        $6$&${\cal C}_3,\ \tilde{\cal C}_3$&$g_1,g_2$\\
        $7$&$\varphi^3{\cal C}_2,\ \varphi^3\tilde{\cal C}_2$&$d_1,d_2$\\
        $7$&$\varphi{\cal C}_3,\ \varphi\tilde{\cal C}_3$&$e_1,e_2$\\
        $8$&$\varphi^2{\cal C}_3,\ \varphi^2\tilde{\cal C}_3$&$f_1,f_2$\\
        $8$&$(\nabla\varphi)^2{\cal C}_2,\ (\nabla\varphi)^2\tilde{\cal C}_2$&$f_3,f_4$\\
        $8$&$({\cal C}_2)^2,\ {\cal C}_2\tilde{\cal C}_2,\ (\tilde{\cal C}_2)^2$&$h_1,h_2,h_3$\\
        \hline\hline
    \end{tabular}
\manuscripttablenote{Dimensions use the canonical scalar $\varphi$, with $[\varphi]=1$. Couplings refer to Eqs.~(\ref{eq: STEFT Lagrangian}) and~(\ref{eq: curvEFT Lagrangian}), in the order shown. The pure curvature coefficients $g_i,h_i$ are independent but are set to zero in our calculation.}
\end{manuscripttableblock}
\end{table}

Finally, we introduce the dimensionless scalar used in the main text,
\begin{align}
    \phi := \frac{\varphi}{\mpl},\quad [\phi] = 0,\label{eq: appendix canonical scalar}
\end{align}
and factor $\mpl^2$ out of the Lagrangian. Assigning the coefficient normalization of Sec.~\ref{sec: theory} then reproduces Eq.~(\ref{eq: total Lagrangian}), with the interaction terms in Eqs.~(\ref{eq: STEFT Lagrangian}) and~(\ref{eq: curvEFT Lagrangian}) multiplied by $\epsilon$. Since $\mpl$ is independent of $\epsilon$, this constant rescaling leaves all perturbative orders unchanged. The pure curvature coefficients $g_i$ and $h_i$ are independent basis coefficients. Setting them to zero therefore restricts the theory after the field redefinitions. A specified theory written in other variables must first be matched consistently to this basis before the restriction is imposed.

\section{Lagrangian as low-energy EFT}\label{sec: LEEFT}

The normalization of the EFT coefficients depends on the low-energy power counting~\cite{deRham:2020ejn}. Here the mass dimension of each operator fixes the power of its suppression scale. Using the canonical scalar $\varphi=\mpl\phi$ of Eq.~(\ref{eq: appendix canonical scalar}), we count $[\varphi]=[\nabla]=1$ and $[R_{\mu\nu\rho\sigma}]=2$. We choose the powers of the suppression scales so that each term in the Lagrangian density has mass dimension four. We allow independent scales $\Lambda_i$ for the operator families in Table~\ref{tab: appendix operator basis}.

We use the same bookkeeping convention as in Sec.~\ref{sec: theory} and write
\begin{align}
    {\cal L} &= {\cal L}_{\rm free} +\epsilon{\cal L}_{\rm EFT} +\cdots,\label{eq: LEEFT canonical action}
\\
    {\cal L}_{\rm free} &= \frac{\mpl^2}{2}R -\frac{1}{2}(\nabla\varphi)^2 -\frac{1}{2}m^2\varphi^2,
\\
    {\cal L}_{\rm EFT}
    &= \frac{1}{\Lambda_{\rm b}}\Big(b_1\varphi{\cal C}_2
    +b_2\varphi\tilde{\cal C}_2\Big)
    +\frac{1}{\Lambda_{\rm c}^2}\Big(c_1\varphi^2{\cal C}_2
    +c_2\varphi^2\tilde{\cal C}_2\Big)
    +\frac{1}{\Lambda_{\rm d}^3}\Big(d_1\varphi^3{\cal C}_2
    +d_2\varphi^3\tilde{\cal C}_2\Big)
\notag\\
    &\quad +\frac{1}{\Lambda_{\rm e}^3}\Big(e_1\varphi{\cal C}_3
    +e_2\varphi\tilde{\cal C}_3\Big)
    +\frac{1}{\Lambda_{\rm f}^4}\Big[f_1\varphi^2{\cal C}_3
    +f_2\varphi^2\tilde{\cal C}_3
    +f_3(\nabla\varphi)^2{\cal C}_2
    +f_4(\nabla\varphi)^2\tilde{\cal C}_2\Big]
\notag\\
    &\quad +\frac{1}{\Lambda_{\rm g}^2}\Big(g_1{\cal C}_3
    +g_2\tilde{\cal C}_3\Big)
    +\frac{1}{\Lambda_{\rm h}^4}\Big[h_1({\cal C}_2)^2
    +h_2{\cal C}_2\tilde{\cal C}_2
    +h_3(\tilde{\cal C}_2)^2\Big].
    \label{eq: LEEFT canonical interactions}
\end{align}
We can set $\epsilon=1$ after the expansion. Here $\mathcal L_{\rm free}$ denotes the canonically normalized free sector. The operator basis and curvature contractions are those of App.~\ref{sec: construction of Lagrangian}. The ellipsis denotes omitted interactions. We include the pure curvature terms with $g_i,h_i$ for comparison, although these coefficients are set to zero in the calculations in the main text.

This normalization assigns a separate suppression scale to each family of operators. Translating a fixed theory between Eq.~(\ref{eq: LEEFT canonical interactions}) and Eqs.~(\ref{eq: STEFT Lagrangian}), (\ref{eq: curvEFT Lagrangian}) requires matching the full coefficients after the scalar rescaling. For example, keeping the same $b_i$ requires $\Lambda_{\rm b}^{-1}=\mpl/\Lambda^2$. For the estimates in this appendix, we place the suppression in the separate scales $\Lambda_i$ and take the dimensionless couplings to be of order unity. In the main text, we instead use a common $\Lambda$ and leave the coupling-weighted strengths independent, without assuming that the dimensionless couplings are of order unity. Choosing comparable $\Lambda_i$ below is an additional benchmark hierarchy, not an assumption underlying the results in the main text. The regime of validity must also satisfy the considerations in App.~\ref{sec: validity and causality}.

We estimate derivatives and curvature on the black hole scale as $\nabla\sim r_{\rm g}^{-1}$ and $R_{\mu\nu\rho\sigma}\sim r_{\rm g}^{-2}$. Using the canonical kinetic normalization for the scalar and gravitational perturbations gives the dimensionless coefficient combinations
\begin{align}
    \epsilon_{\rm b} &:= \frac{1}{\Lambda_{\rm b}\mpl r_{\rm g}^2},\quad
    \epsilon_{\rm c} := \frac{1}{\Lambda_{\rm c}^2r_{\rm g}^2},\quad
    \epsilon_{\rm d} := \frac{1}{\Lambda_{\rm d}^3r_{\rm g}^3},\quad
    \epsilon_{\rm e} := \frac{1}{\Lambda_{\rm e}^3\mpl r_{\rm g}^4},
    \label{eq: LEEFT scalar parameters}
\\
    \epsilon_{\rm f} &:= \frac{1}{\Lambda_{\rm f}^4r_{\rm g}^4},\quad
    \epsilon_{\rm g} := \frac{1}{\Lambda_{\rm g}^2\mpl^2r_{\rm g}^4},\quad
    \epsilon_{\rm h} := \frac{1}{\Lambda_{\rm h}^4\mpl^2r_{\rm g}^6}.
    \label{eq: LEEFT remaining parameters}
\end{align}
The combinations $\epsilon_{\rm c}$, $\epsilon_{\rm d}$, and $\epsilon_{\rm f}$ characterize scalar interactions without Planck suppression. For $b_i$ and $e_i$, canonical normalization of the metric perturbation supplies a factor of $\mpl^{-1}$, giving $\epsilon_{\rm b}$ and $\epsilon_{\rm e}$ for the mixing between scalar and gravitational perturbations. The pure curvature parameters $\epsilon_{\rm g}$ and $\epsilon_{\rm h}$ measure corrections relative to the Einstein--Hilbert term. Actual corrections also contain the corresponding couplings and any background scalar factors. In particular, the $d_i$ contribution to the quadratic scalar perturbation operator carries an additional factor $r_{\rm g}\bar\varphi$, consistent with Eq.~(\ref{eq: appendix scalar power counting}).

Table~\ref{tab: LEEFT scales} evaluates the parameters in Eqs.~(\ref{eq: LEEFT scalar parameters}) and~(\ref{eq: LEEFT remaining parameters}) for a $30M_\odot$ black hole, taking $r_{\rm g}\sim10^2\,{\rm km}$ and setting the bookkeeping parameter $\epsilon=1$. The middle column uses $(\Lambda_i r_{\rm g})^{-1}=0.1$ for each family. The last column estimates the scale at which each $\epsilon_i$ becomes ${\cal O}(1)$. These scales are illustrative. They do not represent observational bounds or enter as assumptions in the perturbative predictions. The actual corrections must remain small in applying the truncated solutions.

\begin{table}[!htbp]
\centering
\begin{manuscripttableblock}{0.72\textwidth}
\caption{EFT parameters and suppression lengths for $r_{\rm g}\sim10^2\,{\rm km}$.}
\label{tab: LEEFT scales}
\normalsize
\setlength{\tabcolsep}{18pt}
\renewcommand{\arraystretch}{1.25}
\begin{tabular}{ccc}
\hline\hline
Couplings & At $\Lambda_i^{-1}=0.1r_{\rm g}$ & At $\epsilon_i\sim1$\\
\hline
$b_1,b_2$ & $\epsilon_{\rm b}\sim10^{-40}$ & $\Lambda_{\rm b}^{-1}\sim10^{18}\,{\rm Gpc}$\\
$c_1,c_2$ & $\epsilon_{\rm c}\sim10^{-2}$ & $\Lambda_{\rm c}^{-1}\sim10^2\,{\rm km}$\\
$d_1,d_2$ & $\epsilon_{\rm d}\sim10^{-3}$ & $\Lambda_{\rm d}^{-1}\sim10^2\,{\rm km}$\\
$e_1,e_2$ & $\epsilon_{\rm e}\sim10^{-42}$ & $\Lambda_{\rm e}^{-1}\sim10\,{\rm pc}$\\
$f_1,\ldots,f_4$ & $\epsilon_{\rm f}\sim10^{-4}$ & $\Lambda_{\rm f}^{-1}\sim10^2\,{\rm km}$\\
$g_1,g_2$ & $\epsilon_{\rm g}\sim10^{-80}$ & $\Lambda_{\rm g}^{-1}\sim10^{18}\,{\rm Gpc}$\\
$h_1,h_2,h_3$ & $\epsilon_{\rm h}\sim10^{-82}$ & $\Lambda_{\rm h}^{-1}\sim10^{-1}\,{\rm Gpc}$\\
\hline\hline
\end{tabular}
\manuscripttablenote{\small Entries are order-of-magnitude estimates for each coupling group $i=b,\ldots,h$, evaluated separately. The parameters $\epsilon_i$ in Eqs.~(\ref{eq: LEEFT scalar parameters}) and~(\ref{eq: LEEFT remaining parameters}) exclude dimensionless couplings and background scalar factors. We use $\mpl^{-1}=\sqrt{8\pi G}$.}
\end{manuscripttableblock}
\end{table}

For comparable suppression scales, the coefficient combinations $\epsilon_{\rm c}$, $\epsilon_{\rm d}$, and $\epsilon_{\rm f}$ have no explicit Planck suppression. Their contributions to frequencies also depend on background scalar factors, perturbative order, and the symmetry of the background. In the massless, static calculation, $c_1,f_1,f_3$ shift scalar-led frequencies at ${\cal O}(\epsilon)$, while $d_1$ enters at ${\cal O}(\epsilon^2)$ through the background scalar, as shown in Eqs.~(\ref{eq: delomega^s_1st}) and~(\ref{eq: scalar QNM}). The complementary coupling information and the effects of rotation are discussed in Sec.~\ref{sec: observational implications}. Observational constraints additionally require excitation and matter coupling, as discussed in Sec.~\ref{sec: scalar-led observability}. Table~\ref{tab: LEEFT scales} by itself does not establish detectability.

\section{Master variables and coefficient functions}\label{sec: ex master eqs}

This appendix specifies the master variables and coefficient functions in Eq.~(\ref{eq: master eq}) for multipoles $\ell\geq2$ on the static backgrounds with a massless scalar. We use $dr_*=dr/f$ with $f=1-r_{\rm H}/r$, as in Sec.~\ref{sec: master eqs}. The factors $\epsilon_2$ and $\epsilon_4$ are included in the coefficient functions, while the explicit powers of $\epsilon$ give the interaction order in Eq.~(\ref{eq: EFT potential}).

\subsection{Definition of the master variables}\label{sec: appendix master variables}

Using the perturbation amplitudes $\Pi,h_1,H_1$, and $K$ defined in Eqs.~(\ref{eq: scalar field}) and~(\ref{eq: metric perturbations}), we introduce the following variables, defined as in the Schwarzschild case
\begin{align}
    \Psi_{\rm s}^{\rm GR} &= r\Pi,\label{eq: master var scalar GR}
\\
    \Psi_{-}^{\rm GR} &= \left(1-\frac{r_{\rm g}}{r}\right)\frac{i h_1}{r\omega},\label{eq: master var - GR}
\\
    \Psi_{+}^{\rm GR} &= \frac{1}{\lambda r+3r_{\rm g}}\left[-r^2K+\left(1-\frac{r_{\rm g}}{r}\right)\frac{irH_1}{\omega}\right],\label{eq: master var + GR}
\end{align}
where $\lambda=(\ell-1)(\ell+2)$, as in Sec.~\ref{sec: light mass limit}. In the GR limit, our master variables $(\Psi_{\rm s},\Psi_-,\Psi_+)$ reduce to
\begin{equation}
    \left(
    \frac{2\sqrt{2}}{\sqrt{\lambda(\lambda+2)}}\,\Psi_{\rm s}^{\rm GR},
    \quad 2\Psi_-^{\rm GR},
    \quad -2\Psi_+^{\rm GR}
    \right).\label{eq: master variables GR limit}
\end{equation}
The supplementary Mathematica file~\cite{STEFT:Mathematica} gives the master variables through ${\cal O}(\epsilon^2)$ in terms of the variables in Eqs.~(\ref{eq: master var scalar GR})--(\ref{eq: master var + GR}) and their first radial derivatives. It also gives the scalar and metric perturbations in terms of the master variables.

\subsection{Coefficient of \texorpdfstring{$\omega^2$}{omega squared} and Schwarzschild potentials}\label{sec: appendix frequency and GR potentials}

For the radial coordinate and master variables used here, the common coefficient of $\omega^2$ in Eq.~(\ref{eq: speed and VGR}) satisfies
\begin{align}
    \frac{1}{Z(r)^2} = \frac{f^2}{F^2} =\frac{\left(1-\frac{r_{\rm H}}{r}\right)^2}{A(r)B(r)} = 1 +\epsilon^2\Delta c^2+{\cal O}(\epsilon^3).
    \label{eq: geometric frequency coefficient}
\end{align}
Here $A(r)$ and $B(r)$ are the static metric functions in Eq.~(\ref{eq: BG metric}), with the corrections in Eqs.~(\ref{eq: 2nd a}) and~(\ref{eq: 2nd b}). The coefficient of $\omega^2$ alone does not determine a coordinate-independent propagation speed. The explicit correction is
\begin{align}
    \Delta c^2 &= \frac{r_{\rm H}}{6930 r^{12}} \Big[
    1848 b_{1}^2\epsilon_{2}^2 r^6(147 r^5+117 r^4 r_{\rm H}+97 r^3 r_{\rm H}^2-98 r^2 r_{\rm H}^3-122 r r_{\rm H}^4-140 r_{\rm H}^5)
\notag\\
    &\quad +44 b_{1} e_{1}\epsilon_{2}\epsilon_{4} r^3(30229 r^8+28969 r^7 r_{\rm H}+28129 r^6 r_{\rm H}^2+23719 r^5 r_{\rm H}^3+22963 r^4 r_{\rm H}^4
\notag\\
    &\quad +22375 r^3 r_{\rm H}^5-62345 r^2 r_{\rm H}^6-41510 r r_{\rm H}^7-36960 r_{\rm H}^8)
\notag\\
    &\quad +e_{1}^2 \epsilon_{4}^2 (765161 r^{11}+751301 r^{10} r_{\rm H}+742061 r^9 r_{\rm H}^2+735131 r^8 r_{\rm H}^3+729587 r^7 r_{\rm H}^4
\notag\\
    &\quad  +724967 r^6 r_{\rm H}^5 -1132273 r^5 r_{\rm H}^6-668458 r^4 r_{\rm H}^7-563738 r^3 r_{\rm H}^8-536634 r^2 r_{\rm H}^9
\notag\\
    &\quad -535626 r r_{\rm H}^{10}-545160 r_{\rm H}^{11})\Big].
\end{align}
With the common factor $f$ outside the potential matrix as in Eq.~(\ref{eq: master eq}), the diagonal entries of $\bm V_{\rm GR}$ have the Schwarzschild form
\begin{align}
    V_{\rm KG} &= \frac{(\lambda+2)r +r_{\rm H}}{r^3},
\\
    V_{\rm RW} &= \frac{(\lambda+2)r-3r_{\rm H}}{r^3},
\\
    V_{\rm Z} &= \frac{9r_{\rm H}^3 +9\lambda r_{\rm H}^2r +3\lambda^2r_{\rm H}r^2 +\lambda^2(\lambda+2)r^3}{r^3(\lambda r+3r_{\rm H})^2}.
\end{align}
They are written using the corrected horizon radius $r_{\rm H}$. The remaining corrections associated with this choice are included in $\bm V_{\rm EFT}$, as described in Sec.~\ref{sec: master eqs}.

\subsection{EFT corrections to the potential}\label{sec: appendix EFT potentials}

At first order, Eq.~(\ref{eq: EFT potential}) gives the scalar diagonal correction and the terms that mix scalar and gravitational perturbations
\begin{align}
    V^{\rm s}_{\rm 1st} &=
    -c_{1}\epsilon_{2} \frac{24 r_{\rm H}^4}{r^6} -f_{1}\epsilon_{4} \frac{24 r_{\rm H}^7}{r^9} -f_3\epsilon_4\frac{72r_{\rm H}^6(5r -6r_{\rm H})}{r^9},
\\
    V^{\rm s-}_{\rm 1st} &= -\frac{12\sqrt{
    2\lambda(\lambda +2)} r_{\rm H}^3}{r^8}(2b_{2}\epsilon_{2} r^3 +3e_{2}\epsilon_{4} r_{\rm H}^3),
\\
    V^{\rm s+}_{\rm 1st} &= -\frac{2\sqrt{2\lambda(\lambda +2)} r_{\rm H}^3}{r^8 (\lambda  r+3 r_{\rm H})^2}\Big[2b_{1}\epsilon_{2} r^3(3 \lambda ^2 r^2+13 \lambda  r r_{\rm H}+15 r_{\rm H}^2)
\notag\\
    &\quad +e_{1}\epsilon_{4} r_{\rm H}^3(9 \lambda ^2 r^2+46 \lambda  r r_{\rm H} +60 r_{\rm H}^2)\Big].
\end{align}
At second order, parenthesized superscripts such as $(b_1d_1)$ label the product factored out of each radial coefficient function. The supplementary Mathematica file~\cite{STEFT:Mathematica} gives their explicit forms, including their dependence on $\ell$. The diagonal entries are
\begin{align}
    V^{\rm s}_{\rm 2nd} &= 
    \epsilon_2^2\Big[ b_1^2V^{(b_1b_1)}_{\rm s,2nd} 
    +b_1d_1 V^{(b_1 d_1)}_{\rm s,2nd}
    +b_2^2 V^{(b_2b_2)}_{\rm s,2nd} 
    \Big]
\notag\\   
    &\quad   
    +\epsilon_2\epsilon_4 \Big[b_1e_1V^{(b_1e_1)}_{\rm s,2nd}
    +b_2e_2 V^{(b_2 e_2)}_{\rm s,2nd}
    +c_1f_3 V^{(c_1 f_3)}_{\rm s,2nd}
    +d_1e_1 V^{(d_1 e_1)}_{\rm s,2nd}
    \Big]
\notag\\
    &\quad 
    +\epsilon_4^2\Big[e_1^2 V^{(e_1e_1)}_{\rm s,2nd}
    +e_2^2 V^{(e_2e_2)}_{\rm s,2nd}
    +f_1f_3 V^{(f_1 f_3)}_{\rm s,2nd}
    +f_3^2 V^{(f_3 f_3)}_{\rm s,2nd}\Big],
\\
    V^{-}_{\rm 2nd} &= b_1^2\epsilon^2_{2} V^{(b_1b_1)}_{\rm -,2nd} +b_1e_1\epsilon_{2}\epsilon_{4} V^{(b_1e_1)}_{\rm -,2nd} +e_1^2\epsilon_{4}^2 V^{(e_1e_1)}_{\rm -,2nd},
\\
    V^{+}_{\rm 2nd} &= b_1^2\epsilon^2_{2} V^{(b_1b_1)}_{\rm +,2nd} +b_1e_1\epsilon_{2}\epsilon_{4} V^{(b_1e_1)}_{\rm +,2nd} +e_1^2\epsilon_{4}^2 V^{(e_1e_1)}_{\rm +,2nd}.
\end{align}
The following off-diagonal entries couple the scalar to each gravitational sector and the two gravitational sectors to each other.
\begin{align}
     V^{\rm s-}_{\rm 2nd} &=  b_1c_2\epsilon_2^2 V^{(b_1c_2)}_{\rm s-, 2nd} +\epsilon_2\epsilon_4\Big[b_1f_2 V^{(b_1f_2)}_{\rm s-, 2nd} +b_1f_4 V^{(b_1f_4)}_{\rm s-, 2nd} +b_2f_3 V^{(b_2f_3)}_{\rm s-, 2nd}+c_2e_1 V^{(c_2e_1)}_{\rm s-, 2nd}\Big]
\notag\\
    &\quad  
    +\epsilon_4^2\Big[e_1f_2 V^{(e_1f_2)}_{\rm s-, 2nd} +e_1f_4 V^{(e_1f_4)}_{\rm s-, 2nd} +e_2f_3 V^{(e_2f_3)}_{\rm s-, 2nd}\Big],
\\
    V^{\rm s+}_{\rm 2nd} &=  b_1c_1\epsilon_2^2 V^{(b_1c_1)}_{\rm s+, 2nd} +\epsilon_2\epsilon_4\Big[b_1f_1 V^{(b_1f_1)}_{\rm s+, 2nd} +b_1f_3 V^{(b_1f_3)}_{\rm s+, 2nd} +c_1e_1 V^{(c_1e_1)}_{\rm s+, 2nd} \Big]
\notag\\
    &\quad +\epsilon_4^2 \Big[e_1f_1V^{(e_1f_1)}_{\rm s+, 2nd} +e_1f_3 V^{(e_1f_3)}_{\rm s+, 2nd}\Big],
\\
    V^{\rm -+}_{\rm 2nd} &=  b_1b_2\epsilon_2^2V^{(b_1b_2)}_{-+,\rm 2nd} +\epsilon_2\epsilon_4\Big[b_1e_2V^{(b_1e_2)}_{-+,\rm 2nd} +b_2e_1V^{(b_2e_1)}_{-+,\rm 2nd}\Big] +e_1e_2\epsilon_4^2V^{(e_1e_2)}_{-+,\rm 2nd}.
\end{align}
The remaining matrix entries follow by symmetry. These expressions specify the coupling structure of every entry of $\bm V_{\rm EFT}$ through ${\cal O}(\epsilon^2)$. The $c_2,f_2,f_4$ terms in $V^{\rm s-}_{\rm 2nd}$ enter the coupled equations at second order. Their leading contribution to the static QNM frequencies also involves a first-order mixing term, so they do not enter the shifts in Eqs.~(\ref{eq: scalar QNM}) and~(\ref{eq: grav QNM}).

\section{Prefactors in the series solutions and the purely outgoing condition in QNM calculations}
\label{sec: prefactor and purely outgoing condition}

In Sec.~\ref{sec: light mass limit}, we calculate the QNMs of the coupled scalar and odd-parity gravitational perturbations in the regime of small scalar mass, $mr_{\rm g}\ll1$. The scalar field is massive and the gravitational perturbation is massless. For each mode we use a common prefactor for the two fields, so that the series coefficients satisfy recurrence relations with finitely many terms. We examine how the truncation condition ${\bf a}_{j_{\max}}=0$ underlying Eq.~(\ref{eq: QNM condition}) selects the purely outgoing solution as $j_{\max}$ increases.

In Eqs.~(\ref{eq: scalar dCS master eq}) and~(\ref{eq: odd dCS master eq}), the coupling between the two fields falls off as $r^{-5}$. The fields therefore decouple at leading order at infinity. In the scalar equation, $fm^2=m^2-m^2r_{\rm g}/r$ changes the wave number and gives the scalar wave an additional logarithmic phase. With ${\bf\Psi}=(\Psi_{\rm s},\Psi_{\rm -})^{\rm T}$ and $r_{\rm H}=r_{\rm g}$, the four asymptotic waves have the leading forms
\begin{align}
    {\bf\Psi} &\sim
    C^{\rm in}_{\rm s} e^{-iK r_\ast/r_{\rm H}}
    \left(\frac{r}{r_{\rm H}}\right)^{-i\alpha/(2K)}\begin{pmatrix}1\\0\end{pmatrix}
 + C^{\rm out}_{\rm s} e^{iK r_\ast/r_{\rm H}}
 \left(\frac{r}{r_{\rm H}}\right)^{i\alpha/(2K)}\begin{pmatrix}1\\0\end{pmatrix}
 \notag\\
    &\quad + C^{\rm in}_{\rm -} e^{-i\Omega r_\ast/r_{\rm H}}\begin{pmatrix}0\\1\end{pmatrix}
 + C^{\rm out}_{\rm -} e^{i\Omega r_\ast/r_{\rm H}}\begin{pmatrix}0\\1\end{pmatrix},\label{eq: GR exaxt massive solution at infinity}
\\
    K &= \sqrt{\Omega^2-\alpha},\quad \Omega=\omega r_{\rm H},\quad \alpha=m^2r_{\rm H}^2 .\notag
\end{align}
Here $C^{\rm in,out}_{\rm s}$ and $C^{\rm in,out}_{\rm -}$ are the complex amplitudes of the incoming and outgoing waves at infinity in the scalar and odd-parity gravitational sectors, respectively. We choose the branch for which $K\to\Omega$ as $\alpha\to0$. The purely outgoing condition is $C^{\rm in}_{\rm s}=C^{\rm in}_{\rm -}=0$.

\subsection{Scalar-led case}

For the scalar-led mode, Eqs.~(\ref{eq: massive ansatz1}) and~(\ref{eq: massive ansatz2}) use the massive scalar prefactor for both fields,
\begin{align}
    {\bf\Psi} &= P_{\rm s}(r)\sum_{j=0}^{\infty}{\bf a}_j f^j,\quad
    {\bf a}_j = \begin{pmatrix}a_j^{\rm s}\\ \epsilon a_j^{\rm -}\end{pmatrix},\label{eq: massive series ansatz}
\\
    P_{\rm s}(r) &= e^{iK r_\ast/r_{\rm H}}\left(\frac{r}{r_{\rm H}}\right)^{i\alpha/(2K)}f^{-i(K+\Omega)}.\notag
\end{align}
We now divide Eq.~(\ref{eq: GR exaxt massive solution at infinity}) by $P_{\rm s}$ and use $r/r_{\rm H}=1/(1-f)$. Suppressing powers of $r/r_{\rm H}$ multiplying the exponentials and displaying the leading component of each wave, we obtain
\begin{align}
    \sum_{j=0}^{\infty}{\bf a}_j f^j &\sim 
    C^{\rm in}_{\rm s} e^{-2iK/(1-f)}\begin{pmatrix}1\\0\end{pmatrix} +C^{\rm out}_{\rm s}\begin{pmatrix}1+{\cal O}(1-f)\\0\end{pmatrix}
\notag\\
    &\quad + C^{\rm in}_{\rm -} e^{-i(K+\Omega)/(1-f)}\begin{pmatrix}0\\1\end{pmatrix} +C^{\rm out}_{\rm -} e^{i(\Omega-K)/(1-f)}\begin{pmatrix}0\\1\end{pmatrix}.\label{eq: massless pre massive}
\end{align}
The outgoing scalar phase has canceled, but an exponential factor remains in the outgoing gravitational wave because its wave number is $\Omega$. To see how this term contributes to ${\bf a}_{j_{\max}}$, we compare its contribution to the series coefficients with the contributions from the two incoming waves for large $j$. For an exponential with parameter $A$, we write
\begin{align}
    e^{A/(1-f)} = \sum_{j=0}^{\infty}b_j(A)f^j,\quad
    b_0(A) = e^A,\quad b_j(A) = e^A A\,{}_1F_1(1-j;2;-A)\quad(j\geq1),
\end{align}
where \({}_1F_1\) is the confluent hypergeometric function. For fixed $A\ne0$, the omitted powers of $r/r_{\rm H}$ affect the algebraic dependence on $j$.\footnote{For a factor $(1-f)^n$, the corresponding coefficient is given by the Cauchy integral
\[
    b_j^{(n)}(A) = \frac{1}{2\pi i}\oint_{|f|=\rho}\frac{(1-f)^n e^{A/(1-f)}}{f^{j+1}}\,df,\quad 0<\rho<1 .
\]
For the values of $A$ considered here, the dominant saddle is at $1-f\simeq\sqrt{A/j}$, with $\operatorname{Re}(\sqrt{A})>0$. Thus $b_j^{(n)}(A)\sim (A/j)^{n/2}b_j(A)$ at fixed nonzero $A$.}
Keeping the exponential dependence and the leading component of each contribution, Eq.~(\ref{eq: massless pre massive}) gives schematically
\begin{align}
    {\bf a}_j &\sim 
    C^{\rm in}_{\rm s} b_j(A^{\rm in}_{\rm s})\begin{pmatrix}1\\0\end{pmatrix} 
    +C^{\rm in}_{\rm -} b_j(A^{\rm in}_{\rm -})\begin{pmatrix}0\\1\end{pmatrix}
\notag\\
    &\quad +C^{\rm out}_{\rm s} h_j^{\rm out}\begin{pmatrix}1\\0\end{pmatrix}
    +C^{\rm out}_{\rm -} b_j(A^{\rm out}_{\rm -})\begin{pmatrix}0\\1\end{pmatrix},\label{eq: aj scalar-led}
\\
    A^{\rm in}_{\rm s} &= -2iK,\quad A^{\rm in}_{\rm -}=-i(K+\Omega),\quad
    A^{\rm out}_{\rm -} = i(\Omega-K)=\frac{i\alpha}{K+\Omega}.\notag
\end{align}
Here $h_j^{\rm out}$ denotes the contribution to $a_j^{\rm s}$ from a unit outgoing scalar wave after $P_{\rm s}$ has been factored out. In Eq.~(\ref{eq: massless pre massive}), the scalar component of this wave is $1+{\cal O}(1-f)$. The constant term contributes only to $h_0^{\rm out}$, while the ${\cal O}(1-f)$ terms can also contribute to $h_j^{\rm out}$ for $j\geq1$.\footnote{With the scalar prefactor, the decaying solution of the scalar recurrence behaves at large $j$ as~\cite{Leaver:1985ax}
\begin{equation}
    h_j^{\rm out}\ \propto\ j^\nu e^{-2\sqrt{-2iK}\sqrt j},\quad
    \nu = -\tfrac34-i\Omega+{\cal O}(\alpha).\label{eq: hJout}
\end{equation}
At fixed $K\ne\Omega$, the $r^{-5}$ coupling induces a subleading component with the same radial phase and does not change this leading scalar recurrence. The comparison in the text only requires that this outgoing contribution be subdominant to both incoming contributions.}

For fixed $A\ne0$ and $\operatorname{Re}(\sqrt{A})>0$, the same saddle gives the large-$j$ behavior
\begin{align}
    b_j(A) \sim \frac{e^{A/2}A^{1/4}}{2\sqrt{\pi}}j^{-3/4}\exp\!\left(2\sqrt A\sqrt j\right).\label{eq: bj large J}
\end{align}
At small fixed $\alpha>0$, both $\operatorname{Re}(\sqrt{-2iK})$ and $\operatorname{Re}(\sqrt{-i(K+\Omega)})$ are close to $\operatorname{Re}(\sqrt{-2i\Omega})>0$ near the scalar-led fundamental mode, whereas $\operatorname{Re}(\sqrt{i(\Omega-K)})$ is of order $\sqrt{\alpha}$. Thus the contribution of the outgoing gravitational wave to ${\bf a}_j$ can grow with $j$, but more slowly than either incoming contribution. The two incoming waves also have independent leading components, one scalar and one gravitational. At a root of the frequency condition for finite $j_{\max}$, small incoming amplitudes can therefore cancel the outgoing contributions in the two components of ${\bf a}_{j_{\max}}$. With a fixed normalization at the horizon, the incoming amplitudes needed for this cancellation both tend to zero as $j_{\max}$ increases along this branch. In this sense the condition ${\bf a}_{j_{\max}}=0$ selects the purely outgoing QNM.

\subsection{Gravitational-led case}

For the gravitational-led mode, Eqs.~(\ref{eq: massless ansatz1}) and~(\ref{eq: massless ansatz2}) use the massless gravitational prefactor,
\begin{align}
    {\bf\Psi} = P_{\rm g}(r)\sum_{j=0}^{\infty}{\bf a}_j f^j,\quad
    P_{\rm g}(r) = e^{i\omega r_\ast}f^{-2i\Omega},\quad
    {\bf a}_j = \begin{pmatrix}\epsilon a_j^{\rm s}\\a_j^{\rm -}\end{pmatrix}.\label{eq: massless series ansatz}
\end{align}
Dividing Eq.~(\ref{eq: GR exaxt massive solution at infinity}) by $P_{\rm g}$ and again suppressing the powers multiplying the exponentials gives
\begin{align}
    \sum_{j=0}^{\infty}{\bf a}_j f^j
    &\sim C^{\rm in}_{\rm s}e^{A^{\rm in}_{\rm s}/(1-f)}\begin{pmatrix}1\\0\end{pmatrix}
    +C^{\rm in}_{\rm -}e^{A^{\rm in}_{\rm -}/(1-f)}\begin{pmatrix}0\\1\end{pmatrix}
\notag\\
    &\quad +C^{\rm out}_{\rm s}e^{A^{\rm out}_{\rm s}/(1-f)}\begin{pmatrix}1\\0\end{pmatrix}
    +C^{\rm out}_{\rm -}\begin{pmatrix}0\\1+{\cal O}(1-f)\end{pmatrix},\label{eq: massive per massless}
\\
    A^{\rm in}_{\rm s} &= -i(K+\Omega),\quad A^{\rm in}_{\rm -}=-2i\Omega,\quad
    A^{\rm out}_{\rm s} = i(K-\Omega)=-\frac{i\alpha}{K+\Omega}.\notag
\end{align}
Here the outgoing gravitational phase cancels, and an exponential factor remains in the outgoing scalar wave. The contribution of the outgoing scalar wave to ${\bf a}_j$ again grows more slowly than either incoming contribution in the regime of small scalar mass. The same comparison in the two components therefore gives $C^{\rm in}_{\rm s},C^{\rm in}_{\rm -}\to0$ as $j_{\max}$ increases along the gravitational-led branch.

\section{Matching for a heavy scalar through mass dimension sixteen}\label{sec: complete action matching}

This appendix gives the compact formula used for matching at tree level in Sec.~\ref{sec: action matching} and the resulting action organized by operator mass dimension. We start from the scalar-curvature interactions of Eq.~(\ref{eq: STEFT Lagrangian}) and set the independent pure curvature coefficients in the original action to zero, $\mathcal L_{\rm EFT}^{\rm Curv}=0$, as in Sec.~\ref{sec: theory}. Integrating out the scalar then generates pure gravity operators built from curvature tensors and their covariant derivatives. Completeness through dimension sixteen refers to the action induced by these specified interactions, rather than to a general basis of independent pure gravity operators.

\subsection{General formula for matching at tree level}\label{sec: matching general formula}

We group the scalar-curvature interactions by their dependence on $\phi$. Here $J$ is the linear source, $Q$ and $U_d$ multiply the terms quadratic and cubic in $\phi$, and $Z_f$ gives the curvature-dependent correction to the scalar kinetic term. The Lagrangian is
\begin{align}
    \frac{\mathcal L}{\mpl^2}
    &= \frac12R-\frac12(\nabla\phi)^2-\frac12m^2\phi^2 +\epsilon\left[J\phi+Q\phi^2+U_d\phi^3+Z_f(\nabla\phi)^2\right],\label{eq: LEEFT1}
\\
    J &= J_b+J_e,\quad Q=Q_c+Q_f,
\\
    J_b &= \frac{1}{\Lambda^2}(b_1\mathcal C_2+b_2\tilde{\mathcal C}_2),\quad
    J_e = \frac{1}{\Lambda^4}(e_1\mathcal C_3+e_2\tilde{\mathcal C}_3),\label{eq: matching sources}
\\
    Q_c &= \frac{1}{\Lambda^2}(c_1\mathcal C_2+c_2\tilde{\mathcal C}_2),\quad 
    Q_f = \frac{1}{\Lambda^4}(f_1\mathcal C_3+f_2\tilde{\mathcal C}_3),
\\
    U_d &= \frac{1}{\Lambda^2}(d_1\mathcal C_2+d_2\tilde{\mathcal C}_2),\quad
    Z_f = \frac{1}{\Lambda^4}(f_3\mathcal C_2+f_4\tilde{\mathcal C}_2).
\end{align}
All curvature invariants have the definitions of Sec.~\ref{sec: theory}. The subscripts on $J_b,J_e,Q_c,Q_f,U_d$, and $Z_f$ identify the corresponding coupling groups. The source $J$ agrees with the source in Eq.~(\ref{eq: linear scalar matching source}). The grouped coefficients $J,Q,U_d$, and $Z_f$ exclude the overall bookkeeping factor $\epsilon$, which is explicit in Eq.~(\ref{eq: LEEFT1}) and can be set to unity after the expansion. The truncation by operator dimension includes terms through $\epsilon^4$, whereas the QNM calculation is carried out through ${\cal O}(\epsilon^2)$.

After integration by parts, the terms in $\mathcal L/\mpl^2$ quadratic in $\phi$ become $-\phi(m^2-\mathcal A)\phi/2$. The differential operator $\mathcal A$ acts on a scalar function $u$ as
\begin{align}
    \mathcal A u := \Box u+2\epsilon Qu-2\epsilon\nabla_\mu(Z_f\nabla^\mu u).\label{eq: matching scalar operator}
\end{align}
The scalar equation of motion is
\begin{equation}
    (m^2-\mathcal A)\phi = \epsilon J+3\epsilon U_d\phi^2.
\end{equation}
Substituting the solution for $U_d=0$ into the nonlinear term on the right-hand side gives the formal solution to first order in $U_d$,
\begin{equation}
    \phi = \epsilon(m^2-\mathcal A)^{-1}J +3\epsilon^3(m^2-\mathcal A)^{-1}\!\left[U_d\big((m^2-\mathcal A)^{-1}J\big)^2\right]+\cdots.\label{eq: leading soln}
\end{equation}
As in Eq.~(\ref{eq: linear scalar matching solution}), the inverse is understood as the local expansion $m^{-2}\sum_{n\geq0}(\mathcal A/m^2)^n$ for the sourced low-energy response. This expansion does not describe independent scalar modes with $\omega\sim m$. Substituting Eq.~(\ref{eq: leading soln}) into Eq.~(\ref{eq: LEEFT1}) gives
\begin{align}
    \frac{\mathcal L_{\rm eff}}{\mpl^2}
    &= \frac12R+\left[\frac{\epsilon^2}{2}J\sum_{n=0}^{4}\frac{\mathcal A^n}{m^{2n+2}}J +\frac{\epsilon^4 U_dJ^3}{m^6}\right]_{\mathrm{dim}\leq16}+\cdots.\label{eq: compact STEFT effective action}
\end{align}
The subscript $\mathrm{dim}\leq16$ means that, after expanding the sources and the operator $\mathcal A$, we retain curvature operators of mass dimension at most sixteen. We count two powers of mass for each curvature tensor and one for each covariant derivative, excluding coefficients and mass factors. When expanding $\mathcal A^n J$, derivatives can act on the coefficients as well as on $J$. For example, $\Box(QJ)$ differentiates both $Q$ and $J$. Derivative corrections to $\epsilon^4U_dJ^3/m^6$ and terms of second or higher order in $U_d$ first generate operators of higher dimension. Equations in this appendix are understood up to total derivatives.

\subsection{Expansion by operator mass dimension}\label{sec: matching dimension expansion}

Using the separate $\Lambda^{-2}$ and $\Lambda^{-4}$ sources in Eq.~(\ref{eq: matching sources}), we write
\begin{align}
    \frac{\mathcal L_{\rm eff}}{\mpl^2}
    = \frac12R+\mathcal I_{8}+\mathcal I_{10}+\mathcal I_{12}+\mathcal I_{14}+\mathcal I_{16}+\cdots.\label{eq: degree expansion of effective action}
\end{align}
The subscripts on $\mathcal I$ label the mass dimension of the induced pure gravity operators. Each $\mathcal I$ includes the corresponding coefficients. The contributions $\mathcal I_8$ and $\mathcal I_{10}$ reproduce the interaction terms in Eq.~(\ref{eq: LEEFT2}). The higher-dimensional terms complete the induced action to the stated truncation. Expanding the compact formula, Eq.~(\ref{eq: compact STEFT effective action}), and integrating by parts gives
\begin{align}
    \mathcal I_{8} &= \frac{\epsilon^2J_b^2}{2m^2},
\\
    \mathcal I_{10} &= \epsilon^2\left(\frac{J_bJ_e}{m^2}+\frac{J_b\Box J_b}{2m^4}\right),
\\
    \mathcal I_{12} &= \epsilon^2\left(\frac{J_e^2}{2m^2}
    +\frac{J_b\Box J_e}{m^4}
    +\frac{J_b\Box^2J_b}{2m^6}\right)
    +\frac{\epsilon^3Q_cJ_b^2}{m^4},
\\
    \mathcal I_{14} &= \epsilon^2\left(\frac{J_e\Box J_e}{2m^4}
    +\frac{J_b\Box^2J_e}{m^6}
    +\frac{J_b\Box^3J_b}{2m^8}\right)
\notag\\
    &\quad +\epsilon^3\left(\frac{2Q_cJ_bJ_e+Q_fJ_b^2}{m^4}
    +\frac{2Q_cJ_b\Box J_b}{m^6}
    +\frac{Z_f(\nabla J_b)^2}{m^4}\right),
\\
    \mathcal I_{16} &= \epsilon^2\left(\frac{J_e\Box^2J_e}{2m^6}
    +\frac{J_b\Box^3J_e}{m^8}
    +\frac{J_b\Box^4J_b}{2m^{10}}\right)
\notag\\
    &\quad +\frac{\epsilon^3}{m^4}\left[Q_cJ_e^2+2Q_fJ_bJ_e
    +2Z_f\nabla_\mu J_b\nabla^\mu J_e\right]
\notag\\
    &\quad +\frac{2\epsilon^3}{m^6}\left[Q_c(J_b\Box J_e+J_e\Box J_b) +Q_fJ_b\Box J_b+Z_f\nabla_\mu J_b\nabla^\mu\Box J_b\right]
\notag\\
    &\quad +\frac{\epsilon^3Q_c}{m^8}\left[(\Box J_b)^2+2J_b\Box^2J_b\right] +\frac{\epsilon^4}{m^6}\left(2Q_c^2J_b^2+U_dJ_b^3\right).\label{eq: degree fourteen matching}
\end{align}
All parity combinations enter these expressions through the sources. Applying the $2\epsilon Q_c$ term in $\mathcal A$ twice gives $2\epsilon^4Q_c^2J_b^2/m^6$. This term includes, for example, a contribution proportional to $b_1^2c_1^2\mathcal C_2^4$. The scalar-curvature interaction cubic in $\phi$ gives $\epsilon^4U_dJ_b^3/m^6$, including a contribution proportional to $b_1^3d_1\mathcal C_2^4$. Both have operator mass dimension sixteen. The terms containing $Z_f$ arise from the curvature-dependent corrections to the scalar kinetic term, with coefficients $f_3,f_4$. The omitted terms contain operators of mass dimension greater than sixteen within the chosen STEFT truncation.

\section{Regime of validity and causality}\label{sec: validity and causality}

The estimates in this appendix use the normalization of Sec.~\ref{sec: theory}, with the common interaction scale $\Lambda$, $\epsilon_2=(\Lambda r_{\rm g})^{-2}$, and $\epsilon_4=(\Lambda r_{\rm g})^{-4}$. We set the formal bookkeeping parameter $\epsilon=1$ throughout. The operator families have independent suppression scales $\Lambda_i$ in App.~\ref{sec: LEEFT}. The parameters in Eqs.~(\ref{eq: LEEFT scalar parameters}) and~(\ref{eq: LEEFT remaining parameters}) therefore use a different coefficient hierarchy.

\subsection{Classical expansion criteria}\label{sec: validity classical}

In the EFT expansion, we assume that dimensionless combinations such as $b_i\epsilon_2$ and $e_i\epsilon_4$ are small. Table~\ref{tab: validity conditions} summarizes the conditions used in the black hole and QNM calculations. Operators with additional background scalar factors can be further suppressed on the branch perturbatively connected to GR. We truncate the field equations at second order in $\epsilon$ for the background and QNM calculations. When some coefficients are enhanced, we must check that both the retained and omitted corrections to the background and perturbation equations remain small in the region relevant to the modes.

\begin{table}[!htbp]
\centering
\begin{manuscripttableblock}{\textwidth}
\caption{Validity conditions for the black hole solutions and QNM calculations.}
\label{tab: validity conditions}
\renewcommand{\arraystretch}{1.10}
\setlength{\tabcolsep}{5pt}
\begin{tabular*}{\textwidth}{@{\extracolsep{\fill}}>{\raggedright\arraybackslash}p{0.24\textwidth} >{\raggedright\arraybackslash}p{0.72\textwidth}@{}}
\hline\hline
Approximation & Representative criterion\\
\hline
EFT expansion & $|b_i|\epsilon_2,\ |c_i|\epsilon_2,\ |d_i|\epsilon_2\ll1$\newline
$|e_i|\epsilon_4,\ |f_j|\epsilon_4\ll1$ ($i=1,2$, $j=1,\ldots,4$)\\[3pt]
Slow rotation & $|\chi|/M\ll1$, with $\chi=J/M$ (rotating background only)\\[3pt]
Light scalar & $\alpha=m^2r_{\rm g}^2\ll1$\\[3pt]
Heavy scalar & $\beta=(m^2r_{\rm g}^2)^{-1}\ll1$, with $\omega r_{\rm g}={\cal O}(1)$\\[3pt]
Linear waves & Terms cubic and higher in the perturbations are small compared with the quadratic action.\\[3pt]
Quantum loops & $\ell_b(E)\ll1$ for the sGB example. Other vertices require analogous checks.\\
\hline\hline
\end{tabular*}
\manuscripttablenote{We use the normalization of the main text, with $\epsilon=1$. The rows for light and heavy scalars describe alternative expansions for fixed, low mode numbers and $\omega r_{\rm g}={\cal O}(1)$. EFT corrections must remain small in the region relevant to the modes, including the potential barrier. The row for quantum loops refers to the sGB estimate in Eq.~(\ref{eq: representative loop counting}).}
\end{manuscripttableblock}
\end{table}

With the common scale $\Lambda$ and coefficients of order unity, interactions with six derivatives are more suppressed than those with four, although both carry one power of the formal parameter $\epsilon$ before it is set to unity. On the black hole background, the $b_1$ kinetic mixing scales as $|b_1|r_{\rm g}/(\Lambda^2r^3)$ and must also remain perturbative~\cite{Serra:2022pzl}. Higher overtones or larger $\ell$ can probe shorter scales and require a separate validity check.

\subsection{Quantum loop estimate in the sGB sector}\label{sec: validity loops}

Following the power counting in Ref.~\cite{Serra:2022pzl}, we illustrate the suppression of quantum loop corrections using the $b_1$ scalar Gauss--Bonnet interaction. We consider four-graviton scattering near flat space, at generic external momenta of characteristic energy $E$ with $m\ll E$. We estimate the parametric size of representative diagrams without evaluating the loop integrals explicitly. We suppress numerical factors, tensor factors, and logarithms.

First, canonically normalize the perturbations as $h_c\sim\mpl h$ and $\varphi=\mpl\delta\phi$. The $b_1$ vertex with one scalar and two gravitons takes the schematic form
\begin{align}
    {\cal L}^{(3)}_b\sim\frac{b_1}{\mpl\Lambda^2}\,
    \varphi(\partial^2 h_c)^2.\label{eq: sGB cubic vertex}
\end{align}
Each linearized curvature supplies two derivatives, so the vertex contains four powers of momentum. Its vertex factor $\mathcal V_b(E)$ therefore has mass dimension one and scales as
\begin{align}
    |\mathcal V_b(E)|\sim\frac{|b_1|E^4}{\mpl\Lambda^2}.\label{eq: sGB vertex estimate}
\end{align}
To define the dimensionless strength $g_b(E)$ of the cubic vertex, we factor out one power of $E$ and write
\begin{align}
    |\mathcal V_b(E)|\sim E\,g_b(E),\quad
    g_b(E)\sim\frac{|b_1|E^3}{\mpl\Lambda^2}.\label{eq: sGB dimensionless strength}
\end{align}

At tree level, two such vertices are joined by a scalar propagator, as in Fig.~\ref{fig: sGB loop counting}(a). At generic momenta and $m\ll E$, this propagator scales as $E^{-2}$. The corresponding four-graviton amplitude is therefore
\begin{align}
    \mathcal M_{\rm tree}^{(b)}
    \sim \underbrace{[E\,g_b(E)]^2}_{\text{two vertices}}\,
    \underbrace{\frac{1}{E^2}}_{\text{one propagator}}
    \sim g_b(E)^2.
    \label{eq: sGB tree amplitude}
\end{align}

\begin{figure}[!htbp]
\centering
\begin{manuscripttableblock}{\textwidth}
\setlength{\abovecaptionskip}{6pt}
\setlength{\belowcaptionskip}{0pt}
\begin{minipage}[t]{0.48\textwidth}
\centering
\textbf{(a) Tree level}\par\medskip
\begin{tikzpicture}[x=1cm,y=1cm,
  graviton/.style={decorate,decoration={snake,amplitude=0.55mm,segment length=2.2mm},line width=0.55pt},
  scalar/.style={dashed,line width=0.7pt},every node/.style={font=\small}]
  \useasboundingbox (-2.8,-1.65) rectangle (2.8,1.65);
  \coordinate (L) at (-0.85,0); \coordinate (R) at (0.85,0);
  \draw[graviton] (-2,1) node[above left] {$h_c$} -- (L);
  \draw[graviton] (-2,-1) node[below left] {$h_c$} -- (L);
  \draw[graviton] (R) -- (2,1) node[above right] {$h_c$};
  \draw[graviton] (R) -- (2,-1) node[below right] {$h_c$};
  \draw[scalar] (L) -- node[above=5pt] {$\varphi$} node[below=5pt] {$E^{-2}$} (R);
  \fill (L) circle (1.6pt); \fill (R) circle (1.6pt);
  \node at (-0.85,0.7) {$E g_b$}; \node at (0.85,0.7) {$E g_b$};
\end{tikzpicture}\par
{\small Two vertices, one scalar propagator}\par
\[
    \mathcal M_{\rm tree}^{(b)}\sim
    \frac{(E g_b)^2}{E^2}\sim g_b^2
\]
\end{minipage}\hfill
\begin{minipage}[t]{0.48\textwidth}
\centering
\textbf{(b) One loop}\par\medskip
\begin{tikzpicture}[x=1cm,y=1cm,
  graviton/.style={decorate,decoration={snake,amplitude=0.55mm,segment length=2.2mm},line width=0.55pt},
  scalar/.style={dashed,line width=0.7pt},every node/.style={font=\small}]
  \useasboundingbox (-2.8,-1.65) rectangle (2.8,1.65);
  \coordinate (TL) at (-0.8,0.65); \coordinate (TR) at (0.8,0.65);
  \coordinate (BL) at (-0.8,-0.65); \coordinate (BR) at (0.8,-0.65);
  \draw[scalar] (TL) -- node[above=4pt] {$\varphi$} (TR);
  \draw[scalar] (BL) -- node[below=4pt] {$\varphi$} (BR);
  \draw[graviton] (TL) -- node[left=5pt] {$h_c$} (BL);
  \draw[graviton] (TR) -- node[right=5pt] {$h_c$} (BR);
  \draw[graviton] (-2,1.25) node[above left] {$h_c$} -- (TL);
  \draw[graviton] (-2,-1.25) node[below left] {$h_c$} -- (BL);
  \draw[graviton] (TR) -- (2,1.25) node[above right] {$h_c$};
  \draw[graviton] (BR) -- (2,-1.25) node[below right] {$h_c$};
  \foreach \v in {TL,TR,BL,BR} {\fill (\v) circle (1.6pt);}
  \node at (0,0) {$\displaystyle\frac{E^4}{16\pi^2}$};
\end{tikzpicture}\par
{\small Four vertices, four propagators}\par
\[
    \mathcal M_{\rm 1\mbox{-}loop}^{(b)} \sim
    \frac{E^4}{16\pi^2}\frac{(E g_b)^4}{(E^2)^4}
    \sim\frac{g_b^4}{16\pi^2}
\]
\end{minipage}
\caption{Power counting for four-graviton scattering through the cubic $b_1$ interaction. Wavy lines denote canonical gravitons $h_c$ and dashed lines the canonical scalar $\varphi$. Each dot contributes $E g_b$, and each internal line contributes $E^{-2}$ for $m\ll E$. The factor inside the box denotes the four-dimensional loop integration. These representative diagrams compare the tree amplitude from scalar exchange with a one-loop contribution built from the same cubic vertex.}
\label{fig: sGB loop counting}
\end{manuscripttableblock}
\end{figure}

Next, consider the one-loop box in Fig.~\ref{fig: sGB loop counting}(b). It contains four cubic vertices and four internal propagators, alternating between scalars and gravitons. The four-dimensional loop integration supplies a factor of order $E^4/(16\pi^2)$. Combining these factors gives
\begin{align}
    \mathcal M_{\rm 1\mbox{-}loop}^{(b)}
    \sim \underbrace{\frac{E^4}{16\pi^2}}_{\text{loop integration}}\,
    \underbrace{[E\,g_b(E)]^4}_{\text{four vertices}}\,
    \underbrace{\frac{1}{(E^2)^4}}_{\text{four propagators}}
    \sim \frac{g_b(E)^4}{16\pi^2}.
    \label{eq: sGB loop amplitude}
\end{align}
The two extra vertices in the box give two additional powers of $g_b$.

To estimate the loop correction from the $b_1$ interaction, we take the ratio of the box amplitude to the tree amplitude for scalar exchange in Eq.~(\ref{eq: sGB tree amplitude}) at the same external momenta,
\begin{align}
    \ell_b(E) &:= \left|\frac{\mathcal M_{\rm 1\mbox{-}loop}^{(b)}}
    {\mathcal M_{\rm tree}^{(b)}}\right|
    \sim \frac{g_b(E)^2}{16\pi^2}
    \sim \frac{b_1^2E^6}{16\pi^2\mpl^2\Lambda^4}.\label{eq: representative loop counting}
\end{align}
The perturbative loop condition for this example is $\ell_b(E)\ll1$.

For $b_1={\cal O}(1)$, the scaling $g_b(E)\sim|b_1|(E/\Lambda)^2(E/\mpl)$ gives $g_b\ll1$ and hence $\ell_b\ll1$ when $E\ll\Lambda$ and $E\ll\mpl$. For low-lying black hole modes with $E\sim r_{\rm g}^{-1}$ in the regime of small scalar mass used for this estimate,
\begin{align}
    g_b &\sim \frac{|b_1|\epsilon_2}{\mpl r_{\rm g}},\quad
    \ell_b \sim \frac{b_1^2\epsilon_2^2}{16\pi^2(\mpl r_{\rm g})^2}.
    \label{eq: sGB loop BH estimate}
\end{align}
The classical EFT condition $|b_1|\epsilon_2\ll1$, together with the semiclassical condition $\mpl r_{\rm g}\gg1$, therefore strongly suppresses this quantum correction. The formal scale $g_b(E)\sim1$ is $(\mpl\Lambda^2/|b_1|)^{1/3}$. As emphasized in Ref.~\cite{Serra:2022pzl}, this strong-coupling scale need not coincide with the EFT cutoff. Other interactions, including GR vertices, and the background-dependent kinetic normalization must also be considered when assessing the full theory.

\subsection{IR causality}\label{sec: validity IR causality}

IR causality can impose further restrictions by excluding resolvable time advances within the regime of validity of the EFT~\cite{Camanho:2014apa,deRham:2020zyh,Chen:2021bvg,deRham:2021bll,Serra:2022pzl,Nie:2024pby}. A related condition on QNM decay times was proposed for pure gravity EFTs in Ref.~\cite{Melville:2024zjq}, and its implications under standard EFT power counting were examined in Ref.~\cite{Juliano:2026stability}. Applying these ideas to the coupled scalar and gravitational perturbations studied here requires an analysis of the coupled propagation channels and their scattering time delays. The frequency shifts calculated here alone do not establish such causality bounds.

\bibliographystyle{apsrev}
\bibliography{ref_20261129_1730}

\end{document}